\documentclass[oneside,12pt]{Classes/RoboticsLaTeX}

\usepackage{adjustbox}
\usepackage{xcolor, colortbl}
\definecolor{Gray}{rgb}{0.8,0.8,0.8}
\usepackage{tabularx,ragged2e}
\newcolumntype{C}{>{\Centering\arraybackslash}X} 
\usepackage{numprint}
\usepackage[en-GB,showdow=false]{datetime2}
\DTMlangsetup[en-GB]{ord=raise,monthyearsep={,\space}}
\DTMsavedate{date}{2015-03-12}
\usepackage{amsmath}
\usepackage{amsthm}
\usepackage{amsfonts}
\usepackage{algorithm}
\usepackage{algorithmic}
\usepackage{multirow}
\usepackage{colortbl}
\usepackage{color}
\usepackage{epigraph}
\usepackage{graphicx}
\usepackage{caption}
\usepackage{subcaption}
\usepackage{hyperref}
\usepackage{tabularx}
\usepackage{float}
\usepackage{longtable}
\usepackage[pdftex]{graphicx}
\usepackage{pdfpages}
\usepackage{pdflscape}
\usepackage[acronym,toc]{glossaries}
\usepackage{setspace}

\usepackage[T1]{fontenc}
\usepackage{rotating}
\usepackage{pdfpages}
\title{\Large{Semantic, Efficient, and Secure Search over Encrypted Cloud Data}}

  \author{\textbf{Fateh Boucenna}}
  \collegeordept{\small{Department of Computer Science}}
  \university{\small{University of Science and Technology, Houari Boumediene (Algeria)}}
  \crest{\includegraphics[width=30mm]{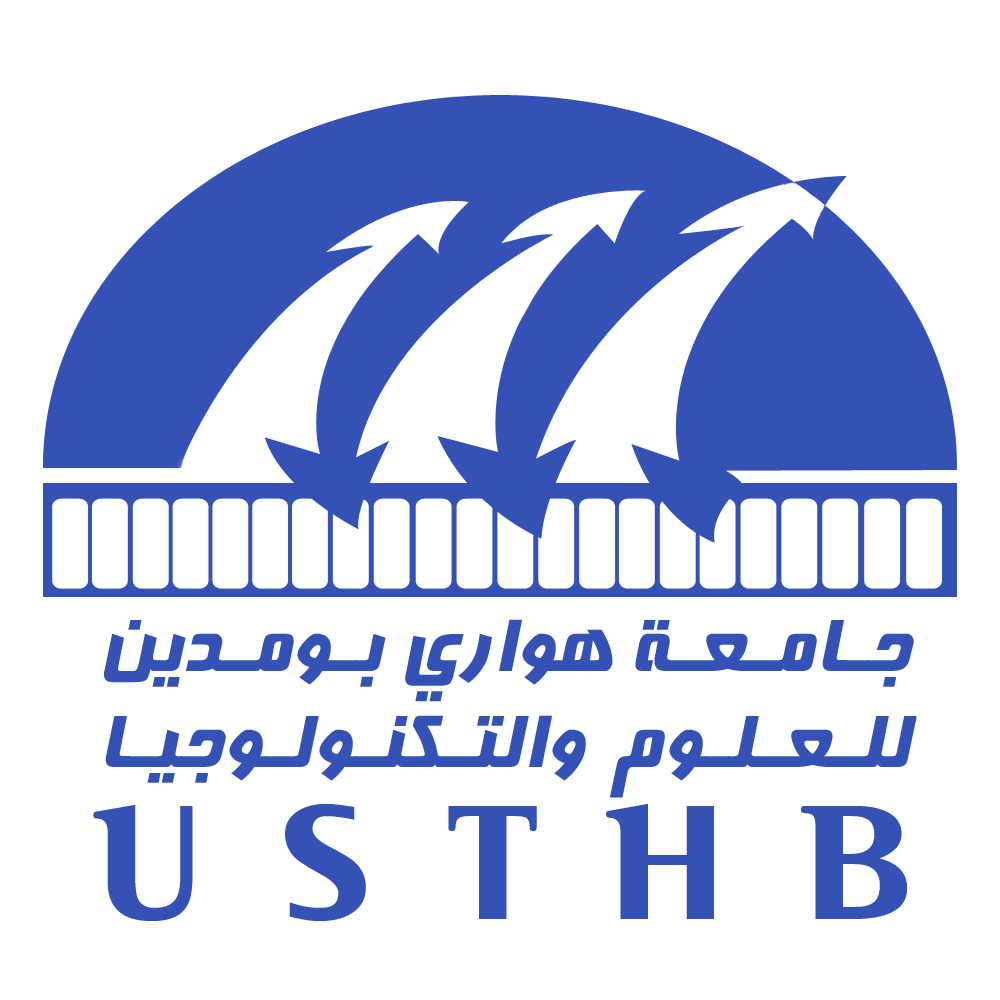}}

\supervisor{Samir Kechid and Omar Nouali}

\renewcommand{\submittedtext}{}
\degree{A thesis submitted for the degree of Doctor of Philosophy}
\degreedate{January 15, 2020}

\makeglossaries

\newacronym{rnn}{RNN}{Recurrent Neural Network}
 
\newacronym{gmm}{GMM}{Gaussian Mixture Model}

\newacronym{hri}{HRI}{Human Robot Interaction}

\newacronym{uav}{UAV}{Unmanned Aerial Vehicle}

\newacronym{imu}{IMU}{Inertial Measurement Unit}

\newacronym{mem}{MEM}{Micro-Electro-Mechanics}

\newacronym{pca}{PCA}{Principal Component Analysis}

\newacronym{dtw}{DTW}{Dynamic Time Warping}

\newacronym{lstm}{LSTM}{Long short-term memory}

\newacronym{narx}{NARX}{Nonlinear AutoRegressive network with eXogenous inputs}

\newacronym{nn}{NN}{Neural Network}

\begin{document}


\includepdf[fitpaper=true, pages=-]{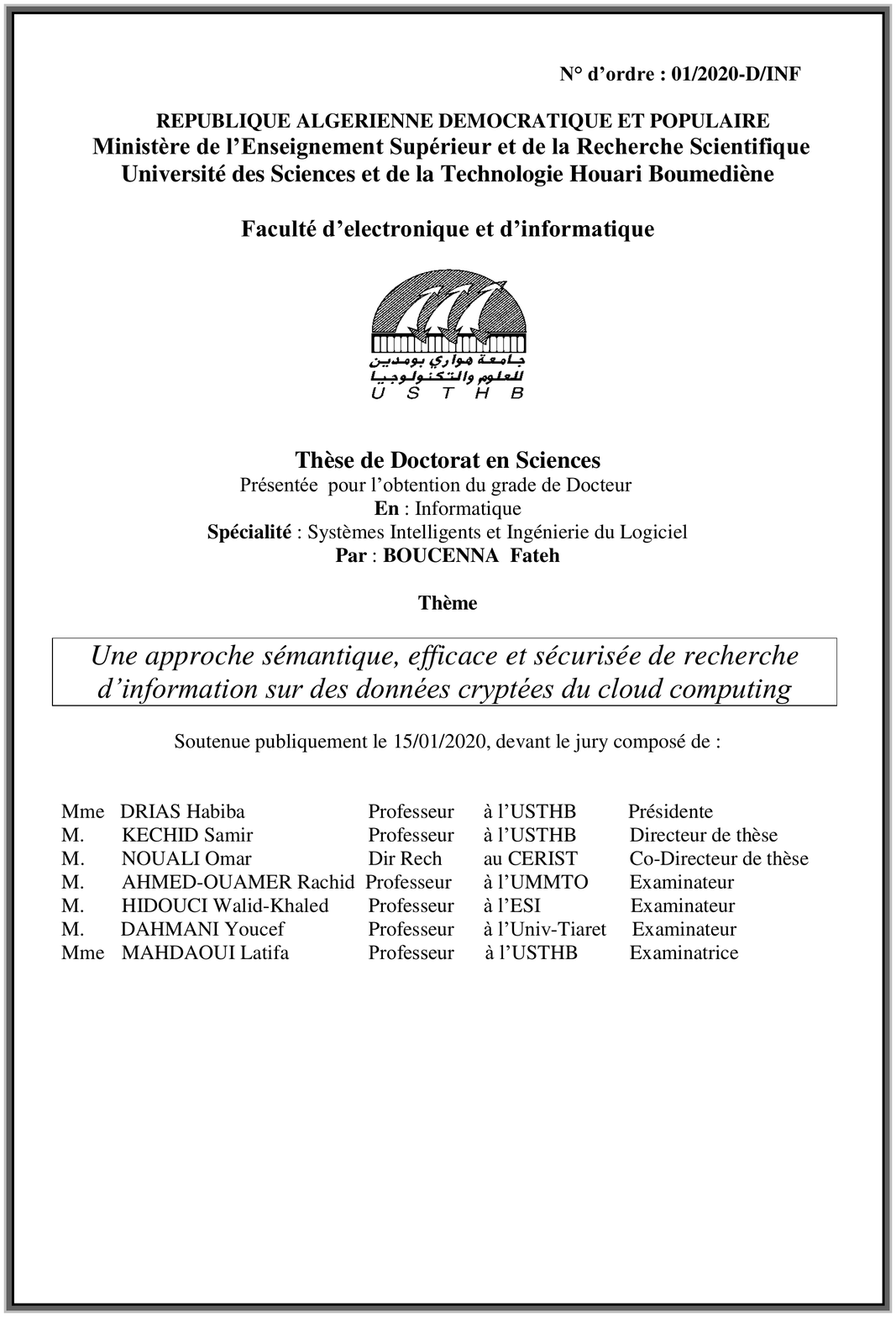}

\begin{spacing}{1}
\maketitle
\end{spacing}


\setcounter{secnumdepth}{3}
\setcounter{tocdepth}{3}

\frontmatter


\begin{acknowledgements}

First of all, I would like to express my gratitude to my thesis supervisors Dr. \textit{Omar NOUALI} and Prof. \textit{Samir KECHID}, for their availability, their guidance, and their valuable pieces of advice.

Exceptional recognition belongs to Prof. \textit{Habiba DRIAS} for agreeing to chair the jury of this thesis.

I wish to express my deepest gratitude to the rest of my thesis committee: Prof. \textit{Latifa MAHDAOUI}, Prof. \textit{Rachid AHMED-OUAMER}, Prof. \textit{Walid-Khaled HIDOUCI}, and Prof. \textit{Youcef DAHMANI}, for kindly agreeing to judge this work.

My sincere gratefulness also goes to Prof. \textit{Kamel ADI}, and Prof. \textit{Tahar KECHADI}, who allowed me to collaborate with them as a guest researcher.

Finally, I would like to thank my family, my friends, and my CERIST colleagues for their help and encouragement.

\end{acknowledgements}

\includepdf[fitpaper=true, pages=-]{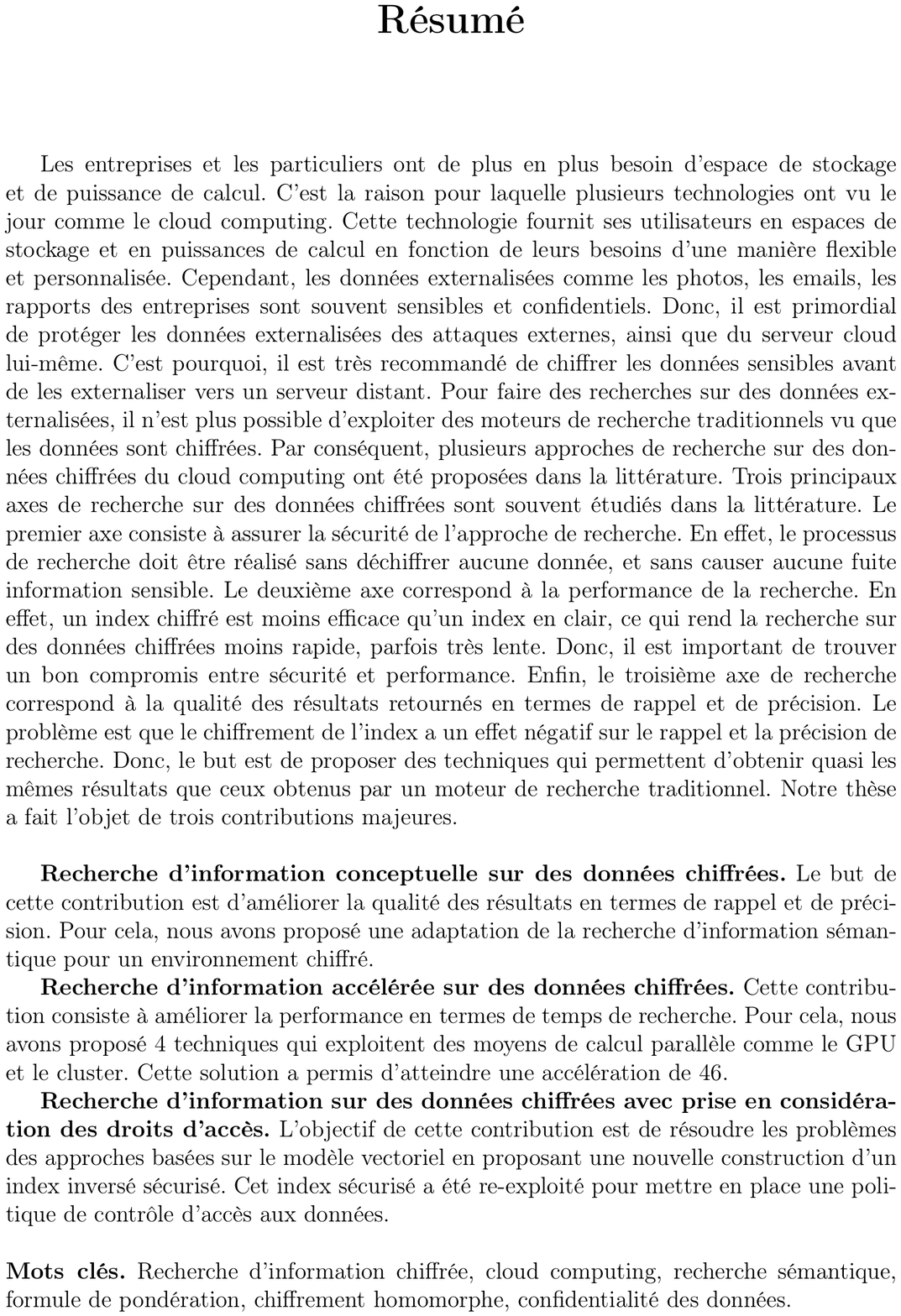}
\include{Abstract/resume}


\begin{abstractslong}

Companies and individuals demand more and more storage space and computing power. For this purpose, several new technologies have been designed and implemented, such as the cloud computing. This technology provides its users with storage space and computing power according to their needs in a flexible and personalized way. However, the outsourced data such as emails, electronic health records, and company reports are sensitive and confidential. Therefore, It is primordial to protect the outsourced data against possible external attacks and the cloud server itself. That is why it is highly recommended to encrypt the sensitive data before being outsourced to a remote server. To perform searches over outsourced data, it is no longer possible to exploit traditional search engines given that these data are encrypted. Consequently, lots of searchable encryption (SE) schemes have been proposed in the literature. Three major research axes of searchable encryption area have been studied in the literature. The first axis consists in ensuring the security of the search approach. Indeed, the search process should be performed without decryption any data and without causing any sensitive information leakage. The second axis consists in studying the search performance. In fact, the encrypted indexes are less efficient than the plaintext indexes, which makes the searchable encryption schemes very slow in practice. More the approach is secure, less it is efficient, thus, the challenge consists in finding the best compromise between security and performance. Finally, the third research axis consists in the quality of the returned results in terms of relevance and recall. The problem is that the encryption of the index causes the degradation of the recall and the precision. Therefore, the goal is to propose a technique that is able to obtain almost the same result obtained in the traditional search.

Three major contributions are proposed in our thesis.

\begin{enumerate}
    \item \textbf{Concept-based Semantic Search over Encrypted Cloud Data.} The aim of this contribution is to improve the quality of the search result in terms of recall and precision. For this, we propose an adaptation of the semantic search for an encrypted environment \cite{boucenna2016concept}.
    \item \textbf{Accelerated search over encrypted cloud data.} This contribution consists in improving the scheme performance in terms of search time. For this, we propose 4 techniques that exploit high performance computing (HPC) architectures, such as a GPU and a cluster. This solution achieves an acceleration of 46x \cite{boucenna2017accelerated}.
    \item \textbf{Secure Inverted Index Based Search over Encrypted Cloud Data with User Access Rights Management.} The aim of this contribution is to solve the problems of the approaches based on the vector space model by proposing a new construction of a secure inverted index. This secure index is also used to set up a data access control policy \cite{boucenna2019secure}.
\end{enumerate}

\textbf{Keywords.} Searchable encryption, cloud computing, semantic search, homomorphic encryption, data confidentiality, weighting formula.

\end{abstractslong}

\tableofcontents
\listoffigures
\listoftables

\mainmatter

\chapter{Introduction}
\label{chap:introduction}
\ifpdf
    \graphicspath{{Introduction/Figures/PNG/}{Introduction/Figures/PDF/}{Introduction/Figures/}}
\else
    \graphicspath{{Introduction/Figures/EPS/}{Introduction/Figures/}}
\fi




\section{Motivations}

Nowadays, companies and individuals demand more and more storage space for their data, and computing power for their applications. For this purpose, several new technologies have been designed and implemented, such as the cloud computing. The latter provides its users with storage space and computing power according to their needs in a flexible and personalized way. However, the outsourced data such as emails, health records, and company reports are sensitive and confidential. Therefore, It is primordial to protect the outsourced data against possible external attacks and the cloud server itself. For this reason, it is highly recommended to encrypt the sensitive data before being outsourced to a remote server.

To perform a search over these data, it is no longer possible to exploit traditional search engines given that the outsourced data are encrypted. Consequently, lots of searchable encryption (SE) schemes have been proposed in the literature \cite{wang:privacy, cao2014privacy, li:efficient, xia2016secure}. The common point between these approaches is that the data owner starts by encrypting the data collection and the generated index before outsourcing them into the cloud. Upon receiving a trapdoor (encrypted query), the cloud server performs a search through the encrypted index in order to retrieve and return a list of top-$k$ documents to the user. Furthermore, the search process should be performed without decrypting any data and without causing any sensitive information leakage.

\section{Context of the Study}

The first work \cite{song:practical} that has been proposed for SE in the literature only support single keyword search which is performed directly on the encrypted data without using any index. However, the encryption of data and queries is deterministic which reduces the security of the scheme. After that, several works \cite{cao2014privacy, curtmola:searchable, kamara2013parallel} have proposed SE schemes based on different kinds of secure index structures (binary tree, vector representation, inverted index) which are encrypted in different ways. A secure index allows to improve the search performance, and the security of the scheme since the data collection is no longer exploited during the search process (the encrypted index is used instead).

Three major research axes of searchable encryption area have been studied in the literature. The first axis consists in ensuring the security of the search approach. Indeed, the data collections, the indexes, and the users' queries should be encrypted in an effective manner (not necessarily in the same way). In addition, the search process should not leak any sensitive information about the content of documents and queries. The second axis consists in studying the search performance. In fact, the encrypted indexes are less efficient to process than the indexes in clear, which makes the searchable encryption schemes very slow in practice. Generally, more the approach is secure, less it is efficient. Therefore, the challenge consists in finding the best compromise between security and performance. Finally, the third research axis consists in the quality of the returned results in terms of relevance and recall. The problem is that the encryption of the index causes the degradation of the recall and the precision. Therefore, the goal is to propose a technique that is able to obtain almost the same result obtained by the search through an unencrypted index.

\section{Objectives and Contributions}

In this thesis, we deal with the three research axes of the searchable encryption area. Our first contribution consists in proposing an approach that allows to improve the quality of the search result in terms of recall and precision. The proposed approach allows adapting the concept-based search to deal with encrypted data. For this purpose, we build from Wikipedia an ontology which is responsible for the mapping operation between terms and concepts in order to provide a search which is based on the meaning of documents and queries. Then, we propose a new weighting formula that enables to perform a concept-based search through an encrypted index, this new formula is called the double score weighting (DSW) formula. The conducted experimental study shows that our proposed scheme brings 60\% of improvement compared to one of the most cited approaches in the literature \cite{cao2014privacy}.

Our second contribution consists in reducing the search time when the vector space model is applied.  This model is the most exploited in the searchable encryption area for security reasons. However, an approach that is based on the vector model is a time consuming process, which makes it impractical. In fact, in the vector space model, each document and query is represented by an encrypted vector\footnote{The S$K$NN method is used to encrypt the vectors.} of size equal to the total number of terms belonging to the data collection. During the search process, the query vector should be compared (by calculating the similarity\footnote{The similarity score corresponds to the scalar product between the document vector and the query vector.} scores) with each document of the data collection. Unfortunately, this makes the search process very slow as demonstrated by our experimental study (it may last for hours). To overcome this problem, we propose 4 solutions, where each of them exploits a different high performance computing (HPC) architecture (multi-core processor, GPU, computer cluster). Each solution is able to deal with several queries at the same time (which means that multiple searches can be performed simultaneously). Furthermore, the similarity scores between a set of documents and a received query are calculated in a parallel way (which means that even one search operation is performed in a parallel manner). Our experimental study shows that the proposed solution can reach an acceleration around a factor of 46.

Our third contribution consists in solving the problems of the SE schemes that are based on the vector model, namely, their inefficiency in practice, and their inability of updating the encrypted indexes. These problems are even more pronounced, since, for security reasons, the vector space model is the most exploited in the searchable encryption area. Therefore, to solve these problems we propose a new construction of a secure inverted index, which provides the most appropriate and efficient way to perform a search. Our solution consists in building a secure inverted index from an unencrypted inverted index by using two techniques, namely, homomorphic encryption, and the dummy documents technique. Homomorphic encryption is used to encrypt the scores within the index, whereas, the dummy documents technique consists in adding random document IDs to each entry of the index to hide the relationship between documents and concepts (each entry of the index corresponds to a concept that points to a set of documents in which it belongs). Nevertheless, these two techniques present two major problems. First, homomorphic encryption has the disadvantage of generating encrypted indexes which are three orders of magnitude larger than the original ones. The second problem is that lots of false positives are returned with the search results because of the dummy documents technique which decreases the accuracy. To solve the first problem, we propose a new technique that allows to compress the secure inverted index. The experimental study shows that this technique allows to decrease the size of the secure index from 161.45 TB to 712.54 GB. Then, to solve the second problem, we propose a similarity computation formula that we call the double score formula. The latter allows to increase the accuracy from 22.45\% to 38.76\%. Then, in order to establish a data access control policy, we exploit a second secure inverted index whose entries correspond to the users' IDs, and each entry leads to the documents accessible by the user corresponding to that entry, in addition to some dummy documents. During the search process, the second index is used to narrow the search space by selecting only documents that the user has the right to access, and the first index is exploited to retrieve the relevant result.

\section{Thesis Organization}

The rest of this thesis is organized as follows. In chapter \ref{Chapter1}, we start by presenting the basic notions of classical information retrieval (IR), with particular focus on the IR process, the IR models, and the semantic IR. In chapter \ref{Chapter7}, we briefly present the basic notions of cloud computing, namely, the essential characteristics, the service models, and the deployment models. Then, we discuss some important security aspects of cloud computing. In the following chapter, we introduce the searchable encryption area. Specifically, we explain the general architecture of a SE scheme, the security notions, and a detailed state of the art. In chapter \ref{Chapter3}, we present different cryptographic methods that are often exploited in the SE area. Particularly, we present the S$K$NN method which is very robust to encrypt vectors. Then, we present the CP-ABE method which allows establishing a data access policy. Finally, we present different variants of homomorphic encryption. In chapter \ref{Chapter4}, we present our first contribution which consists in proposing a semantic search approach on encrypted cloud data. In the following chapter, we present our second contribution that consists in proposing four different techniques that allow to speed up the search process around a factor of 46. Finally, our third contribution is presented in chapter \ref{Chapter6}. In this contribution we propose a new construction of a secure inverted index that allows to perform searches in an appropriate and effective way, and to implement an access control policy over encrypted data. We conclude in chapter \ref{chap:conclusions} by presenting a summary of this thesis and discussing our perspectives.


\chapter{Information Retrieval} 

\label{Chapter1} 


\newtheorem{defn}{Definition}

\section{Motivation}
The recent advancements of data storage technology, the emergence of cloud computing, and the evolution of computer networks have allowed considerable amounts of data to be stored by users and organizations. This evolution has challenged the information retrieval (IR) community to optimize the indexing and search processes by proposing new models more elaborate \cite{sharma2013survey}. 

An information retrieval system (IRS) is an interface between a user and a data collection. Its purpose is to return to the user the most relevant documents in response to a query in an acceptable time frame. For this purpose, two processes must be implemented, namely, the indexing process and the matching process.

The indexing process consists in constructing a representation of documents and queries that allows to optimize the search process time, whereas, the matching process consists in calculating the similarity scores between documents and a query in order to find and sort the most relevant documents.

We present in this chapter the basics of information retrieval. For this, we start by explaining the IR process. Then, we introduce some IR models. After that, we describe the metrics used to the evaluation of an IRS. Finally, we talk about the semantic IR.


\section{Basic IR Concepts}

The information retrieval (IR) is a set of processes, models, and techniques that can be used to select from a data collection, the documents that are considered relevant to a query sent by a data user.

Several important concepts revolve around this definition \cite{baeza2011modern}.

\begin{itemize}
    \item \emph{Data collection.} It corresponds to the set of all documents accessible by the IRS.
    \item \emph{Document.} It is the basic information in a data collection. It can be stored in different formats.
    \item \emph{Query.} It corresponds to the expression of the user's need. A query is the intermediary between the data user and the IRS. It can be expressed by a Boolean expression, natural language, or a graphic.
    \item \emph{Relevance.} It represents the association degree between a given document and a query. 
\end{itemize}

\section{IR Process}

\begin{figure}
\centering
\includegraphics[width=\textwidth,height=\textheight,keepaspectratio]{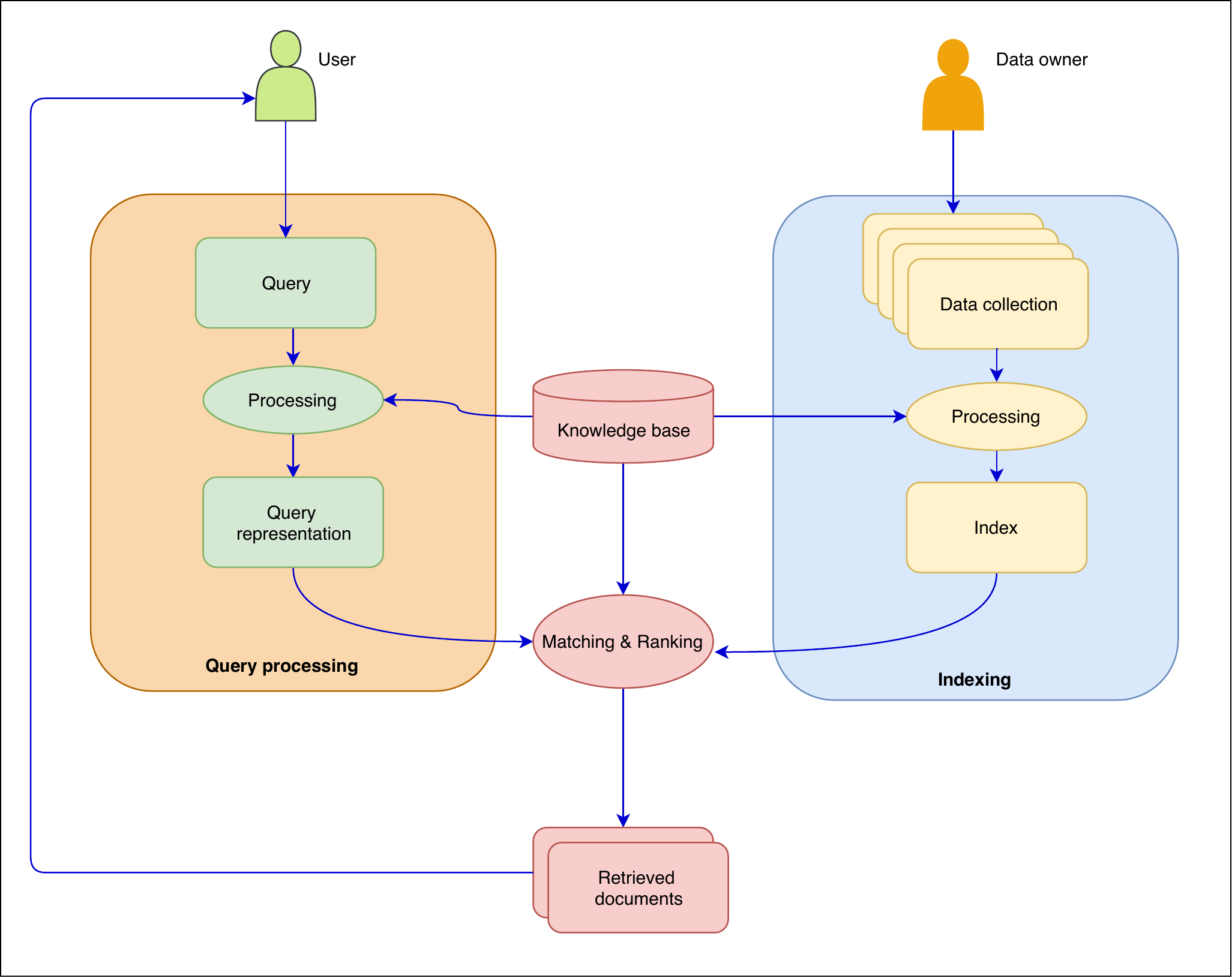}
\decoRule
\caption{Information retrieval process}
\label{fig:irprocess}
\end{figure}

The purpose of the IR process is to deal with the user's need expressed by a query. For this, a set of ranked documents deemed relevant by the IRS should be returned to the user. Figure \ref{fig:irprocess} shows that an IR process is composed of several tasks, such as the indexing, the query processing, the matching, and the ranking. Other tasks such as the query reformulation and the index update may be included in the IR process \cite{singhal2001modern}.

\subsection{Indexing}

The goal of the indexing process is to reduce the time needed to perform a search operation. For this purpose, each document should have a descriptor that contains the most important keywords of the document with associated weights. The indexing process consists of five basic steps: tokenization, dropping stop words, lemmatization, term weighting, and index creation \cite{mogotsi2010christopher}.

\subsubsection{Tokenization}

It consists in the construction of a list of all terms of a document. For this, the spaces, the punctuation, and possibly the numbers should be recognized and ignored \cite{mcnamee2004character}.

\subsubsection{Dropping Stop Words}

The stop words are the terms of a document that are used to present a subject. They are non-thematic words, such as "because" and "through". The stop words can also be personal pronouns, prepositions, etc.

This step consists in dropping all the stop words belonging to a document. For that, a list that contains the set of stop words should be used. This list is called anti-dictionary \cite{wilbur1992automatic}.

\subsubsection{Stemming and Lemmatization}

In the same document, we may encounter a term represented in different forms and having very similar meanings, such as "develop", "develops", and "developer". In order to reduce the size of the index and improve the recall, the terms are replaced by their roots during the indexing process \cite{korenius2004stemming}.

\subsubsection{Term Weighting}\label{term_weighting}

It is one of the most important functions of the indexing process. It consists in assigning a score to each term of a document. Most of the formulas that have been proposed in the literature are based on two factors, namely, the term frequency (TF) and the inverse of document frequency (IDF) \cite{aizawa2003information}. The first one reflects the importance of a term in a document, whereas, the second one represents the importance of a term in a data collection. The combination of these two factors gives a good approximation of the importance of a term in a document of a data collection.

\subsubsection{Index Creation}

In order to optimize the search time, it is essential to save the information previously collected in a particular storage structure. Several structures have been proposed in the literature, such as the inverted index, the suffix tree, and the signature file \cite{baeza2011modern}.

The inverted index is the most exploited storage structure in the literature. It is composed of a set of entries, where each entry corresponds to a term in the data collection. Each entry leads to a set of documents that contain the term corresponding to that entry \cite{croft2010search}.

\subsection{Query Processing}

When a data user formulates his information need under a query, some processing should be made in the same way as the indexing process. For this purpose, the tokenization, the stop words deletion, the lemmatization, and the term weighting tasks should be applied to the user query \cite{alshari2015semantic}.

As the index, the user's query can have different structures. The simplest one which is adapted to the inverted index consists in a set of couples (term, weight). For instance, the query "what is the biggest country" can be represented by the following structure after processing \{(big, $w_1$), (country, $w_2$)\} where $w_1$ and $w_2$ are the weights of the terms "big" and "country", respectively. These weights depend on the applied weighting formula (see Section \ref{term_weighting}). 

\subsection{Matching}

This process consists in attributing a similarity score to each document of the data collection (when using an inverted index, just the documents that contain at least one query term are concerned, the others are ignored). These scores are calculated by applying a specific formula that depends on the model utilized by the IRS (see Section \ref{irmodels}). The similarity score represents the relevant degree of a document with respect to a user's query \cite{nomula2010system}.

\subsection{Ranking}

The relevant documents should be selected and sorted based on the similarity scores which are attributed in the matching process. This step is missing in few IR models, such as the Boolean model where a document is whether relevant or not relevant. The ranking is a very important task because of its ability to place the most relevant documents on the top of the search result (for instance, in the first page). Current approaches try to exploit external knowledge bases in order to appropriately sort the search result \cite{ismalon2014techniques}.

\section{IR Models}\label{irmodels}

\begin{figure}
\centering
\includegraphics[width=\textwidth,height=\textheight,keepaspectratio]{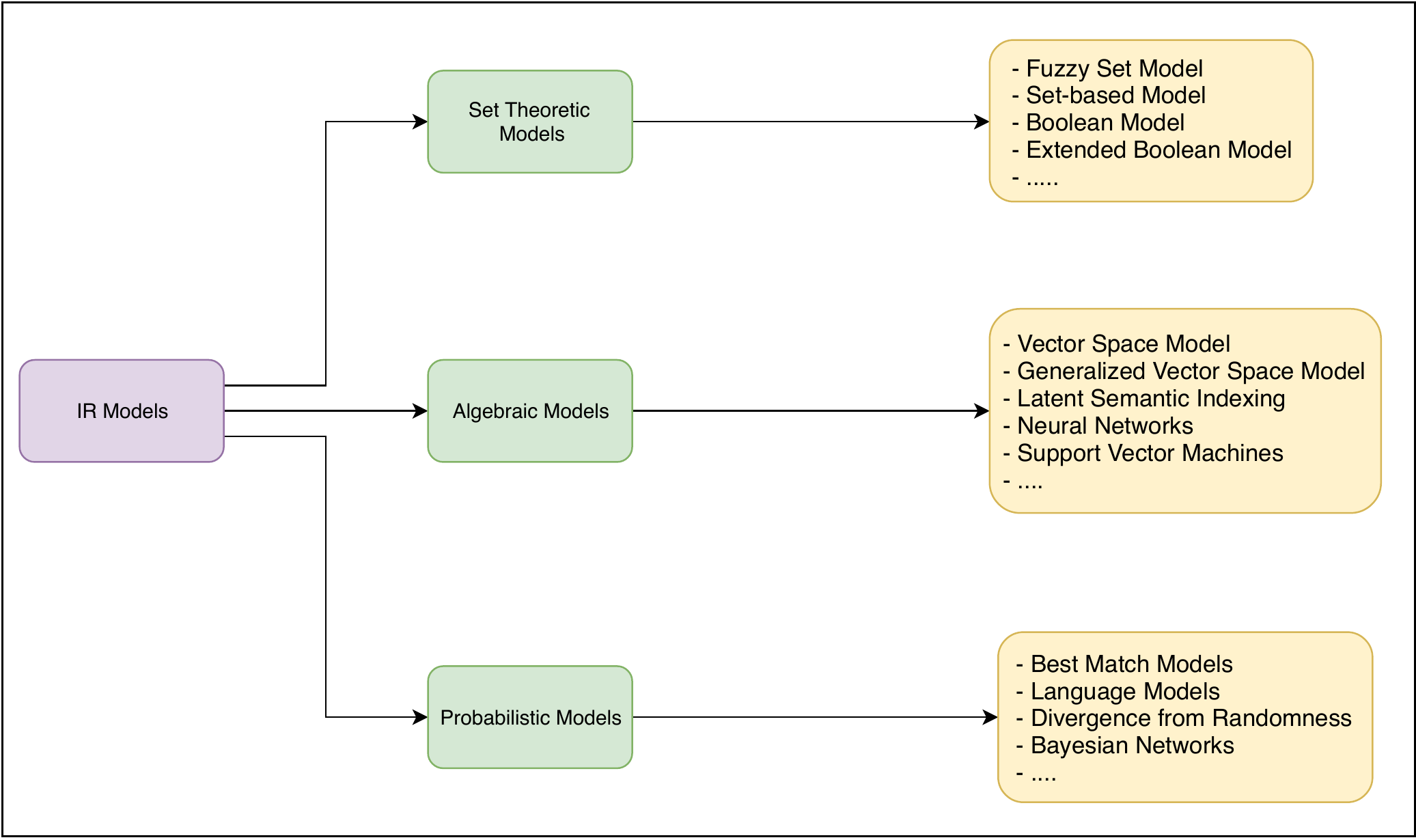}
\decoRule
\caption{Information retrieval models}
\label{fig:irmodels}
\end{figure}

An IR model is composed of two fundamental functions. The first one is the representation of documents and queries which corresponds to the indexing and the query processing steps of the IR process, whereas, the second function is the document-query matching that corresponds to the matching and the ranking steps of the IR process.

Several models have been proposed in the literature \cite{baeza2011modern}. These models are divided into three main categories, namely, the set theoretic models, the algebraic models, and the probabilistic models (see Figure \ref{fig:irmodels}). In the following, one model is presented for each category, namely, the Boolean model, the vector model, and the language model, respectively.

\subsection{Boolean Model}

It is the first IR model that has been proposed in the literature \cite{baeza2011modern}. The Boolean model has been used in the library management systems. In this model, a query is represented as a Boolean expression and its terms are connected by the logical connectors ($\wedge$, $\vee$ and $\neg$). The Boolean model considers that each document is either relevant or non-relevant with regard to a given query. The correspondence score ($CS$) between a document $D$ and a query is calculated by the following formula. 

\begin{equation}
    \begin{cases}
        & CS(D,t) = 1 \text{ if }  t \in D \text{; } 0 \text{ otherwise}\\
        & CS(D, q_i \wedge q_j) = 1 \text{ if } CS(D, q_i) = 1 \text{ and } CS(D, q_j) = 1 \text{; } 0 \text{ otherwise} \\
        & CS(D, q_i \vee q_j) = 1 \text{ if } CS(D, q_i) = 1 \text{ or } CS(D, q_j) = 1 \text{; } 0 \text{ otherwise} \\
        & CS(D, \neg q_i) = 1 \text{ if } CS(D, q_i) = 0 \text{; } 0 \text{ otherwise} \\ 
    \end{cases}
\end{equation}

Where $t$ is a query term, $q_i$ and $q_j$ are Boolean expressions. 

The advantage of the Boolean model is its simplicity and its ability to perform a precise search for an experienced user. Its disadvantage is the missing of a term weighting notion which arises an issue concerning the ranking of documents.

\subsection{Vector Model}

The vector model \cite{salton1975vector} is one of the most  IR models that have been studied in the literature. In this model, the queries and documents are represented in a $n$-dimensional vector space, where $n$ is the total number of the indexing terms. Each document (or query) is represented by a vector of weighted terms. The search process consists in finding the nearest document vectors to the query vector. For this purpose, many similarity functions have been proposed in the literature, such as the dot product, the Jaccard distance, and the cosine similarity \cite{choi2010survey}. The similarity score between a document $D$ and a query $Q$ when using the dot product is calculated as follows. 

\begin{equation}
    Sim(D, Q) = \sum_{i=1}^{n} d_i \times q_i
\end{equation}
Where $d_i$ (resp. $q_i$) is the $i^{\text{eme}}$ dimension of the vector $D$ (resp. $Q$), and $n$ is the dimension of the vector space.

Unlike the Boolean model, the vector model is able to rank the search result based on the similarity measure. Its disadvantage is the independence of the index terms which causes the reduction of the recall.

\subsection{Language Model}

The language model is a probabilistic model that consists in assigning a score to each document during the search process by calculating the probability that a user's query is inferred from that document \cite{ponte1998language}. This model is based on the assumption that the user formulates a query while thinking of few documents that he needs. In the language model, each document is represented by a query that has been inferred from that document. When receiving a query, a score is assigned to each document. This score represents the similarity degree between the received query and that inferred from the document. Finally, the results are sorted according to these scores.

\section{IR Evaluation}

\begin{figure}
\centering
\includegraphics[width=5cm,height=9cm]{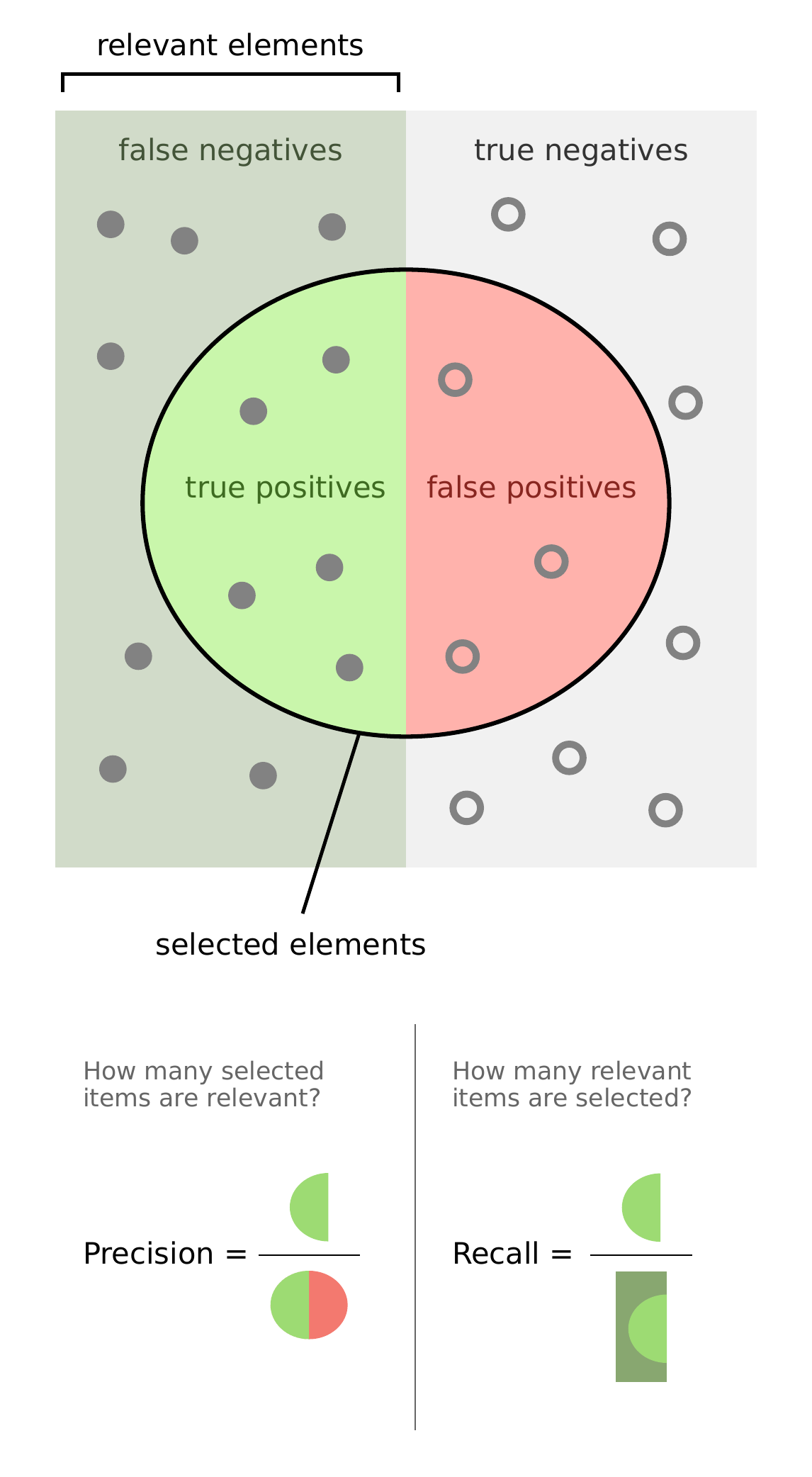}
\decoRule
\caption[Accuracy and recall]{Precision and recall \cite{wiki:xxx}}
\label{fig:precision}
\end{figure}

The evaluation of an IRS is a very important step, since it allows to validate the proposed model by comparing it with other models that have been proposed in the literature \cite{baeza2011modern}. There are two main metrics that are usually used to evaluate an IRS, namely, the recall and the accuracy. Moreover, other metrics can also be exploited such as the index size, the search time, and the security\footnote{The security of a scheme should be guaranteed when the index is outsourced to an external server such as a cloud computing.} of the scheme.

The accuracy (precision) of an IRS corresponds to its ability to reject the irrelevant documents during the search process (see Figure \ref{fig:precision}). It is calculated by the following formula.

\begin{equation}
    \text{Accuracy} = \frac{\text{true positives}}{\text{true positives} + \text{false positives}}
\end{equation}

Where "true positives" is the number of relevant documents that are retrieved by the IRS, and "false positives" is the number of non-relevant documents that are returned by the IRS  

The recall of an IRS corresponds to its ability to retrieve the relevant documents for a given query (see Figure \ref{fig:precision}). It can be calculated as follows. 

\begin{equation}
    \text{Recall} = \frac{\text{true positives}}{\text{true positives} + \text{false negatives}}
\end{equation}
Where, "false negatives" is the number of relevant documents that are rejected by the IRS.

\section{Semantic IR}
\begin{figure}
\centering
\includegraphics[width=12cm,keepaspectratio]{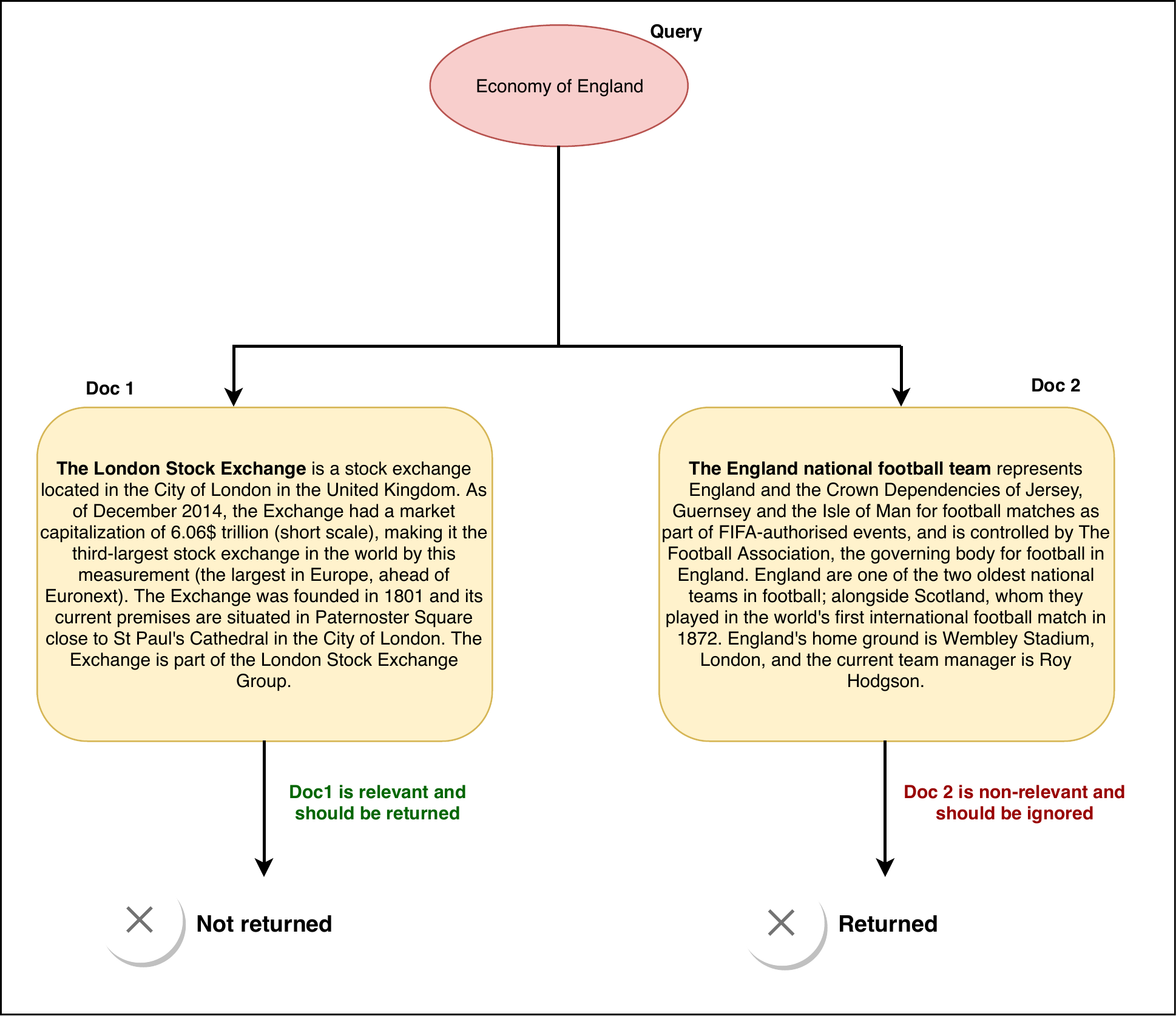}
\decoRule
\caption{The drawbacks of the keyword based search}
\label{fig:keyword_search}
\end{figure}

Classical information retrieval systems perform a keyword based search. Indeed, during the search process, when the IRS receives a query, it tries to find the documents that contain the query terms. The keyword based search has two major drawbacks. The first one is that the documents which do not have any query terms are ignored even if they may be relevant. The second drawback is that the sense of the queries is not taken into consideration even if certain terms may have several meanings. The first drawback decreases the recall of the IRS, whereas, the second one causes the reduction of the accuracy. To illustrate the problem, let us take the example of Figure \ref{fig:keyword_search}.

If a user sends the query "Economy of England", the server will search for the documents that contain the terms "Economy" and/or "England" in the data collection. The server will ignore "Doc 1" because it does not contain any query terms. Contrariwise, it will return the second document that contains the term "England". Nevertheless, if we analyze the content of the two documents, we notice that the first document is relevant, since its meaning is close to that of the query, given that it talks about the London stock exchange which is strongly related to the economy of England, contrary to "Doc 2" that talks about football and hence is supposed to be non-relevant even if it has terms in common with the query.

In order to solve the problem encountered in the syntactic search. The IR community has turned to the use of external knowledge bases such as thesauri and ontologies in order to understand the meaning of documents and queries \cite{muller2004textpresso, castells2007adaptation, fernandez2011semantically}. The goal is to improve the precision and the recall of the search by returning documents that have a meaning close to that of the query rather than relying on the syntax. This area of research is called concept based information retrieval.

\subsection{Basic Notions}

In the following we give some important definitions.

\subsubsection{Term, Concept, and Object}
\begin{figure}
\centering
\includegraphics[width=10cm,keepaspectratio]{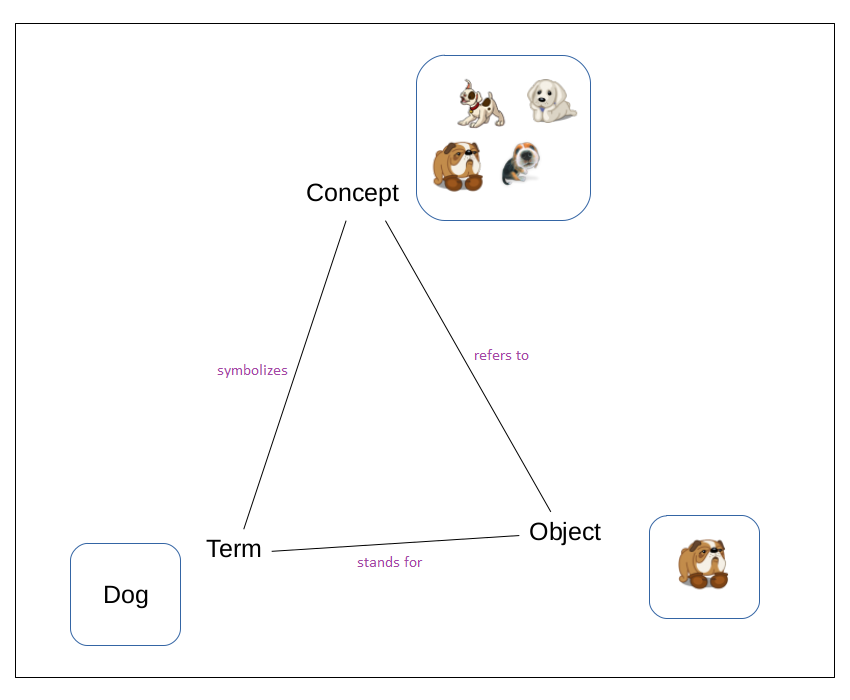}
\decoRule
\caption{Semiotic triangle}
\label{fig:concept}
\end{figure}

A concept is an idea (e.g., sport) grouping in the same category, objects that are semantically close to each other (e.g., football, boxing, golf), whereas, a term is a linguistic representation of a concept or an object. Figure \ref{fig:concept} illustrates the relationship between a term, an object, and a concept using a semiotic triangle \cite{ogden1923meaning}.

\subsubsection{What is Concept Based IR}

Concept based information retrieval is an alternative of classical IR where semantic concepts are used to represent documents and queries instead of using bags of terms during the indexing and matching processes. Its goal is to perform a semantic search rather than relying on the syntax. For this purpose, an external knowledge base should be exploited in order to map between terms and concepts. The exploited knowledge source and the mapping process differ from one approach to another.

\subsection{Concept Based Information Retrieval}

Several kinds of concept based search approaches have been proposed in the literature. Wikipedia ontology and WordNet are the most exploited knowledge bases by those approaches.

\subsubsection{Concept Based IR Guided by Wikipedia}\label{sec_wiki_IR}

In these schemes, a Wikipedia ontology is exploited to map between terms and documents. The construction of an ontology from Wikipedia\footnote{\url{https://en.wikipedia.org/wiki/Main_Page}, Feb. 2019.} can be performed as follows \cite{gabrilovich2006overcoming}.

\begin{enumerate}
\item Each Wikipedia page $p_i$ corresponds to a concept $c_i$.
\item Each concept $c_i$ is represented by a vector of terms $v_i = \{(t_1, w_{i1}), (t_2, w_{i2}), ...,\\ (t_n, w_{in})\}$ extracted from the corresponding Wikipedia page. These terms are weighted by applying TFIDF formula. 

The weight $w_{ij}$ of a term $t_j$ in the vector $v_i$ corresponds to the association degree between the term $t_j$ and the concept $c_i$.
\item In order to accelerate the mapping process, an inverted index $I_{wiki}$ is constructed where each term $t_i$ is represented by a set of concepts $v'_i$ to which it belongs, $v'_i$ = \{($c_1$, $s_{i1}$), ($c_2$, $s_{i2}$), ..., ($c_m$, $s_{im}$)\}.

Where $s_{ij}$ is the association score between the term $t_j$ and the concept $c_i$.
\item The inverted index $I_{wiki}$ = \{$v'_1$, $v'_2$, $v'_3$, ...,$v'_n$\} that is composed of the set of concept vectors corresponds to the Wikipedia ontology.
\end{enumerate}

Before calculating the similarity between a document and a query, each of them should be represented by a vector of concepts as follows.

\begin{enumerate}
\item  A vector of terms $d_i$ = \{($t_1$, $w'_{i1}$), ($t_2$, $w'_{i2}$), ..., ($t_n$, $w'_{in}$)\} is constructed for each document and query using TFIDF formula. Note that $w'_{ik}$ corresponds to the weight of a term $t_k$ in the document $d_i$
\item From the vector $d_i$, a vector of concepts $d'_i = \{(c_1, s'_{i1}), (c_2, s'_{i2}), ..., (c_m, s'_{im})\}$ is calculated by mapping between terms and concepts through the Wikipedia ontology.

The score $s'_{ij}$ assigned to a concept $c_j$ of the vector $d'_i$ is calculated by the following formula.
\begin{equation}\label{eqTFIDF}
    s'_{ij} = \underset{t_k \in d_i}{\sum} w'_{ik} . s_{kj}
\end{equation}
\item Finally, the similarity score between a document and a query is calculated by applying the scalar product.
\end{enumerate}

\subsubsection{Concept Based IR Guided by WordNet}

WordNet\footnote{\url{https://wordnet.princeton.edu/}, Feb. 2019.} is a lexical database that represents English words in a semantic graph. These words correspond to nouns, verbs, adjectives, and adverbs. The words that have the same meaning (e.g., car and automobile) are grouped in the same category called "synset". The synsets correspond to semantic concepts which are related to other synsets by several relations, such as the "hyponymy" ("is a" relation) and the "meronymy" ("part-whole" relation) \cite{miller1998wordnet}. 

Several search approaches have exploited WordNet to perform a concept based search have been proposed in the literature \cite{finkelstein2002placing, liu2004effective, kruse2005clever}. "Varelas et al." proposed an approach based on the vector model as follows \cite{varelas2005semantic}.
\begin{enumerate}
    \item First, Each document and query is represented by a vector of weighted terms using TFIDF formula.
    \item After that, the weight of each query term is adjusted based on the semantic relationship with the other query terms. This allows to increase the weight of the query terms that are semantically similar (e.g., "railway", "train", "metro"), whereas, the weights of the non-similar terms remain unchanged (e.g., "train", "phone").
    \item Then, the query is expanded by the synonym of its terms. Hyponyms (instances of a term) and hypernyms (generic terms) that are semantically similar to the query terms are also added to the query. A weight is assigned to each added term.
    \item After expanding and re-weighting the query, the similarity is calculated between each document of the data collection and the new query.
\end{enumerate}

\section{Summary}

The aim of this chapter is to introduce the basic notions of information retrieval. For this, we started by explaining the IR process which is composed of four main tasks, namely, the indexing, the query processing, the matching, and the ranking. An IR model indicates how the search process should be done. It is composed of three main categories, which are the set theoretic models, the algebraic models, and the probabilistic models. One model of each category is presented in this chapter. After that, the IR evaluation metrics, which are the recall and precision are briefly presented. Finally, the importance of the semantic search is shown with particular focus on the concept based information retrieval.

\chapter{Cloud Computing}

\label{Chapter7}

\section{Motivation}

Cloud computing is a technology that consists in providing users with storage space and computing power through remote servers. This technology allows companies and individuals to take advantage of IT resources (data, network, storage, computing) as a service. According to the U.S. National Institute of Standards and Technology (NIST) \cite{mell2011nist}, cloud computing is a model that enables on-demand network access for authorized users to a shared pool of IT resources. These resources which correspond to networks, servers, storage space, applications, processing, and services, can be provided rapidly with minimal management and interaction efforts. The cloud computing model is composed of five essential characteristics, three service models, and four deployment models.

We present in this chapter the different features of the cloud model. Then, we discuss the security aspect in cloud computing.

\section{Essential Characteristics}

In the following, we present the main characteristics of cloud computing that have been defined by the NIST organization \cite{mell2011nist}.

\begin{itemize}
    \item \textit{On-demand self-service}. Cloud's customers are able to provision computing resources, such as computing power, storage, networks, and software in a flexible way without requiring any human interaction.
    \item \textit{Broad network access}. It refers to the fact that customers can access cloud resources using a wide range of devices, such as tablets, PCs, Macs, and smartphones, as well as a wide range of locations.
    \item \textit{Resource pooling}. The cloud computing model is a multi-tenant model, which means that the cloud resources are shared between the different customers with scalable services. The cloud resources are dynamically assigned and reassigned based on the customer's need.
    \item \textit{Rapid elasticity}. It refers to the cloud ability to provide customers with scalable services. The rapid elasticity allows clients to automatically obtain additional resources when needed.
    \item \textit{Measured service}. It refers to the ability of both customers and providers to control, monitor, and report the cloud resources (e.g., storage, processing, and bandwidth) in order to ensure a transparency of the utilized services.
\end{itemize}

\section{Service Models}

\begin{figure}
\centering
\includegraphics[width=10cm,keepaspectratio]{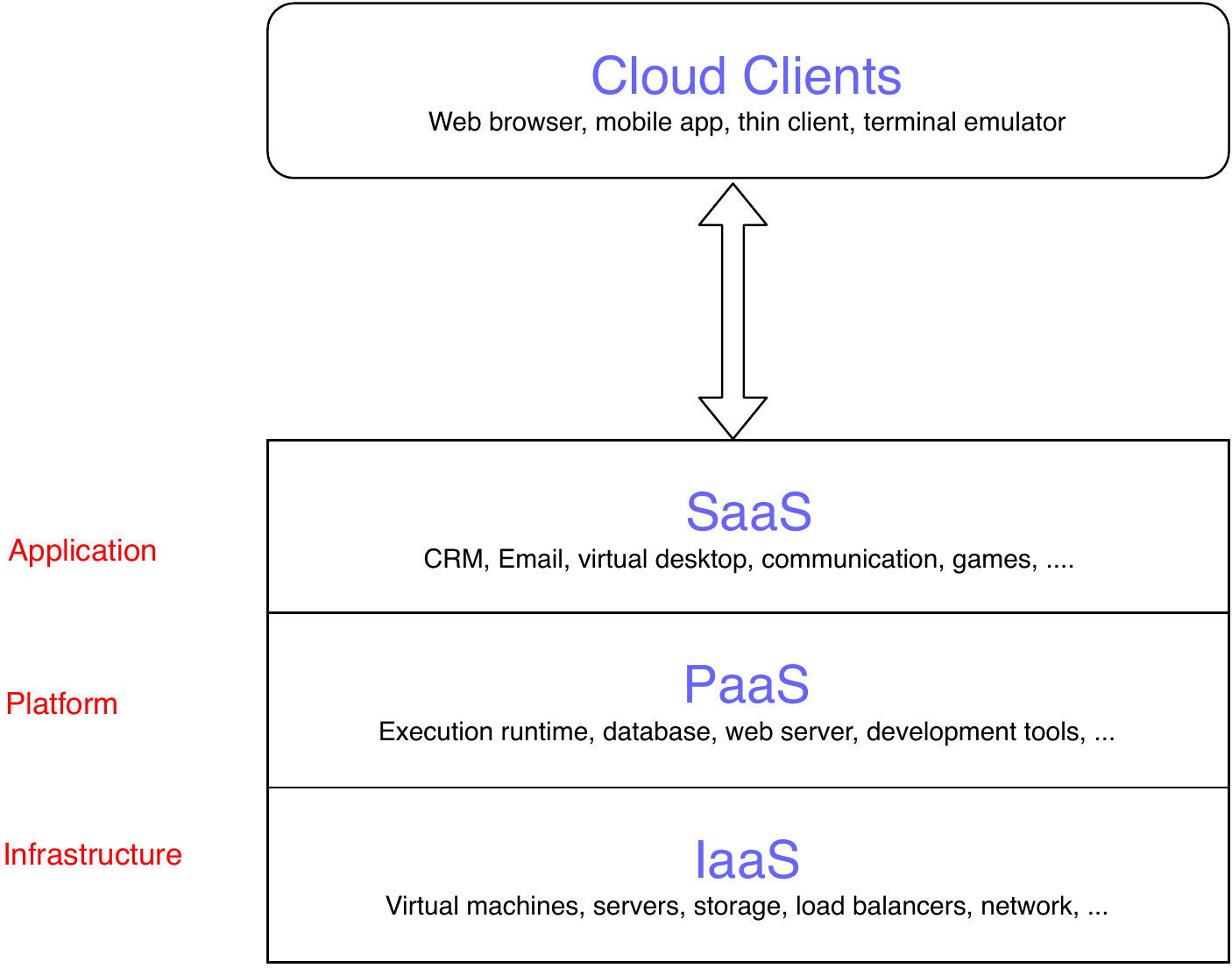}
\decoRule
\caption{Cloud computing layers}
\label{fig:cloud_layers}
\end{figure}

Three main service models are proposed by the cloud providers \cite{mell2011nist} (see Figure \ref{fig:cloud_layers}).

\begin{itemize}
    \item \textit{Software as a Service (SaaS)}. In this model, the applications which are installed and running in the cloud side are provided to the consumers. The letter can access these applications through web browsers or interface programs. The users don't have to manage or control anything about the network, the servers, the storage, the operating systems (OS), or even individual application capabilities.
    
    \item \textit{Platform as a Service (PaaS)}. In this model, several tools such as the operating systems, the libraries, the programming languages, and the databases are provided to the cloud's consumers in order to develop and execute applications. The users don't have to control the cloud infrastructure such as the network, the servers, the storage, and the operating systems, but have to manage the application deployment and the configuration setting.
    
    \item  \textit{Infrastructure as a Service (IaaS)}. In this model, several resources such as processing, storage space, and network are provided to the cloud's users in order to deploy and run their software (e.g., OS and applications). The consumers have control over the OS, the storage space, the deployed applications, and a limited networking components (e.g., firewalls).  
\end{itemize}

\section{Deployment Models}

In the following, we present the four main deployment models of clouds \cite{mell2011nist}.

\begin{itemize}
    \item \textit{Private cloud}. It consists in a set of resources (e.g., servers, applications, and processing) exclusively dedicated to a single organization. The maintenance of the services is always performed through a private network.
    
    \item \textit{Community cloud}. It consists in a set of resources and data shared between several organizations (e.g., hospitals, and government organizations) that have common interests. The cloud infrastructure may be managed by one or more of the organizations, or by a third party.
    
    \item \textit{Public cloud}. It consists in a set of resources provided to the general public. In a public cloud, the hardware, the software, and all the infrastructure are owned and managed by the cloud service provider.
    
    \begin{figure}
    \centering
    \includegraphics[width=10cm,keepaspectratio]{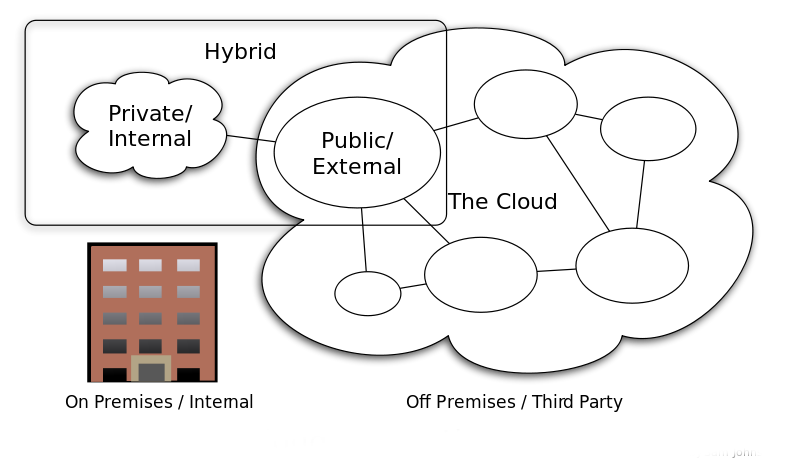}
    \decoRule
    \caption[Hybrid cloud]{Hybrid cloud \cite{wiki:cloud}}
    \label{fig:hybrid_cloud}
    \end{figure}
    
    \item \textit{Hybrid cloud}. It is a composition of two or more distinct private, public, and community cloud in order to take advantage of all of them. For flexibility purpose, an organization can move its data and applications between the different cloud infrastructures (see Figure \ref{fig:hybrid_cloud}).
\end{itemize}

\section{Security Aspect}

To secure an information system, it is important to identify the threats and challenges in order to implement the appropriate countermeasures. The architecture and features of cloud computing bring some security benefits, such as the centralization of security. However, cloud computing introduces new aspects that require risk assessment, such as the availability, confidentiality and privacy, and data integrity \cite{zissis2012addressing}.

\begin{itemize}
    \item \textit{Confidentiality and Privacy.} Data confidentiality consists in accessing data only by authorized users. When using a cloud computing, the data would be exposed to a great number of parties, devices, and applications, which increases the vulnerability of those externalized data. On the other hand, the data privacy corresponds to the user's desire to control the disclosure of his personal information. The cloud providers should be able to comply with the international laws regarding the protection of the personal data. In cloud computing, several users often share the same resources (applications, memory, network, data). However, data sharing brings several confidentiality and privacy threats. These security aspects should be carefully considered when building systems and hosting data.

    Data confidentiality in cloud computing is strongly related to user authentication. Therefore, to prevent unauthorized data access, a robust authentication system should be implemented. Moreover, software confidentiality is as important as confidentiality and privacy of data. Given that data and applications are hosted in the cloud server side, a lack of software security may lead to a vulnerability on data confidentiality. For this reason, it is highly recommended to pay attention to the software security, and make sure that no information is leaked when applications are run by the cloud.
    
    \item \textit{Integrity.} It means that resources can only be modified by authorized users. Therefore, data integrity is the protection of data against unauthorized changes, deletions, and fabrications. The cloud provider should guarantee that the externalized data are not modified, abused, or stolen. For this, an authentication system should be implemented in order to determine for each user his access rights to the secured resources.
    
    Software integrity consists in protecting applications from unauthorized modifications, deletions, fabrications, or thefts. The cloud providers implement software interfaces that allow users to access and manage cloud services. Therefore, the protection of software integrity is transferred to the protection of these interfaces that are managed by the software owners or administrators.
    
    \item \textit{Availability.} It corresponds to the fact that resources are available and usable upon demand by an authorized client. System availability means that the system is always ready to operate even if some authorities misbehave. The cloud provider must maintain the information and data processing available upon demand by an authorized user even if the number of clients increases.
\end{itemize}

\section{Summary}

In this chapter, we introduced some cloud computing concepts. For that, we started by defining the five essential characteristics of cloud computing. Then, we presented and clarified the difference between the three layers of cloud computing (SaaS, PaaS, and IaaS). After that, we presented the four types of cloud computing (public, private, community, and hybrid cloud). Finally, we discussed the three important security aspects in the cloud, namely, confidentiality and privacy, integrity, and availability.


\chapter{Searchable Encryption} 

\label{Chapter2} 

\section{Motivation}

Companies and individuals demand more and more storage space and computing power for their applications. This need led to the development of new technologies such as cloud computing. This technology enables outsourcing data in order to be stored and processed remotely according to the users' needs. In addition, users pay only for the storage space and the computing power. Therefore, this model is much more flexible and offers very good value for money (hardware, software, and maintenance).

However, the outsourced data, such as photos, emails, financial documents, and medical reports are often sensitive and confidential. Therefore, to protect these datasets against possible attacks on the cloud system, it is highly recommended to encrypt them before they are outsourced.

When a user performs a search over the outsourced and encrypted data, the cloud server cannot use any standard search approaches since these data are encrypted. To overcome this issue, many researchers have worked on this problem by proposing approaches more or less effective and secure \cite{song:practical, curtmola:searchable, xu:two, yu:toward, wang:privacy, cao2014privacy, xia2018towards}.

We introduce in this chapter the searchable encryption area. For this, we start by presenting the general architecture of a searchable encryption scheme. Then, we introduce several notions of security in searchable encryption. Finally, we present a detailed related work.

\section{General Architecture}

\begin{figure}
\centering
\includegraphics[width=11cm,keepaspectratio]{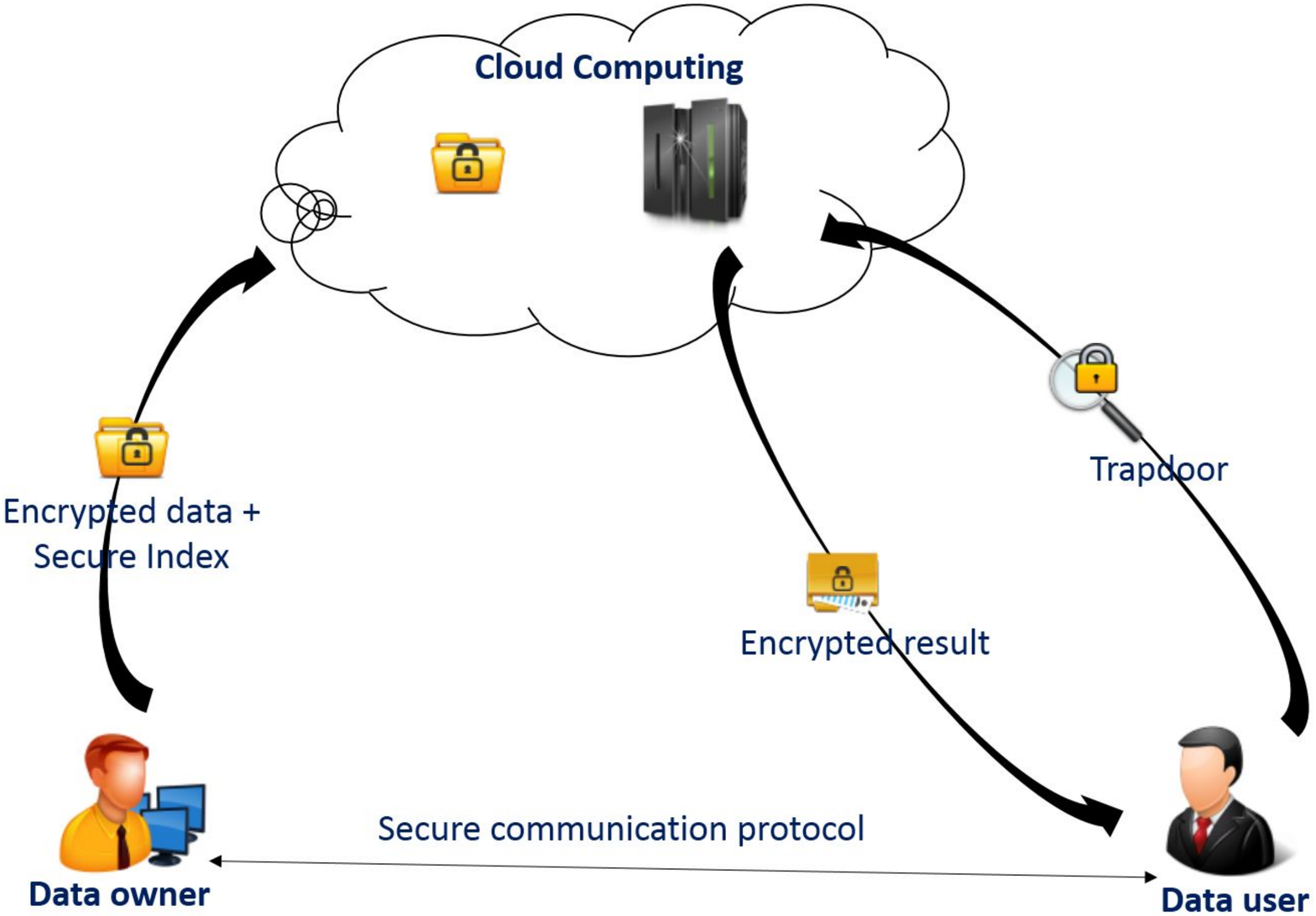}
\decoRule
\caption[General SE architecture]{General searchable encryption architecture}
\label{architecture}
\end{figure}

The search process in a cloud environment is different from traditional search engines. This difference comes from the fact that the outsourced cloud data are encrypted and the search process should not cause any sensitive information leakage. The architecture of a searchable encryption (SE) scheme is composed of three main entities, namely, the "data owner", the "cloud server" and the "authorized users" (Figure \ref{architecture}). 

The data owner is responsible for the creation of an index from the data collection and encrypting both of them before outsourcing them to the cloud. The index should be encrypted using an appropriate cryptosystem that keeps the ability to make calculations on the index such as "homomorphic encryption" \cite{gentry2009fully} and the secure $k$ nearest neighbor (S$k$NN) algorithm \cite{wong2009secure}, whereas, the data collection should be encrypted using a robust cryptosystem such as "AES". Then, the encryption/decryption keys are shared by the data owner with each authorized user using a secure communication protocol. To perform a search, an authorized user formulates a query, encrypts it using the secret key, and sends the trapdoor (the encrypted query) to the cloud server. Upon receipt of the trapdoor, the server initiates the search process over the secure index and returns to the user an encrypted result. Finally, the user decrypts the result and requests a set of relevant documents from the server.

\section{Security in Searchable Encryption}\label{sec_security}
In this section, we present the threat models, the security constraints, and the security definitions in searchable encryption.

 \subsection{Threat Models}\label{sec_threat}
The cloud server is considered "honest-but-curious" in most SE schemes proposed in the literature. This means that the server is honest when applying the different functions, but is curious to learn as much as possible about the content of the data collection by doing statistical analyzes. Based on the information held by the cloud server. The threat models are divided into two main categories. Each category has different attack capabilities \cite{cao2014privacy}.

\begin{itemize}
    \item \textit{Known ciphertext model.} In this model, the cloud server is supposed to only know the encrypted data collection and the secure index sent by the data owner, in addition to the trapdoors sent by the authorized users.
    \item \textit{Known background model.} This model is stronger than the previous one. In addition to the information that can be accessed in the known ciphertext model, the server may hold some background knowledge that can be exploited to perform statistical analyses, for instance, the probabilities of term co-occurrence \cite{islam2012access}.
\end{itemize}

  \subsection{Security Constraints}\label{sec_constraintes}
 When designing a searchable encryption scheme on cloud data, it is necessary to take into account the security constraints elaborated in the literature \cite{islam2012access, li:efficient, cao2014privacy, liu2014search, stefanov2014practical}.
 
 \begin{itemize}
     \item \textit{Keyword privacy.} This constraint makes sure that the proposed scheme is able to hide from the server the "term distribution" and the "inter-distribution" of a document. The term distribution indicates the frequency distribution of a given term in each document of the data collection, whereas, the inter-distribution indicates the distribution of scores of terms in a given document. Hiding these two features allows to prevent the server from making any link between terms and documents. 
     \item \textit{Search pattern.} This constraint (also called "query unlinkability") guarantees that the proposed scheme can prevent the server from deducing the relationship between a given set of encrypted queries. For that, the cryptosystem should be non-deterministic.
     \item \textit{Access pattern.} This constraint guarantees that the proposed scheme is able to hide from the server the sequence of results returned to a user during the search.
     \item \textit{Forward privacy.} This constraint makes sure that the server cannot learn that a new document contains keywords that have been searched for in the past. For instance, if a user searched for the term $w$ and later a new document $d$ containing the term $w$ has been inserted in the data collection, forward privacy constraint guarantees that the server cannot deduce that the term $w$ belongs to the document $d$.
     \item \textit{Backward privacy.} This constraint guarantees that the server is not able to perform searches on deleted documents.  
 \end{itemize}
 
 \subsection{Security Definitions}
The first SE scheme proposed in the literature \cite{song:practical} does not use any notions of security definition. Nevertheless, this scheme is indistinguishable against chosen plaintext attacks (IND-CPA). This means that an adversary \textit{A} cannot distinguish between two ciphertexts even if their corresponding plaintexts are chosen by that adversary. However, this definition does not take into consideration the trapdoors and the indexes which are the main source of leakage in searchable encryption. Consequently, IND-CPA is not sufficient to prove the security of SE schemes \cite{bosch2015survey}.

The first security definition in the context of searchable encryption was proposed by "Goh et al." who defined a semantic security against adaptive chosen keyword attacks (IND1-CKA) which is suitable for indexes \cite{goh2003secure}. This security definition guarantees that given two documents of equal size and an index, an adversary \textit{A} cannot decide which document corresponds to the index. Nevertheless, the trapdoors are not taken into consideration which makes IND1-CKA not effective to prove the security in the context of SE.

"Chang and Mitzenmacher" introduced a new simulation based IND-CKA which is a stronger version than IND1-CKA \cite{chang2005privacy}. Their security definition takes into consideration the trapdoors and makes sure that an adversary \textit{A} cannot link a document to its index even for unequal size documents. Nevertheless, "Curtmola et al." proved that this security definition is incorrect given that it can be satisfied by an insecure SE scheme \cite{curtmola2011searchable}.

After that, "Goh et al." introduced IND2-CKA where an adversary \textit{A} cannot distinguish indexes from two unequal size documents as done by "Chang and Mitzenmacher" \cite{bosch2015survey}. However the trapdoors are still not secure and the server may deduce terms from the received encrypted queries. This makes both IND1/2-CKA weak security definitions in the context of searchable encryption.

"Curtmola et al." pointed out that the indexes and the trapdoors should be secure in the context of SE \cite{curtmola2011searchable}. They proposed two adversarial models that are used as the standard security definitions for searchable encryption to date. The first one (IND-CKA1) is a non-adaptive model, it guarantees the security of a scheme only if all queries are sent at once, whereas, the second one (IND-CKA2) which is an adaptive model guarantees the security even if a user query is based on the prior search. Both IND-CKA1/2 guarantee that no information is leaked to the server except the search pattern and access pattern. IND-CSK2 is considered a strong security definition for SE schemes.

In the case of asymmetric SE, schemes do not guarantee the security of the trapdoors given that they are encrypted using public keys. "Boneh et al." introduced PK-CKA2 as security definition for asymmetric SE \cite{boneh2004public}. It guarantees that an adversary \textit{A} cannot distinguish between ciphertexts of two challenge keywords even if it is able to get the ciphertext of any other keywords. PK-CKA2 makes sure that no information about keywords are leaked until the trapdoor of that word is available.

 \section{Related Work}
 Many works on information retrieval over encrypted cloud data have been carried out in the last few years. This section presents the major works that have been published recently. We classify these works in 9 major categories according to their contributions.
 
\subsection{Single Keyword SE}
 
"Raykova et al." proposed an anonymous searchable encryption scheme that exploits a trusted authority called "query router" \cite{raykova2009secure}. The latter acts as an intermediate between the clients who send the queries and the server which stores the encrypted index. The query router has two major tasks. The first one consists in checking and hiding the identity of the users, whereas, the second task consists in re-encrypting the queries and the search results in order to be exploitable by the server and the users, respectively. For efficiency purposes, the authors exploited the bloom filter \cite{bloom1970space} and the deterministic encryption in the public key setting proposed by " Bellare et al." \cite{bellare2007deterministic}.

"Boldyreva et al." proposed a SE scheme called ranked searchable symmetric encryption (RSSE) \cite{wang2010secure}. An inverted index is exploited where its entries are encrypted using a collision resistant hash function, and each entry points on a list of couples (file id, score) encrypted by a pseudo-random function. Their contribution consists in encrypting the scores using an order preserving symmetric encryption (OPSE) in order to provide ranked results. In addition, to avoid more information leakage than the access pattern and the search pattern, the authors proposed a one-to-many\footnote{One-to-many OPSE guarantees that the encryption of the scores is non-deterministic.} OPSE instead of using the traditional one-to-one OPSE \cite{boldyreva2009order}.
 
 \subsection{Multi Keyword (Ranked) SE}
"Yu et al." proposed a searchable encryption scheme called two-round searchable encryption (TRSE) \cite{yu:toward}. Their goal was to propose a scheme that solves the problem of information leakage caused by statistical analyzes done on the cloud server. Their technique consists in hiding from the server the similarity scores between documents and queries by exploiting an encryption method called fully homomorphic encryption over the integers (FHEI) \cite{van2010fully}. However, this encryption method requires $\Omega(\lambda^{10})$ per-gate\footnote{$\lambda$ is a security parameter.} computation which makes TRSE scheme inefficient in practice.
 
"Cao et al." proposed a searchable encryption scheme called multi-keyword ranked search over encrypted cloud data (MRSE) \cite{cao2014privacy}. They have contributed to the improvement of the S$k$NN algorithm \cite{wong2009secure} by adding dummy dimensions and random parameters to the document and query vectors in order to protect the data privacy.
 
"Elmehdwi et al." proposed a searchable encryption scheme based on the S$k$NN algorithm \cite{elmehdwi2014secure}. Their aim was to preserve the privacy of data and queries and hide the data access pattern from the cloud server during the search process. They exploited the "Paillier" cryptosystem \cite{paillier1999public} to encrypt the database and the queries. Their proposed approach exploits two non-colluding semi-honest cloud service providers, denoted by $C_1$ and $C_2$, where, the encrypted database is outsourced to $C_1$, whereas, the secret key is stored in $C_2$. In addition, $C_1$ and $C_2$ are exploited in order to perform other operations not supported by the Paillier cryptosystem such as the "secure multiplication", the "secure minimum", and the "secure bit-decomposition". Their approach is able to perform a search operation over encrypted cloud data while hiding the data access pattern from both $C_1$ and $C_2$. However, this approach is not applicable in large scale data because of its poor performance.

"Wang et al." proposed a SE scheme that exploits an inverted index encrypted using an asymmetric cryptosystem \cite{wang2015inverted}. Their goal was to propose an efficient scheme which is able to ensure the index privacy and hide the access pattern and the search pattern from the server. The inverted index is composed of a set of inverted lists of document IDs. Each inverted list is represented by a polynomial whose coefficients are encrypted by the Paillier encryption algorithm. This scheme guarantees the index privacy, and can hide the search pattern from the server.

 \subsection{Fuzzy Keyword SE}
 
"Li et al." proposed a fuzzy searchable encryption approach \cite{li2010fuzzy} that is based on some techniques such as the edit distance \cite{levenshtein1966binary} and the deterministic encryption. It consists in grouping the keywords that have an edit distance (from each other) less than a predefined threshold in the same category. The keywords of the same set are related to each other in the index, and provide the same search result. This approach allows to return a correct result even if the documents and queries contain some spelling mistakes. To optimize the storage space and the search performance, a wildcard-based technique was exploited. This technique consists in representing the similar keywords as a regular expression rather than enumerating them one by one.
 
"Wang et al." proposed a searchable encryption scheme over encrypted cloud data which takes into account the possible spelling mistakes in the queries \cite{wang:privacy}. Contrary to most of the fuzzy search schemes over encrypted cloud data proposed in the literature which are based on the extension of the index, this scheme is based on two techniques, namely, the locality sensitive hashing (LSH) technique and the bloom filter. The bloom filter uses the LSH functions to represent terms which are syntactically close in the same representation, such as "network" and "netword".
 
 \subsection{Semantic SE}\label{semantic_se}
 
"Sun et al." proposed a semantic searchable encryption scheme called ranked semantic keyword search (RSS) \cite{sun:secure}. The particularity of this scheme is that the cloud server itself which is responsible for building the index and constructing a semantic graph that is used to extend the trapdoor by adding synonyms of the query terms. However, granting the server the task of building the index and extending the queries may compromise the data privacy.
 
"Fu et al." proposed a stemming based searchable encryption scheme \cite{fu2014semantic}. Their idea consists in grouping each set of words having the same root in the same category (concept) by using the Porter stemming algorithm \cite{porter1980algorithm}. The search index consists in a prefix tree where each node corresponds to a character and each path from the root to a leaf corresponds to a term (root of the word). They used a deterministic cryptosystem to encrypt the index and the queries. Unfortunately, this SE scheme just exploits the root of terms without taking into account the meaning of documents and queries. In addition, the deterministic encryption makes the cloud server task easier when performing statistical analyses. 
 
"Fu et al." proposed a concept based searchable encryption scheme \cite{fu2016enabling}. They exploited some techniques of natural language processing (NLP) such as "text summarization" and "Tregex" to extract the most important sentences from documents in order to be converted into conceptual graphs (CGs). In addition, the authors proposed a technique that maps each CG into vectors. The set of vectors corresponds to the search index, which is encrypted using S$k$NN algorithm and hash functions.

"Fu et al." proposed a concept based semantic search over encrypted cloud data \cite{fu2016towards}. Their approach exploits a concept hierarchy (e.g., WordNet) which is extended by affecting attributes with associated values to the concepts (e.g., the attribute "color" with the value "red" is affected to the concept "car"). Each document and query is represented by two vectors, the first one is based on the concepts and the second vector is based on the attributes. Furthermore, each query is extended by candidate concepts that are similar to the concepts of that query. The S$k$NN algorithm is used to encrypt the vectors and a binary tree is proposed to accelerate the search process.

 \subsection{Dynamic SE}
"Kamara et al." proposed a new searchable encryption scheme that takes into account the dynamic aspect of data (adding and deleting files from the data collection) \cite{kamara2012dynamic}. Their construction is based on an inverted index in the form of a chained list. Each node of the chained list corresponds to a term-file association. The nodes that contain the same term are linked together. The chained list should be reconstructed each time a file is added or deleted from the data collection. The drawback of this dynamic SE scheme is that the ranking of results is not considered, and the access pattern and search pattern are leaked.

"Hahn and Kerschbaum" proposed an information retrieval system on encrypted and dynamic cloud data \cite{hahn2014searchable}. Their construction is based on deterministic encryption of the index entries and the queries. The inverted index which is empty at the beginning is built while searches are performed. When a keyword\footnote{The keyword is encrypted using a deterministic cryptosystem.} is searched for the first time, it would be added to the inverted index as an entry that leads to the documents in which that keyword belongs. When updating the data collection, the index should also be updated by adding/deleting the document IDs in the appropriate entries.

"Xia et al." proposed a SE scheme that supports multi-term queries, and takes into account the dynamic aspect of data \cite{xia2016secure}. Each document of the data collection is represented by a vector encrypted using the S$k$NN algorithm. A binary tree $T$ is built to provide an efficient search. The leaves of $T$ correspond to the document vectors and the intermediate nodes are calculated according to the leaves. during the search process, a scalar product is computed between the root and the query vector. If the similarity score is positive, the search process recursively continues on the child nodes until reaching the leaves. When updating a document $d$, the sub-tree from the root to the leaf corresponding to $d$ should be recalculated.

 \subsection {Multi-Owner Based SE}
 
"Bao et al." proposed a searchable encryption scheme in a multi-user setting \cite{bao2008private}. The authorized users are able to insert encrypted data records in the outsourced database, and each user can search on the whole records using his distinct secret key. The user manager who is a trusted entity can dynamically enroll or revoke any user without any need of updating the encrypted database or distributing new secret keys.
 
"Rhee et al." proposed a public key searchable encryption in a multi-owner environment \cite{rhee2010trapdoor}. The aim of their approach is to enhance the security model proposed by "Baek et al." \cite{baek2008public} by preventing the keyword-guessing attacks \cite{byun2006off}. For that, they introduced the concept of "trapdoor indistinguishability" to their public key SE scheme by adding a random variable to the trapdoor computation.
 
"Li et al." proposed an authorized keyword search over encrypted cloud data \cite{li2011authorized}. Their scheme allows several data owners to encrypt and share data, while enabling a large number of users to search over multiple owners' data.  For that, a trusted authority and several local trusted authorities should be exploited to determine the users' search privileges. The hierarchical predicate encryption \cite{okamoto2009hierarchical} was used to encrypt the index and the queries.
 
 \subsection{SE with Access Right Managements}
 
"Zhao et al." proposed a fine-grained access control aware multi-user keyword search scheme \cite{zhao2011multi}. The search process is performed only on the subset of data accessible by the user. The authors exploited an attribute based cryptosystem \cite{bethencourt2007ciphertext} to apply an access control policy. In addition, the file update and the user revocation are taken into consideration in this scheme. 
 
"Bouabana-Tebibel and Kaci" proposed a searchable encryption approach on cloud data \cite{bouabana2015parallel} that takes into account the users access rights which is possible due to the use of attribute based encryption methods \cite{bethencourt2007ciphertext, goyal2006attribute}. This approach enables the cloud server to perform a search only on data that are accessible by the user rather than searching on the whole data collection. For this purpose, an authorized user has to include the encrypted private key in the trapdoor in order to allow the server to check if the user has the right to access the document.
 
"Yuan et al." proposed an approach to search over encrypted images while taking into account the users access rights to the data \cite{yuan2015seisa}. Their construction is based on the S$k$NN algorithm where every image is represented by a vector. To accelerate the search process, a tree index is constructed by classifying the images into several groups using the $k$-means clustering algorithm. In addition, the tree index stores the access right values in each tree node. A user is able to perform a search via an encrypted query just in the parts of the tree where he has the right to access.

"Deng et al." proposed a multi-user searchable encryption scheme with keyword authorization in a cloud storage \cite{deng2016multi}. This scheme can handle multiple data owners where each owner is able to construct his own data collection and outsource it to the cloud server. Each private key contains some properties that indicate to the cloud server the set of terms that the user has the right to search for. Furthermore, each data owner is able to grant or revoke any keyword authorization to any user.
 
\subsection{Parallel and Efficient SE}

"Li et al." proposed a searchable encryption scheme called Johnson-Lindenstrauss transform based scheme \cite{li:efficient}. Their goal was to achieve high performance and low cost search by reducing the index size. For that, they used a method called Johnson-Lindenstrauss transform (J-L transform) \cite{johnson1984extensions}. The principle of the J-L transform is that, if the points of a vector space are projected into a subspace that is randomly selected and suitably sized, then, the Euclidean distance between the points is approximately preserved. Therefore, each document and query is represented in a new vector space with a much smaller number of dimensions than the original one.

"Kamara and Papamanthou" proposed a parallel searchable encryption scheme which is based on the keyword red black (KRB) tree \cite{kamara2013parallel}. This binary tree is useful because it allows a keyword search by following paths from the root to the leaves and a file update by following paths from the leaves to the root. This scheme supports parallel keyword search as well as parallel addition and deletion of files. The drawback of this scheme is that the search and access patterns are leaked because of the deterministic encryption of the queries.
  
"Fu et al." proposed an SE approach that combines the vector space model with a binary tree structure \cite{fu2015achieving}. Each document of the data collection is encrypted using the S$k$NN algorithm. The index corresponds to a binary tree where each node is encrypted using a static hash table. The authors proposed a parallel search algorithm to efficiently browse through the binary tree.

"Xia et al." proposed an approach to search over encrypted images \cite{xia2017epcbir}. For that, they constructed an index of two layers in order to speed up the search process. The first layer is a pre-filter table constructed by locality sensitive hashing in order to classify the similar images in the same category. The pre-filter table is used to discard the dissimilar images. The second layer is a one-one map index that is used to sort the pertinent images by exploiting the S$k$NN algorithm.
  
\subsection{Verifiable SE}
 
"Van Liesdonk et al." proposed a keyword searchable encryption scheme over dynamic data \cite{van2010computationally}. Their approach exploits an inverted index in which the entries (keywords) are encrypted using a pseudo-random function and each entry points on an index table decryptable using an appropriate trapdoor received from the client. The authors proposed two versions of their approach. The first one is more efficient, but the search should be made interactively with the client, whereas, the second one does not need the client to be interactive. Both approaches are proven IND-CKA2 secure.

"B\"osch et al." proposed a selective document retrieval scheme over encrypted data \cite{bosch2012selective}. They combined the vector based index proposed by "Chang and Mitzenmacher" \cite{chang2005privacy} with the somewhat homomorphic encryption scheme proposed by "Brakerski and Vaikuntanathan" \cite{brakerski2011efficient} to index the documents and queries. During the search process, an encrypted similarity score is calculated by applying the inner product between each document vector and query vector. An additional round is needed in this scheme since the encrypted result should be decrypted in the user side. The advantage of this scheme is its ability to hide the search pattern from the server, but the search efficiency is not optimal given that no inverted index or binary tree is exploited. This scheme is proven fully secure under the assumption that the homomorphic encryption scheme is IND-CPA secure.

"Cash et al." developed a sublinear conjunctive search solution for arbitrarily structured data \cite{cash2013highly}. Their solution exploits the inverted index structure proposed by "Curtmola et al." \cite{curtmola:searchable}. Their contribution consists in adapting the conjunctive search for large datasets. For this purpose, the search process starts with the estimated least frequent keyword and then filtering the result by exploiting the remaining query terms. Their solution is proven IND-CKA2 secure.
 
\section{Summary}

In this chapter, we introduced the basic notions of searchable encryption. The general architecture of a SE scheme is composed of three entities, which are the cloud server, the data owner, and the authorized users. The cloud server should perform a search process over an encrypted index without causing any sensitive information leakage. For that, several security constraints should be respected when designing a searchable encryption scheme. Several SE approaches that have been recently proposed in the literature were presented in this chapter.

\chapter{Cryptographic Tools}

\label{Chapter3}

\section{Motivation}

Any searchable encryption scheme needs to encrypt the index while keeping the ability to exploit it without decryption. For that, traditional encryption schemes (e,g., AES) are not effective to encrypt the indexes, and thus, other more appropriate cryptosystems should be exploited, such as the S$K$NN algorithm and homomorphic encryption.

In this chapter, we start by presenting the S$K$NN algorithm. After that, we introduce the ciphertext-policy attribute based encryption which is the most successful variant of the attribute based encryption. Finally, we present the major homomorphic encryption schemes that have been proposed in the literature.

\section{Secure $k$-NN Algorithm}\label{sec_sknn}
The secure $k$ nearest neighbor (S$k$NN) algorithm \cite{wong2009secure} is used in the searchable encryption area to encrypt documents and queries that are represented in a vector space model of size $m$. S$k$NN algorithm allows to calculate the similarity (the dot product) between an encrypted document vector and an encrypted query vector without any need of decryption. The S$k$NN algorithm is composed of three functions.

\begin{itemize}
    \item \textbf{KeyGen.} This function consists in generating the encryption key that is composed of one vector $S$ of size $(m+u+1)$ and two $(m+u+1) \times (m+u+1)$ invertible matrices $M_1$ and $M_2$. Where $m$ is the size of the original vector space and ($u+1$) is the number of dummy dimensions.
    \item \textbf{Enc.} The encryption process is done in three steps as follows.
        \begin{enumerate}
            \item \textbf{Extension.} At first, ($u+1$) dimensions are added to each document vector $d_i$ of size $m$. The value 1 is assigned to the  $(m+1)^{th}$ dimension, whereas, a random small value $\epsilon_{ij}$ is assigned to each $(m+j+1)^{th}$ dimension (where $j \in[1, u]$). The $u$ last dimensions correspond to dummy keywords.
            \begin{equation}
                D_i = \{ d_i, 1, \epsilon_{i1}, \epsilon_{i2}, \epsilon_{i3}, ..., \epsilon_{iu}\}
            \end{equation}
            Moreover, a query vector $q$ of size $m$ is multiplied by a random parameter $r$. Then, a dimension with a random value $t$ is added to the obtained vector. After that, $u$ dimensions are added to this vector. a value $\alpha_j$ is assigned to each $(m+j+1)^{th}$ dimension (where $\alpha_j \in\{0, 1\}$).
            \begin{equation}
                Q = \{ r . q, t, \alpha_1, \alpha_2, \alpha_3, ..., \alpha_u\} / \alpha_j \in\{0, 1\}
            \end{equation}
            \item \textbf{Splitting.} After that, each document vector $D_i$ is split into two vectors \{$D'_i, D''_i\}$, and each query vector $Q$ is split into two vectors \{$Q', Q''\}$. The vector $S$ is used as a splitting indicator. Indeed, if the $j^{th}$ element of $S$ is equal to 0, then $D'_i[j]$ and $D''_i[j]$ will have the same value as $D_i[j]$, and each of the two elements $Q'[j]$ and $Q''[j]$ will have random values such that their sum is equal to $Q[j]$. In the case where the $j^{th}$ element of $S$ is equal to 1, we follow the same principle, except that the document vector and the query vector are switched.
            \item \textbf{Multiplication.} Finally, both $M_1$ and $M_2$ matrices are used to finalize the encryption of each document vector and query vector as follows.
            \begin{equation}
                \begin{cases}
                    & I_i = \{M_1^T \cdot D'_i, M_2^T \cdot D''_i\} \\
                    & T_q = \{M_1^{-1} \cdot Q', M_2^{-1} \cdot Q''\} \\
                \end{cases}
            \end{equation}
        \end{enumerate}
        \item \textbf{Eval.} When applying the scalar product between a document vector and a query vector we obtain:
                \begin{align*}
                    I_i \times T_q &= \{M_1^T \cdot D'_i, M_2^T \cdot D''_i\} \times \{M_1^{-1} \cdot Q', M_2^{-1} \cdot Q''\} \\ 
                    &= D'_i \times Q' + D''_i \times Q'' \\
                    &= \{ d_i, 1, \epsilon_{i_1}, \epsilon_{i_2}, \epsilon_{i_3}, ..., \epsilon_{i_u}\} \times \{ r . q, t, \alpha_1, \alpha_2, \alpha_3, ..., \alpha_u\} \\ 
                    &= r \cdot d_i \cdot q + \sum_{j=1}^U \epsilon_{i_j} \cdot \alpha_j + t   
                \end{align*}

                The random parameters $\{\epsilon_{i_j}, \alpha_j, t, r\}$ are used to hide the real similarity score between a document and a query. However, the alternative similarity scores are still useful to sort the documents by relevance as was proved by "Cao et al." \cite{cao2014privacy}.
\end{itemize}

\section{Attribute Based Encryption}

The attribute based encryption (ABE) is an encryption method used to apply an access control policy in order to control the access to a data collection \cite{sahai2005fuzzy}. It consists in adding some attributes to the encrypted data as well as to the users' private key. During the decryption process, the ciphertext can be decrypted only if the number of matching attributes between the ciphertext and the private key exceeds a certain threshold.

Later on, "Goyal et al." proposed an encryption method called "key-policy attribute based encryption" (KP-ABE) \cite{goyal2006attribute} which consists in storing an access structure in the user's private key and some attributes in the ciphertext. This method is able to achieve a fine-grained access control and brings more flexibility in the management of the users than the previous method. Nevertheless, KP-ABE has the disadvantage of not being intuitive. 

To solve this problem, "Bethencourt et al." proposed an alternative method called "ciphertext-policy attribute based encryption" (CP-ABE) \cite{bethencourt2007ciphertext} that works in the same way as the KP-ABE method except that the access structure is stored in the ciphertext, whereas, the attributes are stored in the user's private key. During the decryption process, if some attributes of the private key satisfy the access policy of the data, the ciphertext may be decrypted. Otherwise, the user does not have the right to access this data and his private key cannot decrypt the ciphertext. For example, if the access policy in the data is "Pediatrics $\land$ (Doctor1 $\lor$ Doctor2)" and the user's secret key contains the attributes "Pediatrics" and "Doctor1", then the user has the right to access the data and is able to decrypt the ciphertext using his private key. After that, several approaches based on the CP-ABE method have been proposed in the literature \cite{emura2009ciphertext, ibraimi2009efficient}.

CP-ABE consists of five algorithms: \textit{Setup}, \textit{KeyGen}, \textit{Enc}, \textit{Dec}, \textit{Delegate}. In the following, we present each of them. We suggest the reader to refer to the original paper \cite{bethencourt2007ciphertext} for details.

\begin{itemize}
\item \textit{Setup} $\rightarrow \{PK, MK\}$. The purpose of this algorithm is the generation of two keys, a public key $PK$ and a master key $MK$.
\item \textit{KeyGen}($MK$, $A_u$) $\rightarrow SK$. This algorithm takes as input the master key $MK$ and a set of attributes $A_u$. It generates a private key $SK$ for a user who is associated with the set of attributes $A_u$.
\item \textit{Enc}($PK$, $M$, $A$) $\rightarrow CT$. This algorithm is executed by the data owner in order to encrypt a message $M$ under an access structure $A$ using a public key $PK$.
\item \textit{Delegate}($SK$, $A^\prime_u$) $\rightarrow SK^\prime$. This algorithm takes as input a set of attributes $A^\prime_u$ and a private key $SK$. The private key is associated with a set of attributes $A_u$, such that $A^\prime_u$ is included in $A_u$. The aim of this algorithm is to produce a new private key $SK^\prime$ which is associated with the set of attributes $A^\prime_u$.
\item \textit{Dec}($SK$, $CT$, $PK$) $\rightarrow M$. This algorithm is executed by the user to decrypt a ciphertext $CT$. It takes as input the ciphertext $CT$ containing an access structure $A$, the user's private key $SK$ containing a set of attributes $A_u$, and the public key $PK$. If the set of attributes $A_u$ satisfies the access structure $A$, the ciphertext $CT$ can be decrypted back to get the original message $M$.
\end{itemize}

\section{Homomorphic Encryption}

Homomorphic encryption (HE) is a cryptosystem that allows mathematical operations to be performed on encrypted data, generating an encrypted result which, after decryption, matches the result of the same operations performed on unencrypted data (see Formula \ref{HE_eq}). 

\begin{equation}\label{HE_eq}
    f\left(Enc(x_1), Enc(x_2), ..., Enc(x_n)\right) = Enc\left(f(x_1, x_2, ..., x_n)\right)
\end{equation}

Several homomorphic encryption methods have been proposed in the literature \cite{gentry2009fully, van2010fully, smart2010fully, brakerski2012leveled}. In the following we present some of the major HE schemes that have been studied in the literature. 

\subsection{Paillier Cryptosystem}

The Paillier cryptosystem is a probabilistic asymmetric encryption algorithm proposed by "Paillier" \cite{paillier1999public} based on the work of "Okamoto and Uchiyama" \cite{okamoto1998new}. It is an addictive homomorphic encryption method which can calculate the addition of two ciphertexts without any need of decryption. The Paillier cryptosystem can be exploited in many fields that need to protect the data privacy such as the searchable encryption area. In the following we present the three functions of the Paillier cryptosystem and its two properties.

\begin{itemize}
    \item \textbf{KeyGen.} This function generates the public and secret keys as follows.
    \begin{enumerate}
        \item Take as input two large primes $p$ and $q$.
        \item Compute $n = p \times q$ and $\lambda = \text{ LCM} (p-1, q-1)$, where "LCM" means the least common multiple.
        \item Select a random integer $g \in \mathbb{Z}^*_{n^2}$ such that $g$ has order multiple of $n$. 
        \item Calculate $\mu = (L(g^\lambda \text{ mod } n^2))^{-1} \text{ mod } n$, where $L(x) = \frac{x-1}{n}$.
        \item Output $pk$ and $sk$ such that $pk = (n, g)$ and $sk = (\lambda, \mu)$.
    \end{enumerate}
    \item \textbf{Enc.} This function encrypts a message $m$ as follows.
    \begin{enumerate}
        \item Take as input a message $m \in \mathbb{Z}_n$ and the public key $pk$.
        \item Select a random integer $r \in \mathbb{Z}^*_n$. This ensures the encryption algorithm to be probabilistic since a plaintext can have several corresponding ciphertexts. 
        \item Output a ciphertext $c \in \mathbb{Z}^*_{n^2}$, where $c = g^m \times r^n \text{ mod } n^2$.
    \end{enumerate}
    \item \textbf{Dec.} This function decrypts a ciphertext $c$ as follows.
    \begin{enumerate}
        \item Take as input a ciphertext $c \in \mathbb{Z}^*_{n^2}$ and the secret key $sk$.
        \item Output a plaintext message $m = L(c^\lambda \text{ mod } n^2) \cdot  \mu \text{ mod } n$.
    \end{enumerate}
    \item \textbf{Add.} Given the public key $pk$ and ciphertexts $c_i$, which are valid encryptions of plaintexts $m_i$, where $c_i = Enc(m_i) = g^{m_i} \cdot r_i^n \text{ mod } n^2$. The encryption of the addition of two plaintexts $m_1$ and $m_2$ is equal to the multiplication of the corresponding ciphertexts $c_1$ and $c_2$.
    \begin{proof}
    \begin{align*}
        c_1 \times c_2 \text{ mod } n^2 &= \left( (g^{m_1} \cdot r_1^n \text{ mod } n^2) \times (g^{m_2} \cdot r_2^n \text{ mod } n^2) \right) \text{ mod } n^2 \\
                       &= g^{m_1} \cdot r_1^n \times g^{m_2} \cdot r_2^n \text{ mod } n^2 \\
                       &= g^{m_1+m_2} \cdot (r_1r_2)^n \text{ mod } n^2 \\
                       &= \text{Enc}(m_1 + m_2)
    \end{align*}
    \end{proof}
    
    \item \textbf{Mult.} Given the public key $pk$, a plaintext $m_1$, and a ciphertext $c_2$ which is a valid encryption of a plaintext $m_2$, where $c_2 = Enc(m_2) = g^{m_2} \cdot r_2^n \text{ mod } n^2$. The encryption of the multiplication of two plaintexts $m_1$ and $m_2$ is equal to the ciphertext $c_2$ to the power of the plaintext $m_1$.
    
    \begin{proof}
    \begin{align*}
        c_2^{m_1} \text{ mod } n^2  &= (g^{m_2} \cdot r_2^n \text{ mod } n^2)^{m_1} \text{ mod } n^2 \\
                  &= g^{m_1 m_2} \cdot (r_2^{m_1})^n \text{ mod } n^2 \\
                  &= \text{Enc}(m_1 \times m_2)
    \end{align*}
    \end{proof}
    
    \end{itemize}

\subsection{Ideal Lattice Based Fully Homomorphic Encryption}

"Gentry and Boneh" proposed the first fully homomorphic encryption (FHE) scheme which is based on ideal lattices \cite{gentry2009fully}. The encryption process consists in hiding the message by adding noise, whereas, the decryption consists in using the secret key to remove the noise from the message.

This scheme can perform homomorphic operations on arbitrary depth circuit. For that, the authors start by constructing a somewhat homomorphic encryption (SWHE) scheme that can perform a limited number of operations. However, the size of the noise grows after each arithmetic operation, especially when it consists of multiplication, until it becomes not possible to decrypt. To avoid this problem, the authors proposed a technique called "bootstrapping" that consists in refreshing the ciphertext by reducing the size of the noise. This technique allows to transform a SWHE scheme into a FHE scheme. Nevertheless, this scheme is still a theoretical model because of its inefficiency. Later on, many more efficient schemes based on Gentry's construction have been proposed in the literature \cite{van2010fully, smart2010fully, gentry2011implementing, naehrig2011can, brakerski2011fully, brakerski2012fully, gentry2013homomorphic}.

\subsection{Fully Homomorphic Encryption over the Integers}

"Van Dijk et al." proposed a homomorphic encryption scheme \cite{van2010fully} that is similar to the scheme proposed by "Gentry and Boneh" \cite{gentry2009fully} except that it is simpler and less efficient since it works with integers instead of ideals. The semantic security of this scheme is based on the hardness assumption of the approximate great common divisor (GCD) problem \cite{howgrave2001approximate}.

\begin{defn}
The approximate GCD problem can be defined as follows.

Given a set of approximate multiples of $p$ $\{x_1, x_2, x_3, ..., x_\tau\}$, where \\$x_i = q_i \cdot p + r_i$ for $1 \leq i \leq \tau$, find $p$.
\end{defn}

In the following, we present the \emph{somewhat} homomorphic encryption construction, we suggest the reader to refer to the original paper for details \cite{van2010fully}.
\begin{itemize}
    \item \textbf{KeyGen.} The public and secret keys are generated as follows.
    \begin{enumerate}
        \item Take the security parameter $\lambda$ as input and calculate the parameters $\tau$, $\gamma$ and $\mu$ based on $\lambda$.
        \item Generate a random odd $\mu$-bits integer $p$.
        \item For $0 \leq i \leq \tau$, generate random $x_i = q_i \cdot p + r_i$, where $q_i \in [0,2^\gamma/p[$ and $r_i \in ]-2^p, 2^p[$
        \item Output $sk = p$ and $pk = \{x_0, x_1, x_2, ..., x_\tau\}$.
    \end{enumerate}
    \item \textbf{Enc.} A message $m \in \{0, 1\}$ is encrypted as follows.
    \begin{enumerate}
        \item Select a random subset $S \subset \{x_1, x_2, x_3, ..., x_\tau\}$.
        \item Select a random $r \in ]-2^p, 2^p[$.
        \item Output $c = (m + 2 \cdot r + 2 \cdot \sum_{i \in S}x_i) \text{ mod } x_0$.
    \end{enumerate}
    \item \textbf{Dec.} A ciphertext $c$ is decrypted as follows. $m = (c \text{ mod } p) \text{ mod } 2$.
    
    \item \textbf{Eval.} Given two ciphertexts $c_1$ and $c_2$ associated with the messages $m_1$ and $m_2$, respectively. The addition (resp., multiplication) of the two ciphertexts $c_1$ and $c_2$ is equal to the encryption of the addition (resp., multiplication) of their associated plaintexts $m_1$ and $m_2$.
    
    \begin{equation*}
      \begin{cases}
        & c_1 + c_2 = \text{Enc}(m_1 + m_2) \\
        & c_1 \times c_2 = \text{Enc}(m_1 \times m_2) \\
      \end{cases}
    \end{equation*}
    
\end{itemize}

In order to obtain a \emph{fully} homomorphic encryption scheme, the authors applied the bootstrapping technique proposed by "Gentry and Boneh" \cite{gentry2009fully}. Furthermore, to resist to the approximate GCD attack, the secret key should be very large, namely, its size should be at least $\tilde{O}(\lambda^{10})$ ($\lambda$ is the security parameter), which is the major drawback of this cryptosystem.

\subsection{Homomorphic Encryption from Learning with Errors}

"Brakerski and Vaikuntanathan" proposed an asymmetric fully homomorphic encryption scheme that operates over bits \cite{brakerski2011fully}. This scheme is based on the ring learning with errors (RLWE) assumption that was introduced by "Lyubashevsky et al." \cite{lyubashevsky2010ideal}. It manipulates polynomials with integer coefficients. In the following, we present the different functions of this scheme.

\begin{itemize}
    \item \textbf{KeyGen.} The public and secret keys are generated as follows.
        \begin{enumerate}
            \item Sample a ring element $s$ from $\mathbb{R}$ where, $\mathbb{R} = \frac{\mathbb{Z}[X]}{F[X]}$ is the ring of all polynomials modulo $F[X]$ which is an irreducible degree $n$ polynomial and $\mathbb{Z}[X]$ is the ring of polynomials over the integers.
            \item Calculates $pk = (a_0 = - (a_1 \cdot s + 2 \cdot e_0), a_1)$, where $a_1 \in \mathbb{R}_q$ is a ring element with coefficients modulo $q$ and $e_0$ is a ring element with small coefficients sampled from the discrete Gaussian distribution.
        \end{enumerate}
    \item \textbf{Enc.} This function encrypts a message $m$ as follows. 
    \begin{enumerate}
        \item Take as input the message $m \in \{0,1\}$ and the public key $pk = (a_0, a_1)$.
        \item Sample $u$, $g$, $h$ from $\mathbb{R}$.
        \item Output $c$, where, $c = (c_0, c_1) = (a_0 \cdot u + 2 \cdot g + m, a_1 \cdot u + 2 \cdot h$).
     \end{enumerate}
    \item \textbf{Dec.} This function decrypts a ciphertext $c$ as follows.
    \begin{enumerate}
        \item Take as input the ciphertext $c = (c_0, c_1)$ and the secret key.
        \item Compute $\tilde{m} = c_0 + c_1 \cdot s$.
        \item Output $\tilde{m}$ mod 2.
    \end{enumerate}
    \item \textbf{Eval.} This scheme supports homomorphic addition and multiplication operations. A noise management (re-linearization technique and dimension modulus reduction technique) should be applied after each multiplication. The tensor product (resp., addition) of the two ciphertexts is equal to the encryption of the multiplication (resp., addition) of their associated plaintexts.
\end{itemize}

\subsection{Leveled Fully Homomorphic Encryption}\label{sec_leveled}

\begin{figure}
\centering
\includegraphics[width=11cm,keepaspectratio]{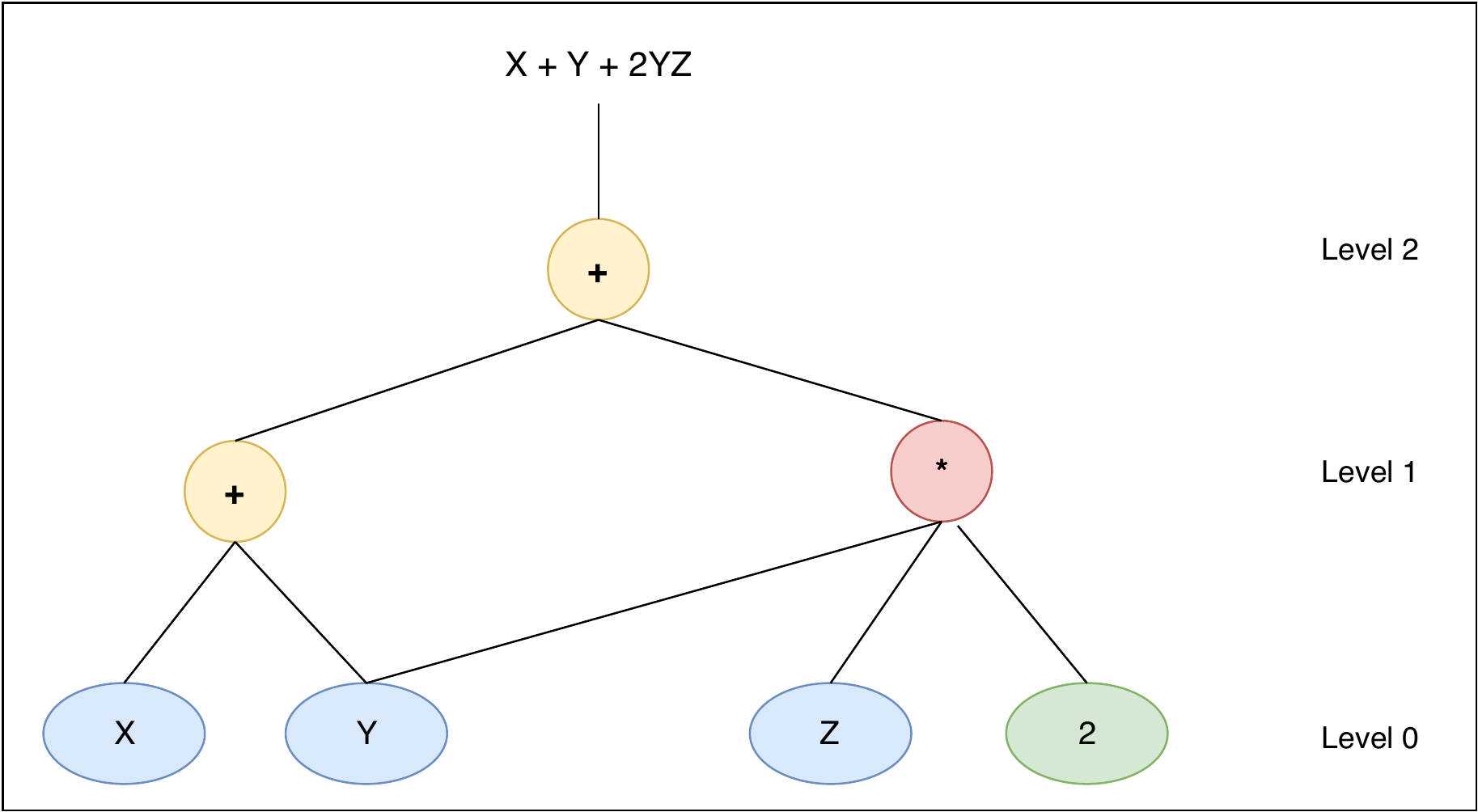}
\decoRule
\caption{Example of an arithmetic circuit}
\label{arithmetic}
\end{figure}

The term "leveled" is used to describe this approach because of the secret key updating performed at each level of a circuit. The latter represents an arithmetic function and is composed of a set of gates. Each gate corresponds to the operation of addition or multiplication (see Figure \ref{arithmetic}). Two main techniques are introduced in the leveled homomorphic encryption\footnote{In the rest of the thesis, the expression "leveled homomorphic encryption" will be used to describe this encryption method for simplicity.} \cite{brakerski2012leveled}, namely, the key switching and the modulus switching. 

\begin{itemize}
    \item \textbf{Key switching.} It is a technique inspired by another technique called  "re-linearization" proposed by "Brakerski and Vaikuntanathan" \cite{brakerski2011efficient, brakerski2011fully}. The key switching consists in the transition to a new key at each level of the circuit. Its main purpose is to reduce the size of ciphertexts in order to preserve the system efficiency during the calculations. This technique consists of two main functions.
\begin{itemize}    
\item \emph{SwitchKeyGen}$(s_1, s_2, q) \rightarrow B$. This function takes as input two private keys $s_1$ and $s_2$, and a modulus $q$. Its aim is to construct a public key (a matrix $B$) which enables the transition from the private key $s_1$ to a new private key $s_2$.
\item \emph{SwitchKey}$(B, c_1) \rightarrow c_2$. This function takes as input the matrix $B$ constructed by the function \emph{SwitchKeyGen}, and a ciphertext $c_1$ encrypted by the private key $s_1$. Its task is to produce a smaller ciphertext $c_2$ that encrypts the same message as the ciphertext $c_1$, and is decryptable by the second private key $s_2$.
\end{itemize}
\item \textbf{Modulus switching.} It is a refinement of the dimension reduction technique proposed by "Brakerski and Vaikuntanathan" \cite{brakerski2011efficient}. This method consists in reducing the modulus from $q$ to $p$ ($p < q$) in order to better manage the error contained in the ciphertext. In fact, the size of the error affects the number of homomorphic operations that is possible, since this size decreases after each multiplication. Thus, the aim of this technique is to switch from an exponential reduction of the error size after a multiplication to a linear reduction, which increases the number of homomorphic operations.
\end{itemize}

\section{Summary}
Three categories of encryption schemes frequently exploited in the searchable encryption area have been presented in this chapter.

First, we introduced the S$K$NN algorithm which is an order preserving symmetric encryption that allows to encrypt vectors while keeping the ability to calculate the similarity between those encrypted vectors by applying the inner product operation. This cryptosystem is the most used in the SE area. 

After that, we briefly explained the attribute based encryption that can be exploited to encrypt datasets when an access control policy should be applied.

Finally, we presented several variants of homomorphic encryption which is an encryption method that allows a third party to apply functions on encrypted data, and to obtain encrypted results without any need of decryption. However, many efficiency issues may be encountered such as the large size of ciphertexts and the slowness of the arithmetic operations. These drawbacks make this encryption scheme very difficult to be exploited in practice.
\chapter{Concept Based Semantic Searchable Encryption}

\label{Chapter4}

\section{Motivation}

The majority of searchable encryption schemes that have been proposed in the literature perform a keyword based search, but few studies \cite{sun:secure, fu2014semantic, yang2015attribute, fu2016enabling, fu2016towards} have exploited a semantic search over encrypted cloud data (see Subsection \ref{semantic_se}). These works are based on the query expansion technique that consists in adding the synonyms of the query terms. The drawback of these schemes is that, except the synonymy, other relationships between terms, such as the associative relationship, the homonymy, the instance-of relationship, and the related terms are not exploited. In other words, these techniques allow to improve the recall, but they are still far from the real semantic search.

Among the fields of semantic information retrieval, there are the contextual IR, the personalized IR, and the conceptual information retrieval\footnote{In this thesis, "conceptual IR" and "concept based IR" refer to the same area.}. Contrary to the other areas of the semantic search, machine learning, search context, and user profile are not utilized in the concept based IR. Therefore, the server can learn nothing, neither about the user interests nor about the content of the data collection. Consequently, the conceptual IR is the most appropriate for the construction of a semantic searchable encryption scheme.

Conceptual IR is based on concepts rather than keywords in the indexing and matching processes. Therefore, it is necessary to use an external knowledge base, such as an ontology or a thesaurus to perform the mapping process between keywords and concepts. Conceptual IR allows to detach from the syntactic aspect and take advantage of the natural language to perform a semantic guided search rather than relying on the syntax of the query.

In the rest of this chapter, we present the problem formulation and our proposed scheme.

\section{Problem Formulation}

In this section, the threat model, the design goals, and the system model are presented.

\subsection{Threat Model}

The security is a crucial aspect in cloud computing given that the outsourced data are often personal or professional. The cloud server is exposed to all kinds of external attacks. Therefore, every data (document, Index, query) should be encrypted before being outsourced. Moreover, we suppose that the cloud server is honest-but-curious, and can collect information about the content of documents by doing statistical analyzes. Hence, the search process should be secure and has to protect the data privacy.

When designing an SE scheme, it is important to take into account the threats discussed below. For this reason, several security constraints were elaborated in the literature (see Section \ref{sec_constraintes}). In this work, we focus on two security constraints, namely, the keyword privacy and the query unlinkability, in order to make our scheme IND-CSK2 secure.

\subsection{Design Goals}

Our goal is to propose a semantic searchable scheme over encrypted cloud data. For that, the Wikipedia ontology is exploited during the indexing and search processes. The documents and queries are represented in a vector space of dimension $n$ (where $n$ is the total number of concepts). Therefore, it is necessary to use an encryption method applicable on vectors such as the S$k$NN algorithm (see Section \ref{sec_sknn}). Moreover, in order to effectively exploit the Wikipedia ontology over encrypted data, a new weighting formula is proposed. Two major contributions are proposed in this work.

\begin{enumerate}
\item Introducing the Wikipedia ontology to the searchable encryption.
\item Proposition of a new weighting formula in order to effectively exploit the Wikipedia ontology in searchable encryption
\end{enumerate}

\subsection{System Model}

\begin{figure}
\centering
\includegraphics[width=11cm,keepaspectratio]{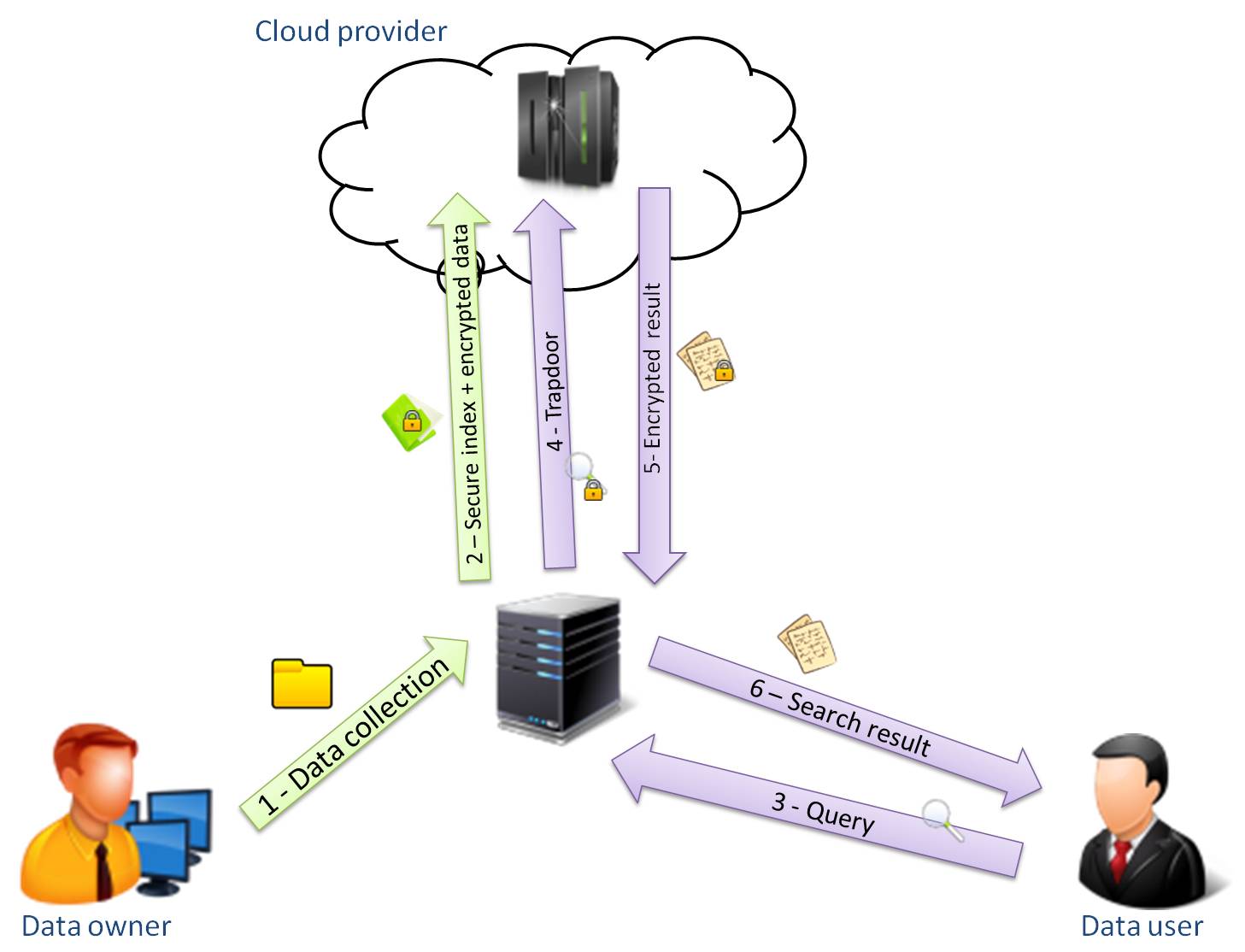}
\decoRule
\caption[General Architecture of SSE]{System Model}
\label{fig:archi_sses}
\end{figure}

Our proposed scheme exploits the Wikipedia ontology during the indexing process of documents and queries. Indeed, after the creation of the keyword based index, each document is represented by a vector of terms, and from those vectors, the data owner can construct a vector of concepts for each document using the Wikipedia ontology which is stored in an internal and trusted server. The set of the conceptual vectors corresponds to the index of the data collection. Then, the data collection is encrypted using AES, whereas, the concept based index is encrypted using the S$k$NN algorithm. After the encryption process is done, both the encrypted data collection and the secure index are outsourced to the cloud server. When an authorized user builds and sends an encrypted query to the server. The latter calculates the scalar product between each document vector and the query vector, and returns the most relevant documents to the user (Figure \ref{fig:archi_sses}).

\section{Proposed Scheme}

In this section, we present in detail our proposed approach that we call "semantic searchable encryption" (SSE) scheme.

\subsection{Wikipedia as Ontology}\label{sec_wiki}

In order to exploit the meaning of queries and documents, many researchers have utilized external resources such as dictionaries, thesauri, semantic graphs, and ontologies. In our work, we opted for the use of an ontology due to its robustness and reliability. More precisely, we decided to use Wikipedia as an ontology. The choice of Wikipedia is guided by its great richness of information given that it contains more than 4 million English pages, in addition, it contains articles in all areas and most languages.

\begin{figure}
\centering
\includegraphics[width=12cm,keepaspectratio]{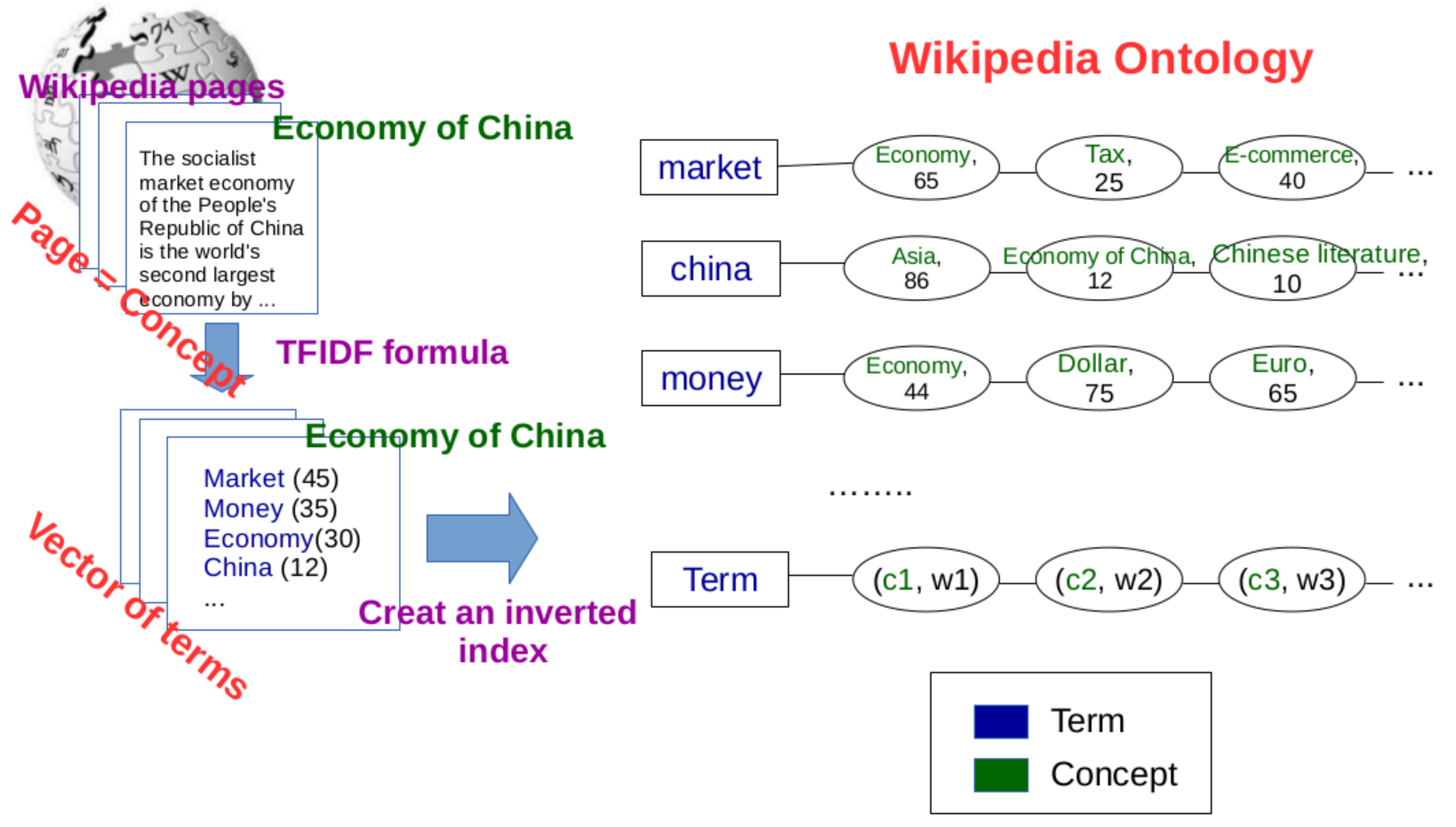}
\decoRule
\caption{Wikipedia ontology}
\label{fig:wiki}
\end{figure}

The Wikipedia ontology is constructed as follows \cite{gabrilovich2006overcoming, egozi:concept} (see Figure \ref{fig:wiki} and Subsection \ref{sec_wiki_IR}). 
\begin{enumerate}
    \item First, we suppose that each Wikipedia page corresponds to a concept (e.g., "Economy of China").
    \item After that, each concept is represented by a set of weighted terms extracted from the corresponding page (e.g., Economy of China = \{(market, 45),  (money, 35),  (economy, 30),  (China, 12), ...\}, where "market", "money", "economy", and "China" are terms and "Economy of China" is a concept).
    \item Finally, in order to accelerate the mapping process, an inverted index is created where each term is represented by a set of weighted concepts (e.g., China = \{(Economy of China, 12),  (Chinese literature, 10), ...\}, where "Economy of China" and "Chinese literature" are concepts, and "China" is a term).
\end{enumerate}

To implement our proposed scheme, we have constructed an ontology based on a version of Wikipedia dated \DTMusedate{date} containing \numprint{4828395} English pages. We have exploited \numprint{4342689} pages given that the special pages and those containing less than 100 words were ignored. The Wikipedia ontology that we have constructed contains \numprint{3965848} entries (terms), and each entry points on a set of at most \numprint{5000} concepts.

\subsection{Double Score Weighting Formula}\label{sec_DSW}

Concept based IR allows users to find relevant documents even if they do not contain query terms or their synonyms. This is explained by the fact that the search process is guided by the meaning through the use of an ontology.

Let us take the example given by "Egozi et al." \cite{egozi:concept}. Suppose that a user sends the query "shipwreck salvaging treasure" and the data collection contains the document entitled "Ancient artifacts found" (see Table \ref{2doc}). A keyword based search approach cannot retrieve that document since it has no term in common with the query. However, with the conceptual IR, this document would be returned to the user given that the document vector has few concepts in common with the query vector.

\begin{table*}
\centering
\caption{Two sample documents}
\label{2doc}
\begin{tabularx}{\textwidth}{l|X} \hline
\rowcolor{Gray}
Document title & Content \\ \hline
\textbf{Ancient artifacts found} & Divers have recovered artifacts lying underwater for more than 2,000 years in the wreck of a Roman ship that sank in the Gulf of Baratti, 12 miles off the island of Elba, newspapers reported Saturday. \\
\rowcolor{Gray}
\textbf{Olympic news in brief} & Cycling win for Estonia. Erika Salumae won Estonia's first Olympic gold when retaining the women's cycling individual sprint title she won four years ago in Seoul as a Soviet athlete. \\ \hline
\end{tabularx}
\end{table*}

Unfortunately, it happens that a conceptual search approach returns irrelevant documents that contain some query terms. To illustrate this, an example was given by "Egozi et al." \cite{egozi:concept}, if a user sends the query "Estonia economy" and the data collection contains the document entitled "Olympic news in brief" (see Table \ref{2doc}). As the keyword based search, conceptual search approaches cannot ignore this irrelevant document. This is justified by the high frequency of the term "Estonia" in the document "Olympic news in brief", and thus, the vector representing that document contains many concepts associated with the term "Estonia". Similarly, more than half of the concepts of the vector representing the query "Estonia economy" are associated with the term "Estonia". Therefore, there are many common concepts (34 concepts were found in our experimental study) between the document vector and the query vector. Consequently, a concept based search approach may return the document "Olympic news in brief" in response to the query "Estonia economy", despite it is supposed to be irrelevant.

In order to understand the root cause, we analyzed the top 10 concepts representing the document "Olympic news in brief" and the top 10 concepts representing the query "Estonia economy" (Table \ref{table1}). We also analyzed the top 10 concepts associated with the terms "economy" and "Estonia", separately (Table \ref{table2}).

\begin{table*}
\centering
\caption{Top 10 concepts when applying TFIDF formula}
\label{table1}
\begin{adjustbox}{width=\textwidth}
\begin{tabular}{c|c|c} \hline
\rowcolor{Gray}
Rank & Document (Olympic News In Brief) & Query (Estonia economy) \\ \hline
1 & Estonia national football team 2003 (0.81) & Estonia national football team 2003 (0.40) \\
\rowcolor{Gray}
2 & Estonia national football team 1997 (0.78) & Estonia national football team 1997 (0.39) \\
3 & Estonia national football team 1998 (0.72) & Estonia national football team 1998 (0.36) \\
\rowcolor{Gray}
4 & Estonia national football team 1995 (0.68) & Estonia national football team 1995 (0.34) \\
5 & Estonia national football team 1993 (0.63) & Estonia national football team 1993 (0.32) \\
\rowcolor{Gray}
6 & Estonia national football team 1994 (0.59) & Estonia national football team 1994 (0.30) \\
7 & UEFA Euro 1996 qualifying Group 4 (0.48) & UEFA Euro 1996 qualifying Group 4 (0.25) \\
\rowcolor{Gray}
8 & Energy in Estonia (0.46) & Energy in Estonia (0.23) \\
9 & Estonia at the Olympics (0.41) & UEFA Euro 2004 qualifying Group 8 (0.19) \\
\rowcolor{Gray}
10 & Estonia at the 2012 Winter Youth Olympics (0.41) & 1994 FIFA World Cup qualification (Group 1) (0.19) \\ \hline
\end{tabular}
\end{adjustbox}
\end{table*}

\begin{table*}
\centering
\caption{Top 10 concepts of terms, "Estonia" and "economy"}
\label{table2}
\begin{adjustbox}{width=\textwidth}
\begin{tabular}{c|c|c} \hline
\rowcolor{Gray}
Rank & Term (Estonia) & Term (Economy) \\ \hline
1 & Estonia national football team 2003 (0.40) & Closed household economy (0.16) \\ 
\rowcolor{Gray}
2 & Estonia national football team 1997 (0.39) & Information economy (0.14) \\ 
3 & Estonia national football team 1998 (0.36) & Small open economy (0.13) \\ 
\rowcolor{Gray}
4 & Estonia national football team 1995 (0.34) & Tiger economy (0.12) \\ 
5 & Estonia national football team 1993 (0.32) & Ministry of Economy (Bulgaria) (0.09) \\ 
\rowcolor{Gray}
6 & Estonia national football team 1994 (0.30) & Review of African Political Economy (0.09) \\ 
7 & UEFA Euro 1996 qualifying Group 4 (0.25) & Natural economy (0.09) \\ 
\rowcolor{Gray}
8 & Energy in Estonia (0.23) & Lithium economy (0.09) \\ 
9 & UEFA Euro 2004 qualifying Group 8 (0.19) & Pancasila economics (0.09) \\ 
\rowcolor{Gray}
10 & 1994 FIFA World Cup qualification (Group 1) (0.19) & Socialist-oriented market economy (0.08) \\ 
\end{tabular}
\end{adjustbox}
\end{table*}

On one hand, we notice that 8 of the top 10 concepts representing the document "Olympic news in brief" are part of the top 10 concepts associated with the term "Estonia". This is justified by the high frequency of the term "Estonia" in the document "Olympic news in brief" which increases the scores of the concepts associated with this term when applying Formula \ref{eqTFIDF}. On the other hand, even if the frequencies of the query terms are similar, we notice that the majority of the concepts representing the query are associated with the term "Estonia" instead of the term "economy". This is due to the fact that the concepts associated with the term "Estonia" have greater weights, and thus most of these concepts are selected to represent the query.

In order to represent documents and queries by the most appropriate concepts, we propose a new weighting formula that we call
"double score weighting" (DSW) Formula which allows to represent a document (or a query) by a set of concepts strongly associated with the general meaning of the document rather than representing it by concepts associated with terms that have the highest frequencies.

In Wikipedia ontology, each term is associated with a set of concepts. Thus, to represent a document by the most appropriate concepts, our idea is to select the concepts that are associated with the greatest number of terms of the document. For example, if we have the query "Estonia economy", to represent this query, it is more advantageous to choose a concept associated with both "Estonia" and "economy" than choosing a concept only associated with the term "Estonia", even  if the second concept has a greater score.

The proposed DSW formula allows to represent a document by concepts that are associated with its general meaning. In the following, we present the steps needed to represent a document by the most appropriate concepts using DSW formula.

\begin{enumerate}
\item Construct a weighted term vector for the document by applying the TFIDF formula.
\item Get all concepts associated with each term of the document vector constructed above by using the Wikipedia ontology.
\item Attribute two scores for each of these concepts as follows.
\begin{enumerate}
\item The first score corresponds to the number of distinct terms of the document vector associated with the concept. This score is called the primary score ($S_p$).
\item The second score is the TFIDF weight of the concept in the document. This score is called the secondary score ($S_s$) and is calculated by Formula \ref{eqTFIDF}.
\end{enumerate}
\item Sort the concepts with regard to their primary scores, then based on their secondary scores in the case of equality.
\begin{equation}\label{DSWF}
   (S_{p1}, S_{s1}) > (S_{p2}, S_{s2}) \Rightarrow (S_{p1} > S_{p2})\lor \left((S_{p1} = S_{p2}) \land (S_{s1} > S_{s2})\right) 
\end{equation}

\item Keep the top $X$ concepts with their associated primary and secondary scores to represent the document.
\end{enumerate}
We applied our method on the first example to calculate the similarity between the document "Ancient artifacts found" and the query "shipwreck salvaging treasure". We found that there are 13 common concepts between the top 100 concepts representing the document and the top 100 concepts representing the query (see Table \ref{table3}) rather than one concept when applying the method proposed by "Gabrilovich and Markovitch" \cite{gabrilovich2006overcoming}. Thus, as Gabrilovich's method, our method is able to retrieve relevant documents even if they have no term in common with the query. Besides, our method is more efficient than the Gabrilovich's method in retrieving relevant documents (13 common concepts in our method versus 1 common concept in Gabrilovich's method).

Similarly, we applied our method on the second example to calculate the similarity between the document "Olympic news in brief" and the query "Estonia economy". We did not find any common concepts between the top 100 concepts representing the document and the top 100 concepts representing the query (see Table \ref{table4}) which is correct since this document is not relevant to the query "Estonia economy". On the other hand, 34 common concepts were found when Gabrilovich's method was applied. Therefore, the DSW formula allows to correct the problem encountered when applying Gabrilovich's method, and thus, our proposed formula is able to ignore irrelevant documents even if they have terms in common with the query.

\begin{table*}
\centering
\caption{Top 10 concepts when applying DSW formula (example 1)}
\label{table3}
\begin{adjustbox}{width=\textwidth}
\begin{tabular}{c|c|c} \hline
\rowcolor{Gray}
Rank & Document (Ancient Artifacts Found) & Query (salvaging shipwreck treasure) \\ \hline 
1 & San Esteban (1554 shipwreck) (7, 0.006) & Abandoned Shipwrecks Act (3, 0.11) \\ 
\rowcolor{Gray}
2 & Roosevelt Inlet Shipwreck (6, 0.17) & Treasure hunting (marine) (3, 0.10) \\ 
3 & SS Vienna (1873) (6, 0.12) & 1715 Treasure Fleet (3, 0.09) \\ 
\rowcolor{Gray}
4 & SS Sagamore (1892) (6, 0.11) & Brent Brisben (3, 0.07) \\ 
5 & SS Samuel Mather (1887) (6, 0.10) & Odyssey Marine Exploration (3, 0.06) \\ 
\rowcolor{Gray}
6 & El Nuevo Constante (6, 0.09) & Nuestra Senora de Atocha (3, 0.05) \\ 
7 & Antikythera wreck (6, 0.09) & Urca de Lima (3, 0.05) \\ 
\rowcolor{Gray}
8 & Comet (steamboat) (6, 0.08) & Barry Clifford (3, 0.04) \\ 
9 & Underwater archaeology (6, 0.08) & Comet (steamboat) (3, 0.04) \\ 
\rowcolor{Gray}
10 & Mel Fisher (6, 0.07) & E. Lee Spence (3, 0.04) \\ 
19 & & San Esteban (1554 shipwreck) (3, 0.02) \\ 
\rowcolor{Gray}
63 & & Mel Fisher (2, 0.03) \\ 
73 & & El Nuevo Constante (2, 0.02) \\ \hline
\end{tabular}
\end{adjustbox}
\end{table*}

\begin{table*}
\centering
\caption{Top 10 concepts when applying DSW formula (example 2)}
\label{table4}
\begin{adjustbox}{width=\textwidth}
\begin{tabular}{c|c|c} \hline
\rowcolor{Gray}
Rank & Document (Olympic News In Brief) & Query (Estonia economy) \\ \hline
1 & Jorg Schmidt (7, 0.25) & Baltic Tiger (2, 0.07) \\ 
\rowcolor{Gray}
2 & Chu Mu-yen (7, 0.23) & Estonian European Union membership referendum (2, 0.06) \\ 
3 & Pavel Lednyov (7, 0.22) & Economy of Estonia (2, 0.06) \\ 
\rowcolor{Gray}
4 & Erika Meszaros (7, 0.20) & Tiit Vahi (2, 0.04) \\ 
5 & Estonia at the Olympics (6, 0.41) & Estonian Hound (2, 0.03) \\ 
\rowcolor{Gray}
6 & Chile at the Olympics (6, 0.32) & Baltic states housing bubble (2, 0.03) \\ 
7 & Erwin Keller (6, 0.31) & Post-communism (2, 0.03) \\ 
\rowcolor{Gray}
8 & Portugal at the Olympics (6, 0.28) & Currency board (2, 0.02) \\ 
9 & Georgeta Damian (6, 0.28) & OECD Development Centre (2, 0.02) \\ 
\rowcolor{Gray}
10 & Yuriy Poyarkov (6, 0.27) & Rosario Manalo (2, 0.01) \\ 
\end{tabular}
\end{adjustbox}
\end{table*}

\subsection{Semantic Searchable Encryption Scheme}

Our proposed appraoch called "semantic searchable encryption" (SSE) scheme is composed of five functions (\textit{KeyGen}, \textit{BuildOnto}, \textit{BuildIndex}, \textit{Trapdoor}, and \textit{Search}), and two main phases (\textit{Initialization} and \textit{Retrieval} phases). In the following, we present the five functions of our scheme.
\begin{itemize}
\item \textit{KeyGen ($m$, $u$) $\Rightarrow SK$.} The data owner randomly generates a secret key $SK = \{S, M_1, M_2\}$, where $S$ is a vector of size $(m + u + 1)$ and $(M_1, M_2)$ are two invertible matrices of size $(m + u + 1) \times (m + u + 1)$.
\item \textit{BuildOnto $\Rightarrow I_{wiki}$.} The ontology is built from Wikipedia as follows. First, we suppose that each Wikipedia page corresponds to a concept. Then, the English Wikipedia pages are indexed, where each page is represented by a vector of weighted terms by applying the TFIDF formula. Finally, to accelerate the mapping process between a term an its associated concepts, an inverted index of Wikipedia $I_{wiki}$ is created where each term is represented by a vector of weighted concepts (see Subsection \ref{sec_wiki} for details).
\item \textit{BuildIndex ($F$, $SK$) $\Rightarrow I'$.} The secure concept based index is built as follows. First, a vector of terms is constructed for each document of the data collection $F$ by applying the TFIDF formula. Then, using the Wikipedia ontology and the term vectors, a vector of concepts is built for each document by applying the DSW formula (see Subsection \ref{sec_DSW}). Finally, each vector of concepts is encrypted by applying the S$k$NN algorithm using the secret key $SK$ (see Section \ref{sec_sknn}). The set of the encrypted vectors constitutes the secure index $I'$ of the data collection $F$.
\item \textit{Trapdoor ($W$, $SK$) $\Rightarrow T$.} A trapdoor $T$ (encrypted query) is built from the query terms as follows. First, a vector of terms is constructed from the query keywords $W$, where the $i^{th}$ field of the vector is set to 1 if the query contains the corresponding term, otherwise, it is set to 0. After that, a vector of concepts is constructed to represent the query, by applying the DSW formula that exploits the Wikipedia ontology and the vector of terms (see Subsection \ref{sec_DSW}). Finally, the vector of concepts is encrypted by applying the S$k$NN algorithm that uses the secret key $SK$ (see Section \ref{sec_sknn}).
\item \textit{Search ($T$, $I'$, $k$).} Upon receipt of the trapdoor $T$ (represented by an encrypted vector of concepts), the cloud server calculates the scalar product between each document vector and the query vector. After that, Formula \ref{DSWF} is applied to sort the documents on the basis of their primary scores, and possibly their secondary scores in case of equality. Finally, the top-$k$ document IDs are returned to the user.
\end{itemize}

The search process consists of two main steps.

\begin{itemize}
\item \textit{Initialization phase.} In this phase, the data owner prepares the search environment as follows.
\begin{enumerate}
\item At first, the \textit{KeyGen} function is called to generate a secret key $SK$ that is shared with the authorized users through a secure communication protocol.
\item Then, the \textit{BuildOnto} function is called to construct an ontology from Wikipedia. This ontology is stored in an internal and trusted server, and is accessible by the authorized users.
\item Finally, the \textit{BuildIndex} function is called to construct a secure index from a collection of documents. The secure index and the encrypted data collection\footnote{The data collection is encrypted with a robust encryption algorithm such as \textit{AES}.} are outsourced to the cloud server.
\end{enumerate}
\item \textit{Retrieval phase.} This is the phase where the search process is performed as follows.
\begin{enumerate}
\item First, an authorized user calls the \textit{Trapdoor} function to build an encrypted query that is sent to the cloud server.
\item Upon the server receives the trapdoor, it calls the \textit{Search} function, and returns to the user the top-$k$ document IDs.
\end{enumerate}
\end{itemize}

\section{Result and Comparison}

"Yahoo! Answers"\footnote{\url{https://answers.yahoo.com/}, Feb. 2019.} is a website that allows users to ask, or answer questions that are asked by other users. A data collection is collected from this website to create the "Yahoo! Answers" corpus. This collection is composed of \numprint{142627} questions and \numprint{962232} answers. We performed our experimental study over this corpus. For that, we suppose that the queries correspond to the questions and the documents correspond to the answers.

We tested \numprint{1150} randomly selected queries to compare our proposed scheme with two other schemes. The first one is a keyword based scheme called MRSE \cite{cao2014privacy}, whereas, the second one is a concept based scheme proposed by "Gabrilovich and Markovitch" \cite{gabrilovich2006overcoming}.

\begin{figure}
\centering
\includegraphics[width=9cm,keepaspectratio]{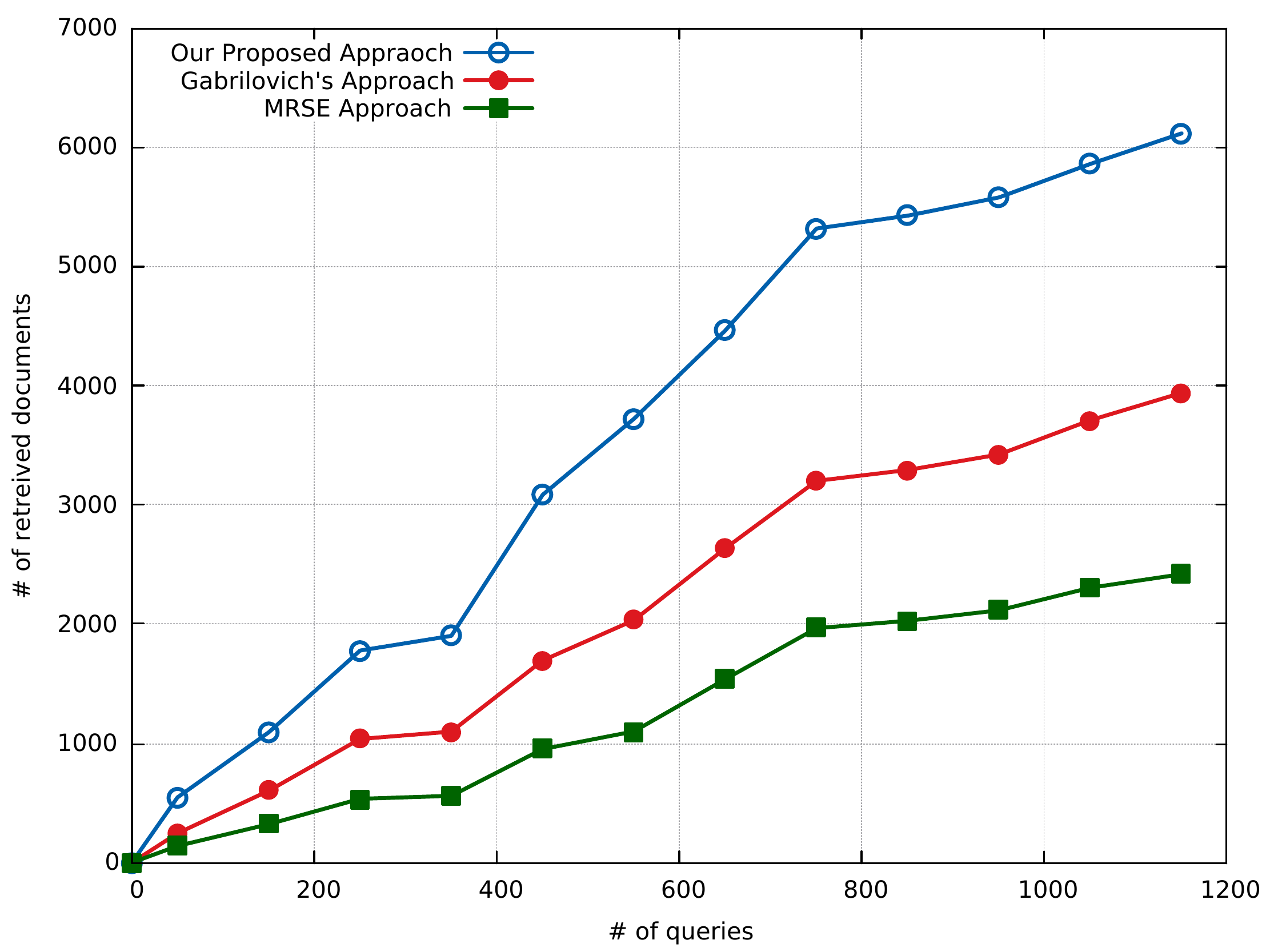}
\decoRule
\caption[Comparison between 3 different approaches]{The number of relevant documents that have been retrieved according to the number of queries in three different approaches.}
\label{exp1}
\end{figure}

Each scheme returns 100 documents in response to a received query. we calculated the total number of relevant documents that have been retrieved in each scheme according to the number of queries. Figure \ref{exp1} shows that the returned results in our proposed scheme have better accuracy than the returned results in the MRSE scheme (60\% of improvement) due to the use of a concept based search in our approach. Moreover, our approach returns better results than the Gabrilovich's scheme (36\% of improvement) due to the use of the proposed DSW formula in our approach. This clearly demonstrates that applying a concept based search (Our approach and Gabrilovich's scheme) increases the recall compared to a keyword based search approach (i.e, MRSE scheme). Moreover, our experiments confirm that the proposed DSW formula is more efficient than TFIDF formula used in the Gabrilovich's scheme.

\begin{figure}
\centering
\includegraphics[width=9cm,keepaspectratio]{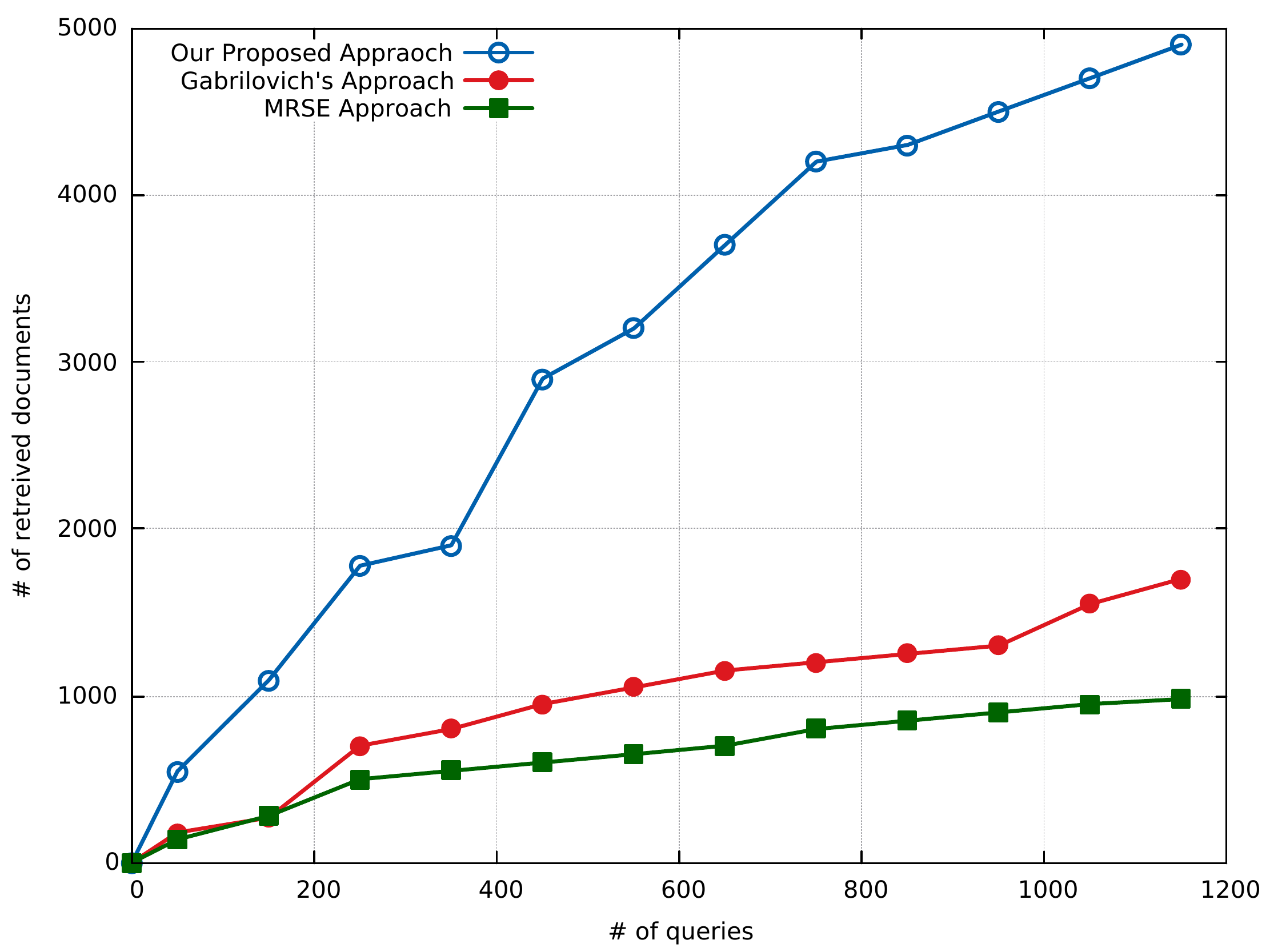}
\decoRule
\caption[Comparison between 3 approaches when applying a filter]{The number of relevant documents that have been retrieved according to the number of queries in three different approaches when applying a filter.}
\label{exp2}
\end{figure}

Then, in order to test the quality of the results that are returned by each scheme, we suppose that the detailed answers are better than the short ones. Thus, to measure the quality of the retrieved documents, we added a filter that ignores the documents which have a size less than a given threshold $\alpha$. We have gradually increased the value of this threshold as follows, $\alpha = 0$ in the first 50 queries, then, $\alpha = 10$ in the 50 queries that follow, then, $\alpha = 20$ in the third group of the 50 queries, and so on. Figure \ref{exp2} shows that the returned results in our proposed scheme have better accuracy than the returned results in both MRSE scheme and Gabrilovich's scheme. Indeed, our proposed scheme improves the accuracy by 67\% compared to the Gabrilovich's scheme and 84\% compared to the MRSE scheme.

\section{Summary}

In this chapter, we identified the problems of keyword based information retrieval that is exploited in most searchable encryption approaches. To fix these problems, we proposed a new semantic searchable encryption scheme. In fact, the use of a concept based search allows a significant enhancement of the \emph{recall} by retrieving relevant documents even if they do not contain any query term. Moreover, the use of the proposed DSW formula rather than TFIDF formula allows to increase the \emph{precision} by ignoring irrelevant documents that contain terms in common with the query. Finally, We validated our scheme by an experimental study, where we have compared our scheme with two other schemes proposed in the literature.


\chapter{Accelerated Searchable Encryption}

\label{Chapter5}

\section{Motivation}
For privacy and feasibility concerns, the vector representation of documents and queries is the most exploited in the SE area, instead of the inverted index \cite{yu:toward, cao2014privacy, li:efficient, wang2010secure, xia2016secure}. Unfortunately, this kind of representation has a negative effect on the search performance in terms of processing time, given that, during the search process, the query vector should be compared with each document vector. For instance, if there are one million documents, the similarity function which is a time consuming operation would be called one million times.

Very few studies \cite{wang:privacy, li:efficient} focused on the search performance by compressing the index in order to accelerate the search process. However, this technique causes the degradation of the quality of results without significantly improving the performance. 

In addition, to the best of our knowledge, all prior works assume that only one query is received at the same time. However, In practice, the server may receive several queries sent from different users simultaneously, making the search performance issue more complicated, since the search process is already slow while treating a single query. To overcome this challenge, we  propose several techniques that enable the server to treat multiple queries simultaneously and respond to each user within a reasonable time.

The goal of this work is to propose some techniques which allow to accelerate any searchable encryption scheme that uses the vector space model without any degradation on the search quality in terms of recall and precision. Our proposed parallelization techniques exploit several high performance computing (HPC) architectures, such as a multi-core CPU processor, a computer cluster, and a graphic processing unit (GPU) in order to accelerate the search process by distributing the work between several processes and treating lots of queries simultaneously. These solutions allow to achieve a speed-up of 46x.

The first challenge of our work is to find the most appropriate way to parallelize the search process and take benefit of each HPC architecture, whereas, the second one consists in reducing the number of servers without degrading the search performance by treating several queries simultaneously.


The rest of the chapter is organized as follows. First, we describe the background information of our work. After that, we give an overview of the problem that we have faced. Then, we present the techniques that we have proposed to accelerate the search over encrypted cloud data. Finally, we present the results of our experimental study.

\section{Background}
In this section, we briefly describe the background information of our work, where a summary of few HPC architectures that we have used in this work are disclosed.

\subsection{Graphic Processing Unit (GPU)}
GPU is a graphic processor that has a highly parallel structure which makes it very efficient for graphic tasks such as video memory management, signal processing, and image compression. Recently, other areas such as the industry and research are interested in using the GPU in parallel with the CPU in order to accelerate the calculations.

\begin{figure}
\centering
\includegraphics[width=7.5cm,height=9cm]{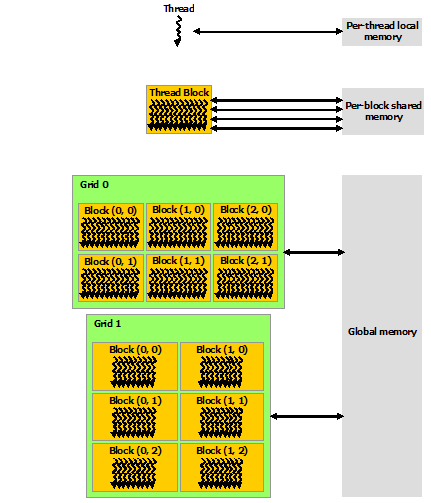}
\decoRule
\caption[Memory hierarchy in a GPU]{Memory hierarchy in a GPU \cite{cuda:programming}}
\label{fig:gpu}
\end{figure}

GPU architecture is based on SIMT paradigm (Single Instruction, Multiple Threads). When a program called "kernel" is launched on the GPU, it is executed in a parallel way by the different threads. GPUs are structured so that each set of threads is grouped in a block and each set of blocks is grouped in a grid. The number of threads and blocks is specified when the kernel is launched. Threads of the same block can access to a cached memory called "shared memory". All threads have access to other slower and larger memories such as the "global memory" and the "constant memory" (see Figure \ref{fig:gpu}).

Unlike CPU that includes a limited number of cores optimized for sequential processing, a GPU includes thousands of cores designed to deal effectively with a huge number of tasks, simultaneously. Therefore, to take advantage of the use of GPU and get the best performance, it is recommended\footnote{\url{http://docs.nvidia.com/cuda/cuda-c-programming-guide/}, Feb. 2019.} to assign to the GPU the heaviest portions of source code that can be calculated in a parallel way such as large matrices. The remainder of the source code that runs in a sequential way must be assigned to the CPU.

\subsection{Computer cluster}
A computer cluster is a grouping of a set of computers called "nodes" that appears as a single computer much more powerful in terms of processing power, storage space, and system memory. A computer cluster is constructed of a set of nodes. Each node can launch several processes simultaneously. Each node has a local memory that is shared between its different processes and is not accessible from other nodes. One process called "master" is chosen to be responsible for the distribution of the processing between the different nodes. A computer cluster architecture allows to perform calculations in a parallel way.

\section{Problem formulation}
Our work aims to improve the search performance. It is based on two main assumptions.

\subsection{Huge Number of Documents}
In searchable encryption area, it is very difficult to represent a data collection by an inverted index while respecting the security constraints and without causing any sensitive information leakage. It is mainly for this reason that the major approaches that have been proposed in the literature preferred to use a vector representation of documents to create a secure index \cite{yang2014secure, cao2014privacy, elmehdwi2014secure, xia2016secure, fu2016enabling, wang2018searchable, li2018personalized}. In addition, it is easier to encrypt vectors while keeping the ability to make calculations over them during the search process. In this representation, each document and query is represented by a vector of size $n$, where $n$ is the total number of terms in the data collection or the total number of concepts in the ontology, depending on the approach.

Cloud computing is often used when a user has a huge number of documents to be hosted or a large amount of data to be processed. During the search process, the server has to calculate the scalar product between each document vector and the query vector, sorts the result, and returns the top-$k$ documents that have the highest scores to the user. Considering the huge number of documents on the one hand, and the non-use of an inverted index on the other hand, the search process may be slower than expected (in our experiments, it took about 32s to perform a search on a collection of \numprint{400000} documents). This drawback can cause users dissatisfaction or worse, the abandonment of cloud computing.

Therefore, our first contribution aims to improve the search performance and make the search process much faster. For this purpose, we propose four techniques that utilize few HPC architectures in order to achieve a parallel computing. The idea is to exploit the parallelism during the calculation of the similarity scores, and during the sort of documents. The proposed techniques are applicable in any approach that exploits the vector representation of documents and queries.

\subsection{Huge Number of Queries}
Another problem that may be encountered in the information retrieval is when many queries are received simultaneously. This problem is even more pronounced in the searchable encryption given that no inverted index is used, and hence the search process is slower. This problem can be solved by providing enough servers in order to simultaneously process all queries that have been received at the same time by the cloud. Unfortunately, this solution is costly for the user who would pay for the use of these servers.

The goal of our second contribution is to manage several queries simultaneously, on condition that these queries are destined to the same data collection. Thus, instead of processing sequentially the queries that have been received at the same time, the cloud server performs several searches simultaneously, and returns to each user a list of top-$k$ documents in response to his request. This new contribution allows the cloud to decrease the number of calculators, without causing any large deterioration in the search performance.

\section{Proposed Techniques}

In this section, we present the four techniques that we propose to accelerate the search process over encrypted cloud data. For that, we start with an overview of the proposed techniques. Then, we present the different functions of the search process that we want to accelerate using our proposed techniques. Finally, we explain in detail the proposed solutions.

\subsection{Overview}

These new techniques aim to accelerate any searchable encryption approach based on the vector model. Indeed, these techniques allow to significantly reduce the search time without causing any degradation on the recall and precision. In other words, our solutions have no effect on the search results, but the search process becomes much faster.

In fact, the acceleration is done at the cloud server where the search process is executed. As mentioned above, there are two assumptions that have been taken into consideration when designing our techniques, namely, a huge number of documents are hosted in the cloud, and lots of queries may be simultaneously received. It is important to remember that in the vector space model, each document is represented by a vector, and the same goes for the queries, and during the search process the query vector is compared with each document vector. Therefore, in order to accelerate the search process, we exploit the parallel computing by distributing the documents between different processes, and treating multiple queries simultaneously.

\subsection{Functions to Accelerate}

As explained in the previous subsection, accelerating the search process is done at the cloud side. Therefore, we propose an accelerated version for each function supposed to be called by the cloud server. These accelerations are achieved by taking into account the two assumptions previously presented (huge number of documents, and lots of queries simultaneously received). We adapt each function used by the cloud server to our assumptions. Indeed, we designed these functions so that they can manage several queries simultaneously (contribution 2). The way these functions are exploited based on the architecture of the calculator (contribution 1) is explained in the next subsection.

The search process is composed of three main functions (\textit{Similarity}, \textit{Sort}, and \textit{Merge}).

\begin{itemize}

\item \textit{Similarity $(d_i, Q) \rightarrow R$.} This function has two parameters, namely, a document vector $d_i$, and a set of query vectors $Q = \{q_1, q_2, ..., q_n\}$. The similarity function computes the scalar product between the document vector $d_i$ and each query vector $q_j$, with $j \in \lbrack 1,n \rbrack$. The result is a list $R$ of size $n$, where the $j^{th}$ field corresponds to the similarity score between the document $d_i$ and the query $q_j$. 

\item \textit{Sort $(d_{ij}, D, k) \rightarrow r$.} This function returns the rank $r$ of a document $d_i$ with regard to a set of documents $D = \{d_{1j}, d_{2j}, ...,d_{mj}\}$, with $d_{ij} = (d_i, s_{ij})$, and $s_{ij}$ corresponds to the similarity score between the document $d_i$ and a query $q_j$. The rank $r$ is calculated by counting the number of documents of the set $D$ that have a higher score than the document $d_i$. During the computing process, if the rank $r$ exceeds a given value $k$, the function is stopped and the value \numprint{-1} is returned (see Algorithm \ref{sort}). This is justified by the fact that only the top-$k$ documents that have the highest scores are important.

\begin{algorithm}
\caption{Sort}
\label{sort}
\begin{algorithmic}
\REQUIRE $d_{ij} = (d_i, s_{ij}); D = \{d_{1j}, d_{2j}, ...,d_{mj}\}; k$
\ENSURE Rank $r$ of the document $d_i$ for a query $q_j$
\STATE $r \leftarrow 0$
\FOR{$x=1$ to $m$}
\IF{$s_{ij} < s_{xj}$}
\STATE $r \leftarrow r + 1$
\ENDIF
\IF{$r > k$}
\STATE $ r \leftarrow {-1}$
\STATE break
\ENDIF
\ENDFOR
\end{algorithmic}
\end{algorithm}

\item \textit{Merge $(D', D'', k) \rightarrow D$.} This function takes as parameters two sorted lists of documents $D'$ and $D''$, and an integer $k$. Its aim is to merge the two lists $D'$ and $D''$ into a single sorted list of top-$k$ documents (see Algorithm \ref{merge}). The merge function is often used in the case where two processors work simultaneously on different documents and each of them returns its own result.

\begin{algorithm}
\caption{Merge}
\label{merge}
\begin{algorithmic}
\REQUIRE k; $D' = \{d'_{1j}, d'_{2j}, ...,d'_{kj}\} / d'_{ij} = (d'_i, s'_{ij}); D'' = \{d''_{1j}, d''_{2j}, ...,d''_{kj}\} / d''_{ij} = (d''_i, s''_{ij})$
\ENSURE $D = \{d_{1j}, d_{2j}, ...,d_{kj}\} / d_{ij} = (d_i, s_{ij})$
\STATE $x \leftarrow 1, y \leftarrow 1, z \leftarrow 1$
\WHILE{$z \le k$}
\IF{$s'_{xj} > s''_{yj}$}
\STATE $d_{zj} \leftarrow d'_{xj} $
\STATE $x \leftarrow x + 1$
\STATE $z \leftarrow z + 1$
\ELSE
\STATE $d_{zj} \leftarrow d''_{yj} $
\STATE $y \leftarrow y + 1$
\STATE $z \leftarrow z + 1$
\ENDIF
\ENDWHILE
\end{algorithmic}
\end{algorithm}

\end{itemize}

\subsection{The proposed Techniques}
In this subsection, we present the different techniques that we have proposed in order to accelerate the search process. Each technique is an adaptation of the different functions of the search process (\textit{Similarity}, \textit{Sort}, and \textit{Merge}) for an HPC architecture (GPU, multi-core CPU processor, and computer cluster). Each technique operates in a parallel way where many processes work simultaneously. The difference between these calculators is mainly at their architectures such as the number of threads that can be executed simultaneously, the shared memory between threads, and the data transmission time from the CPU (the master process in the case of the computer cluster) to the different threads (see Table \ref{tab_comparison}). In the following, we present our four proposed techniques.

\begin{table}[!ht]
\caption{Comparison between three HPC architectures}
\label{tab_comparison}
\setlength\extrarowheight{2pt} 
\begin{tabularx}{\textwidth}{C|C|C|C}
\hline
\rowcolor{Gray}
\textbf{Architecture} & \textbf{Availability of shared memory between threads} & \textbf{Number of threads} & \textbf{Data transmission time to the threads} \\ \hline
Multi-core CPU processor & Yes & Dozens & Fast \\
\rowcolor{Gray}
GPU & Yes (but there is no shared memory between GPU and CPU) & Thousands & Slow \\ 
Computer cluster & Partially (no shared memory is available between the processes of different nodes) & Hundreds & Slow \\ \hline
\end{tabularx}
\end{table}

\subsubsection{Multi-Core Based Solution}

In this technique, several threads are executed simultaneously, where the processing of documents is distributed between them. Moreover, each thread has access to the main memory. The search process in a multi-core CPU processor is accomplished as follows.

\begin{enumerate}
\item First, the main process receives a set of queries destined for the same data collection.
\item Upon receiving the queries, it launches a set of threads (the number of threads is greater than or equal to the number of cores).
\item After that, the main process sends to each thread, the received queries and a subset of documents (the documents are equitably distributed between the different threads).
\item The threads work simultaneously. Each of them calls the \textit{Similarity} function to calculate the similarity score between each of its documents and each query, so that a document has several scores, each of them corresponds to a query.
\item Then, the main process synchronizes the threads. For that, it waits until all threads finish their work before starting the next step.
\item Next, for each query, each thread calls the \textit{Sort} function to assign a rank to each of its documents compared with the whole documents of the data collection, including those of other threads, since the main memory is shared between all threads. Therefore, a document has several ranks where each rank corresponds to a query.
\item Finally, when all threads finish their works, the main process collects for each query, a list of top-$k$ documents.
\end{enumerate}

\subsubsection{GPU Based Solution}

In this technique, we adapted the search process for a GPU. For that, each GPU thread has to deal with only one document. It should be noted that the GPU global memory can be accessed by any thread. The search process in a GPU can be performed as follows.

\begin{enumerate}
\item First, the CPU receives at the same time several queries destined for the same data collection.
\item Upon the queries are received, it transfers both the document vectors and the received queries to the GPU.
\item The GPU launches enough threads so that each thread deals with only one document.
\item After that, each thread calls the \textit{Similarity} function to calculate the similarity score between a document and each received query, so that each document has a score for each query.
\item Each thread waits until all other threads finish their work before it proceeds to the next step.
\item Then, for each query, a thread calls the \textit{Sort} function to assign a rank to its document compared with the documents of other threads given that the global memory is accessible by all threads. That way, each document will have a rank for each query.
\item Finally, the GPU transfers to the CPU a set of lists of to-$k$ documents where each list corresponds to the result of a query.
\end{enumerate}

\subsubsection{Hybrid Solution}

In this technique, the idea is not to let the CPU pending while the GPU is working. For that, we thought of sharing the work between the CPU and the GPU. Each of them deals with a part of the data collection by calling its own threads. Thus, this solution is an aggregation of the two previous solutions. In the following we give the details of this third solution.

\begin{enumerate}
\item First, the CPU receives a set of queries destined for the same collection of documents.
\item Upon the queries are received, it launches a set of threads, and sends to them the received queries and a part of the index (a subset of document vectors).
\item Then, it sends the other part of the index as well as the received queries to the GPU.
\item After that, the "multi-core based solution" (steps 4 to 7) is applied by the CPU to obtain for each query a list of top-$k$ documents.
\item Simultaneously, the "GPU based solution" (steps 3 to 7) is applied by the GPU to obtain for each query a list of top-$k$ documents.
\item Finally, for each query, the CPU calls the \textit{Merge} function to merge the list of documents retrieved by the GPU with the list of documents retrieved by the CPU into a single list of top-$k$ documents.
\end{enumerate}

\subsubsection{Cluster Based Solution}

In this technique, the search process is distributed between the different nodes of the computer cluster. It is important to know that the nodes do not share any memory between them. Thus, each node which is composed of a set of processes, works completely autonomously of other nodes. The search process in a computer cluster is performed as follows.

\begin{enumerate}
\item First, the master process receives several queries destined for the same data collection.
\item Upon receiving the queries, the master process distributed the treatment between different processes. For that, it sends to each process a subset of document vectors and the received queries.
\item Then, For each query, a process calls the \textit{Similarity} function to assign a similarity score to each of its documents. At the end of this step, each document will have a score for each query.
\item The processes operate simultaneously and asynchronously. Therefore, when a process terminates the calculations of the similarity scores, it continues with calculating the ranks of documents. For that, for each query, each process calls the \textit{Sort} function to assign a rank to each of its documents compared with the documents that it treats (rather than with all documents of the collection as is done in both multi-core based solution and GPU based solution given that there is no shared memory between the nodes in a computer cluster). At the end of this step, each document will have a rank for each query.
\item After that, each process returns for each query a list of top-$k$ documents to the master process.
\item Whenever the master process receives the results, it calls the \textit{Merge} function to merge the results of the same query until it obtains a final list of top-$k$ documents for each query.
\end{enumerate}

\section{Results and comparison}

We performed our experiments on "Yahoo! Answers" collection that contains \numprint{142627} queries and \numprint{962232} documents. We applied our techniques on the proposed approach entitled "semantic searchable encryption" (SSE) that performs a semantic search through a secure index based on the vector model (see Chapter \ref{Chapter4}). We performed several tests for each proposed technique. We realized two kinds of experiments for each solution. 
\begin{itemize}
\item The first experiment consists of calculating the time required for the search process in response to a query based on the number of documents (between \numprint{50000} and \numprint{400000}). 
\item The second experiment consists of calculating the time needed to perform a search on a collection of \numprint{200000} documents based on the number of queries that are received simultaneously. 
\end{itemize}

In the following, we analyze the results obtained in the four proposed techniques.

\subsection{Sequential Search}
First, we tested the SSE approach that works in a sequential way without any parallelism. The experiments were performed on a computer equipped with an Intel Xeon 2.13 GHz processor. Table \ref{tab_ch5_table1} shows that the search process takes about 32s to treat \numprint{400000} documents in response to a single query. In addition, concerning the second experiment, table \ref{tab_ch5_table2} shows that it takes longer than 206s to perform a sequential search on a collection of \numprint{200000} documents in response to 20 queries that have been received at the same time.

\begin{sidewaystable}
\Centering
\caption[Results of experiment 1]{The time (ms) needed for the search process based on the number of documents (experiment 1)}
\label{tab_ch5_table1}
\begin{tabular}{c|c|c|c|c|c|c} \hline
\rowcolor{Gray}
\textbf{Solution} & \textbf{100k doc} & \textbf{150k doc} & \textbf{200k doc} & \textbf{250k doc} & \textbf{300k doc} & \textbf{400k doc} \\ \hline
Sequential search & \numprint{8017.80} & \numprint{12398.80} & \numprint{17101.00} & \numprint{20996.20} & \numprint{24285.40} & \numprint{31753.20} \\ 
\rowcolor{Gray}
Multi-core (4 threads) & \numprint{2349.60} & \numprint{3715.60} & \numprint{5135.40} & \numprint{6300.00} & \numprint{7586.80} & \numprint{9983.80} \\ 
Multi-core (8 threads) & \numprint{2311.80} & \numprint{3551.00} &	\numprint{4939.60} & \numprint{5986.60} & \numprint{6854.00} & \numprint{9118.00} \\ 
\rowcolor{Gray}
Multi-core (16 threads) & \numprint{2334.40} & \numprint{3516.00} & \numprint{4790.60} & \numprint{5885.80} & \numprint{6874.20} & \numprint{8885.60} \\ 
GPU & \numprint{1726.80} & \numprint{2750.80} & \numprint{4466.60} & \numprint{4764.60} & - & - \\ 
\rowcolor{Gray}
GPU-CPU (40/60) & \numprint{1814.60} & \numprint{2571.80} & \numprint{3523.00} & \numprint{4375.80} & \numprint{5135.20} & \numprint{6561.20} \\ 
GPU-CPU (50/50) & \numprint{1552.60} & \numprint{2297.20} & \numprint{3122.00} & \numprint{3798.80} & \numprint{4602.40} & \numprint{5837.40} \\ 
\rowcolor{Gray}
GPU-CPU (60/40) & \numprint{1363.80} & \numprint{2310.40} & \numprint{3124.00} & \numprint{3811.00} & \numprint{4593.60} & \numprint{6383.80} \\ 
Cluster (8 processes) & \numprint{1979.20} & \numprint{2966.00} & \numprint{3958.40} & \numprint{4967.80} & \numprint{5951.40} & \numprint{7967.20} \\ 
\rowcolor{Gray}
Cluster (16 processes) & \numprint{1777.60} & \numprint{2660.20} &	\numprint{3545.20} & \numprint{4438.60} & \numprint{5337.80} & \numprint{7104.40} \\ 
Cluster (32 processes) & \numprint{1890.80} & \numprint{2773.80} &	\numprint{3648.80} & \numprint{4531.60} & \numprint{5401.20} & \numprint{7153.40} \\ 
\rowcolor{Gray}
Cluster (64 processes) & \numprint{2210.00} & \numprint{3088.80} &	\numprint{3836.40} & \numprint{4871.40} & \numprint{5887.20} & \numprint{7463.00} \\ 
Cluster (128 processes) & \numprint{2571.40} & \numprint{3709.40} & \numprint{4835.00} & \numprint{6090.60} & \numprint{6278.60} & \numprint{7721.20} \\ \hline
\end{tabular}
\end{sidewaystable}

\begin{sidewaystable}
\Centering
\caption[Results of experiment 2]{The time (ms) needed to perform a search on a collection of \numprint{200000} documents based on the number of queries (experiment 2)}
\label{tab_ch5_table2}
\begin{tabular}{c|c|c|c|c|c} \hline
\rowcolor{Gray}
\textbf{Solution} & \textbf{4 queries} & \textbf{8 queries} & \textbf{12 queries} & \textbf{16 queries} & \textbf{20 queries} \\ \hline
Sequential search & \numprint{46525.20} & \numprint{86356.40} & \numprint{125591.20} & \numprint{165183.60} & \numprint{206659.80} \\ 
\rowcolor{Gray}
Multi-core (16 threads) & \numprint{12267.40} & \numprint{22367.80} & \numprint{32431.80} & \numprint{43036.60} & \numprint{54280.80} \\ 
GPU & \numprint{6993.60} & \numprint{11319.60} & \numprint{15603.20} & \numprint{19901.20} & \numprint{24146.00} \\ 
\rowcolor{Gray}
GPU-CPU (50/50) & \numprint{6949.80} & \numprint{12129.60} & \numprint{17217.00} & \numprint{22287.00} & \numprint{27802.80} \\ 
GPU-CPU (60/40) & \numprint{5747.40} & \numprint{9994.60} & \numprint{14197.00} & \numprint{18046.20} & \numprint{23072.40} \\ 
\rowcolor{Gray}
GPU-CPU (70/30) & \numprint{5477.60} & \numprint{8543.20} & \numprint{11558.40} & \numprint{14515.60} & \numprint{17702.40} \\ 
GPU-CPU (80/20) & \numprint{6167.00} & \numprint{9286.40} & \numprint{12704.40} & \numprint{16131.20} & \numprint{19225.20} \\ 
\rowcolor{Gray}
Cluster (8 processes) & \numprint{4796.60} & \numprint{6016.00} & \numprint{7283.20} & \numprint{8501.80} & \numprint{9666.40} \\ 
Cluster (16 processes) & \numprint{3928.00} & \numprint{4499.60} & \numprint{5082.00} & \numprint{5638.60} & \numprint{6182.40} \\ 
\rowcolor{Gray}
Cluster (32 processes) & \numprint{3835.00} & \numprint{4108.00} & \numprint{4388.20} & \numprint{4658.80} & \numprint{4922.40} \\ 
Cluster (64 processes) & \numprint{3942.60} & \numprint{4072.20} & \numprint{4181.00} & \numprint{4305.00} & \numprint{4474.00} \\ 
\rowcolor{Gray}
Cluster (128 processes) & \numprint{4850.20} & \numprint{4946.60} & \numprint{5006.60} & \numprint{5083.80} & \numprint{5131.60} \\ \hline
\end{tabular}
\end{sidewaystable}

\subsection{Multi-core Based Solution}
We tested our first technique on a computer equipped with an Intel Xeon 2.13 GHz processor (4 cores). For the first experiment where we calculate the time needed to perform a search based on the number of documents in response to a single query, we tested the first solution with different numbers of threads.

\begin{figure}
\centering
\includegraphics[width=9cm,keepaspectratio]{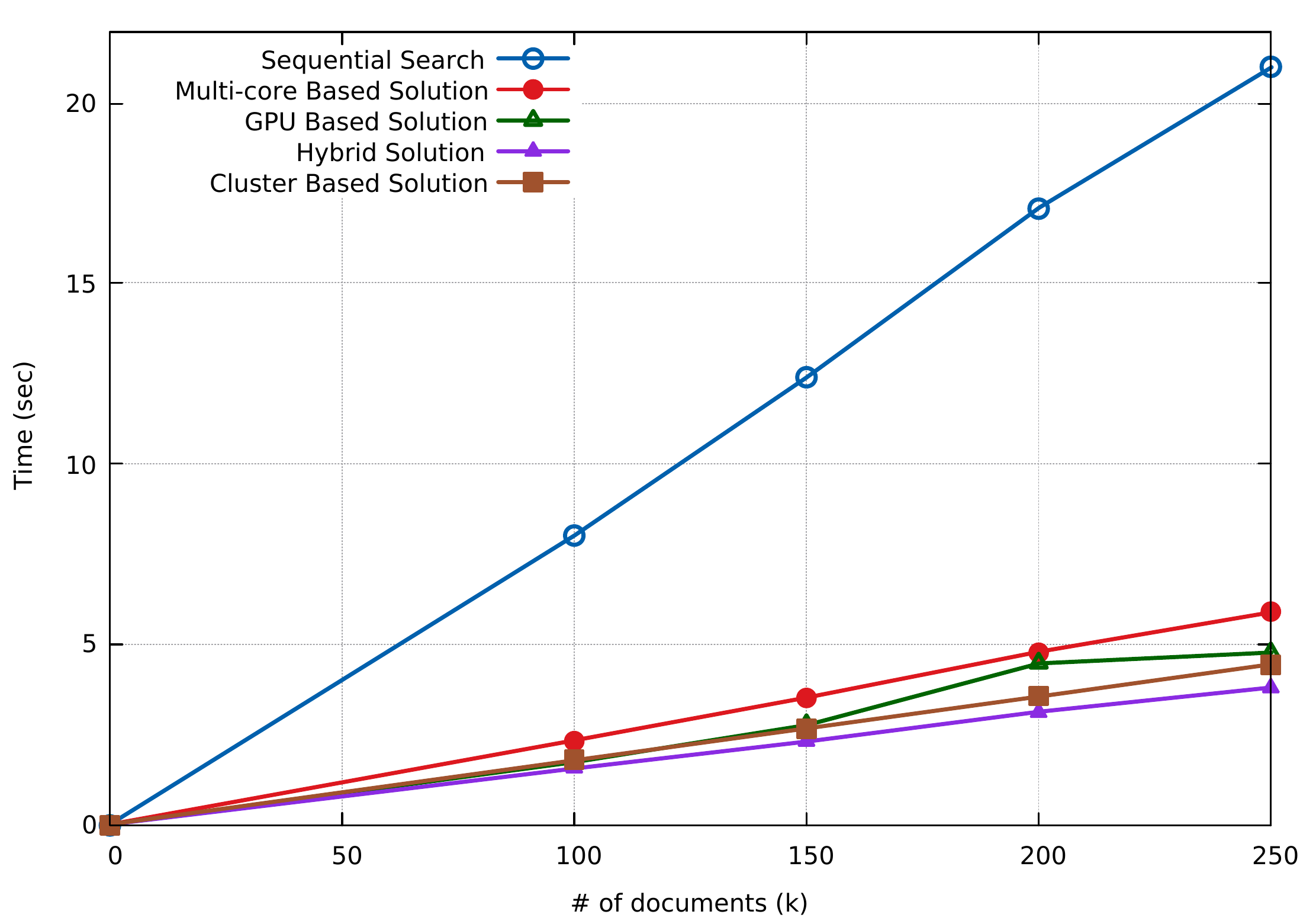}
\caption[Results of experiment 1]{The time (ms) needed for the search process in response to a query based on the number of documents}
\label{diag1}
\end{figure}

\begin{figure}
\centering
\includegraphics[width=9cm,keepaspectratio]{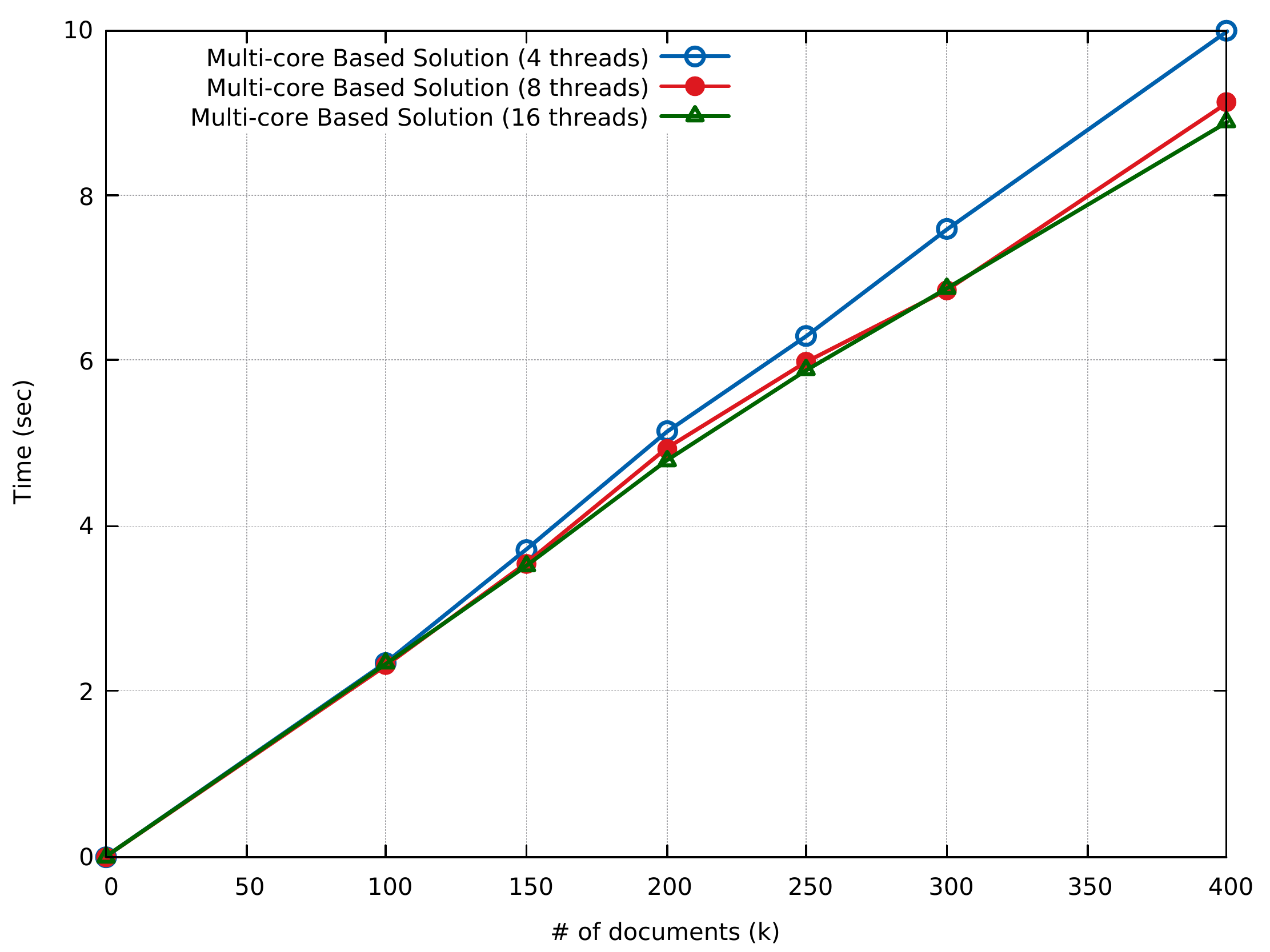}
\caption[Multi-core based solution, experiment 1]{Multi-core based solution tested with different number of threads}
\label{diag3}
\end{figure}

Figure \ref{diag1} shows that the multi-core based solution accelerates the search process compared to the sequential search by reaching a speed-up around 4x. This acceleration is justified by the distribution of the work between threads that run simultaneously. In addition, Figure \ref{diag3} shows that increasing the number of threads may provide an improvement in the search performance. Therefore, it is preferable to use a larger number of threads than the number of cores.

\begin{figure}
\centering
\includegraphics[width=9cm,keepaspectratio]{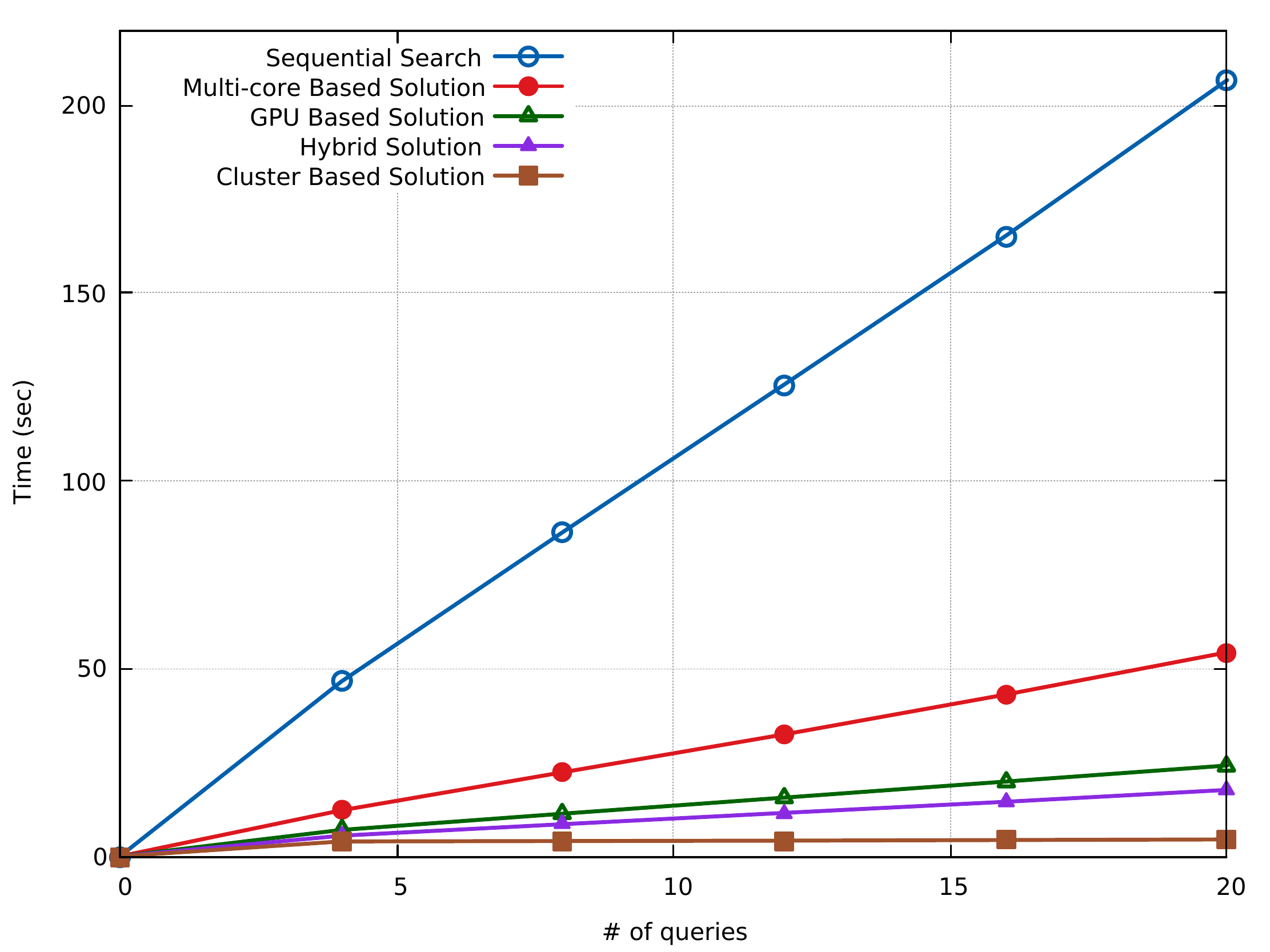}
\caption[Results of experiment 2]{The time (ms) needed to perform a search on a collection of 200,000 documents based on the number of queries}
\label{diag2}
\end{figure}

For the second experiment, we tested this solution using 16 threads with the same processor. Remember that this experiment consists of calculating the time needed to perform a search on a collection of \numprint{200000} documents based on the number of queries. Figure \ref{diag2} shows that this solution brings an improvement in the search performance compared to the sequential search by achieving a speed-up around 4x.

\subsection{GPU Based Solution}
We tested this technique on "NVIDIA Tesla C2075" graphic processor. For the first experiment, we used a number of documents between \numprint{50000} and \numprint{250000} because of the limited memory space of this GPU. Figure \ref{diag1} shows that there is an improvement compared to the sequential search due to an acceleration around 4.5x. However, there is not any significant improvement compared to the previous solution where we have used a multi-core CPU processor. Indeed, even if a GPU contains thousands of threads that operate simultaneously, the data transmission from the CPU memory to the GPU memory takes few additional seconds depending on the number of documents. Consequently, the time saved during the calculation process, is wasted during the data transmission. To overcome this problem, it is important to attribute as much as possible of work to the GPU such as treating several queries simultaneously instead of a single query.

Therefore, in the second experiment which consists in treating multiple queries simultaneously, the GPU is expected to have a better performance given that the threads have more work. Figure \ref{diag2} shows that our assumption is correct given that GPU based solution gives a better performance than both sequential solution and multi-core based solution. Indeed, GPU based solution which consists in treating multiple queries simultaneously by a GPU achieves an acceleration around 8.5x.

\subsection{Hybrid Solution}
\begin{figure}
\centering
\includegraphics[width=9cm,keepaspectratio]{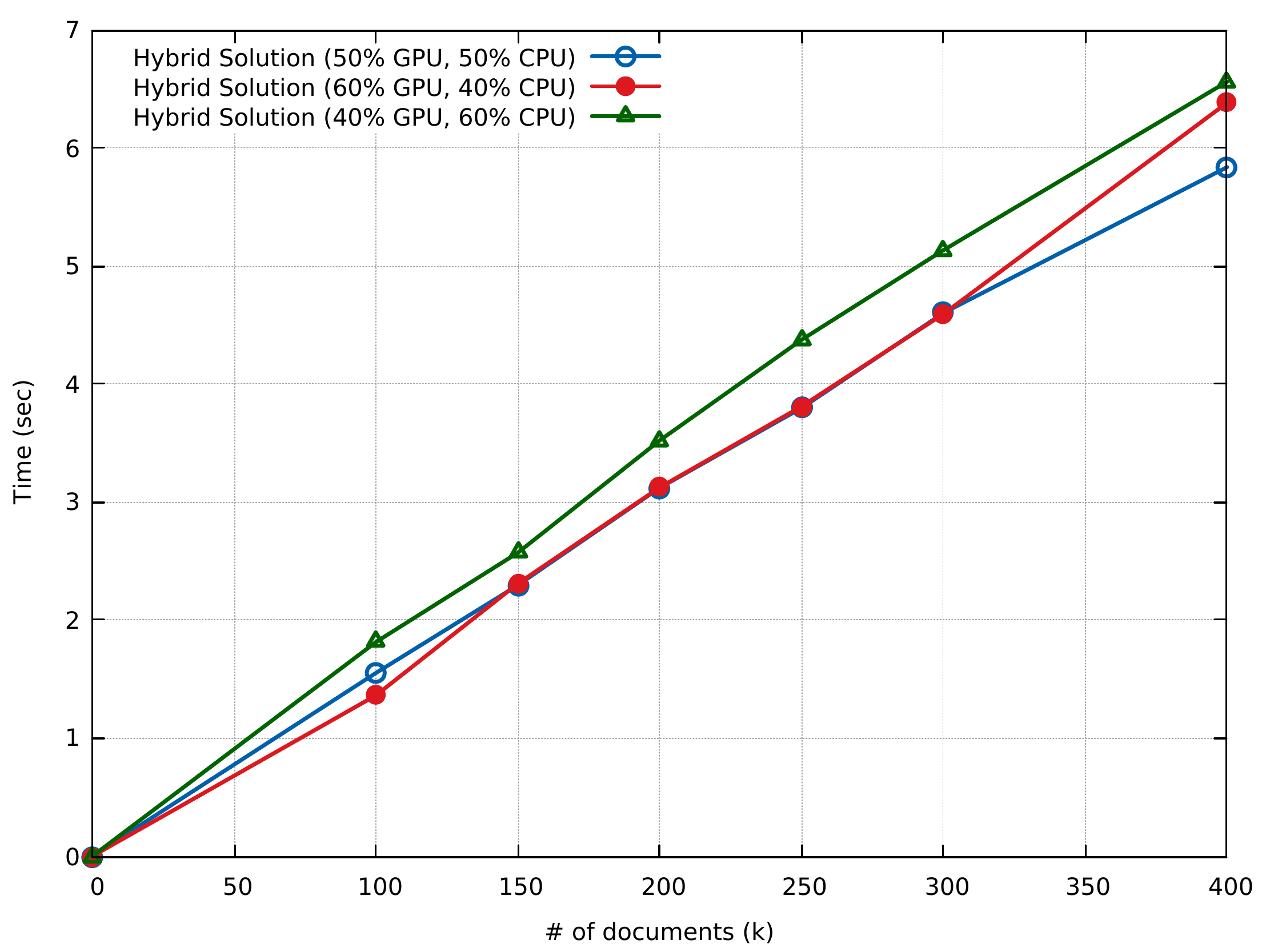}
\caption[Hybrid solution, experiment 1]{Different distributions of documents between GPU ad CPU (experiment 1)}
\label{diag4}
\end{figure}

We tested the third technique using a calculator equipped with an Intel Xeon 2.13 GHz processor (4 cores) and a NVIDIA Tesla C2075 graphic processor. In the first experiment, we tested different distributions of documents between GPU and CPU. The best performance was obtained by assigning 50\% of documents to the GPU and 50\% of documents to the CPU (see Figure \ref{diag4}). However, this distribution is not always the most optimal, since, it depends on several criteria such as the data transmission frequency, the CPU frequency, and the GPU frequency. Figure \ref{diag1} shows that in the case of single query, the hybrid solution gives a better performance than both multi-core based solution and GPU based solution by achieving an acceleration around 5.5x.

\begin{figure}
\centering
\includegraphics[width=9cm,keepaspectratio]{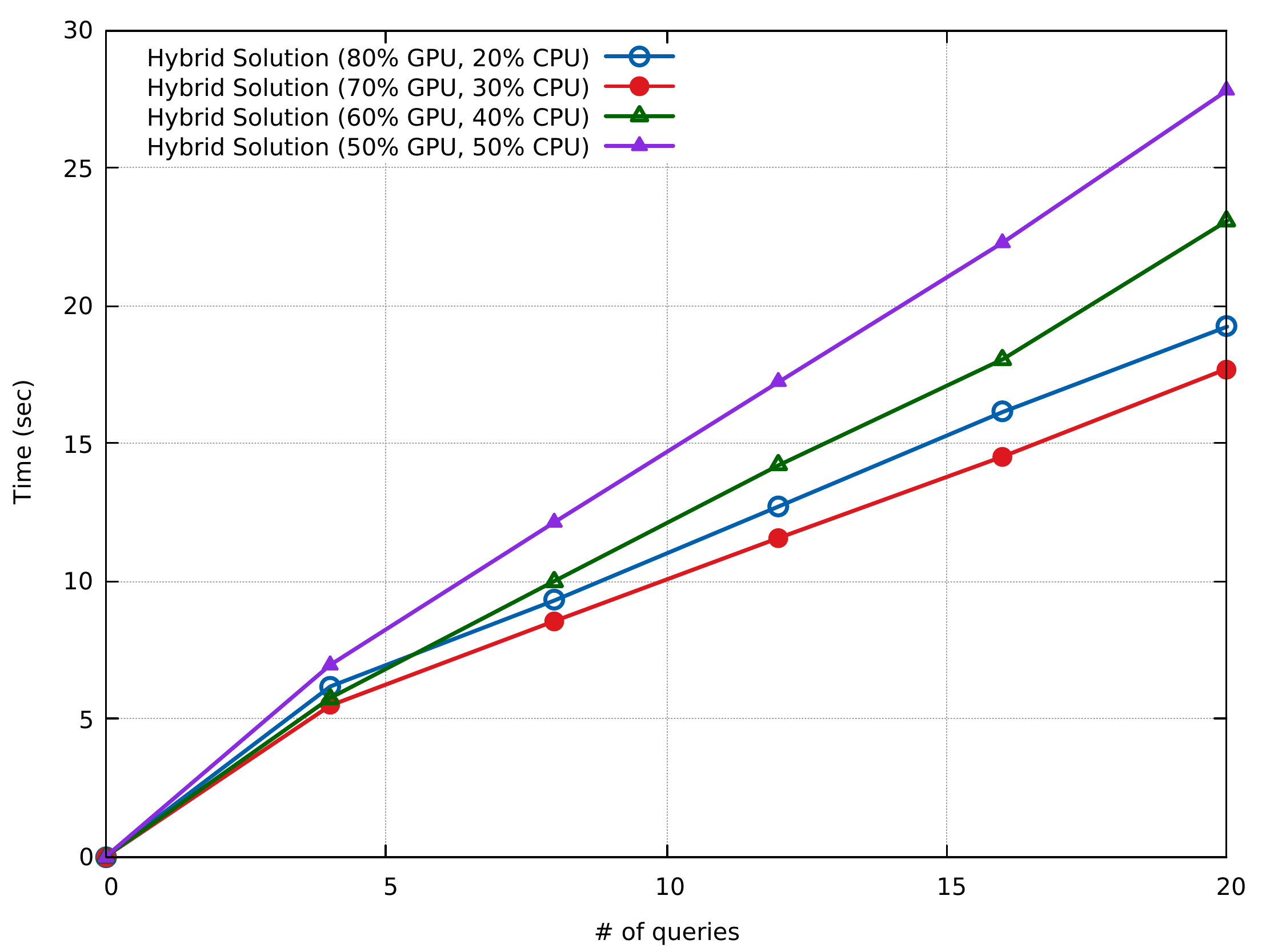}
\caption[Hybrid solution, experiment 2]{Different distributions of documents between GPU ad CPU (experiment 2)}
\label{diag5}
\end{figure}

In the second experiment, we also tested several distributions of documents between GPU and CPU. In this case, we chose to assign the greater part of documents to the GPU given that in the case of multiple queries, a GPU is much faster than a CPU. Figure \ref{diag5} shows that the best performance was obtained by assigning 70\% of documents to the GPU and 30\% of documents to the CPU. Moreover, we notice that in the case of multiple queries, the hybrid solution is better than the two previous solutions by reaching an acceleration around 12x (Figure \ref{diag2}).

\subsection{Cluster Based Solution}
We tested this technique on "CERIST IBNBADIS cluster"\footnote{\url{http://www.rx-racim.cerist.dz/?page_id=207}, Feb. 2019.} which is composed of 32 compute nodes where each node is equipped with an "Intel Xeon SandyBridge" bi-processor of 16 cores. Therefore, IBNBADIS cluster has a total of 512 cores and a theoretical power of 8 TFLOPS.

\begin{figure}
\centering
\includegraphics[width=9cm,keepaspectratio]{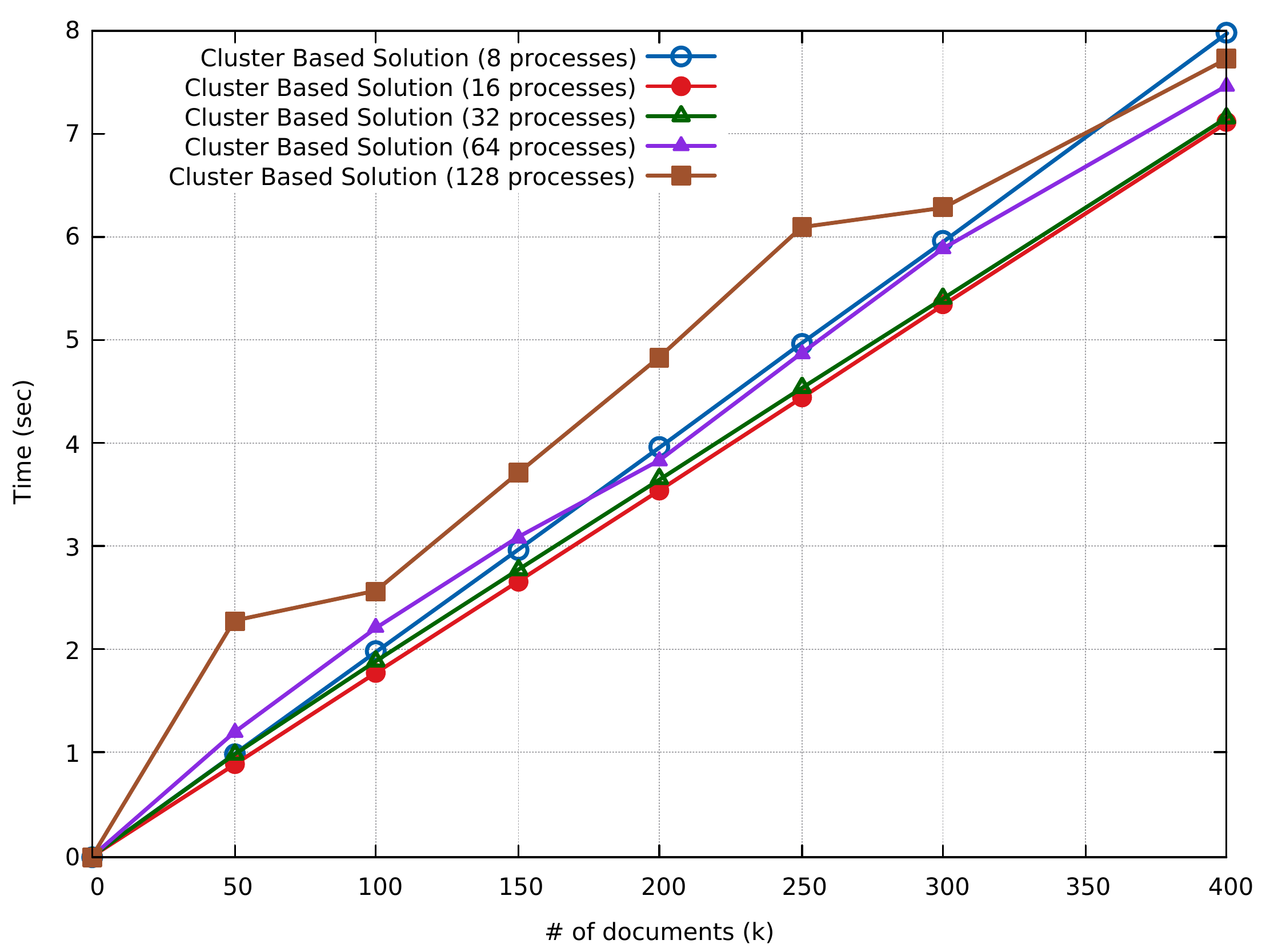}
\caption[Cluster based solution, experiment 1]{Different number of processes used in cluster based solution (experiment 1)}
\label{cluster1}
\end{figure}

In the first experiment, we tested several numbers of processes, namely, 8, 16, 32, 64, and 128 processes that are executed simultaneously. Figure \ref{cluster1} shows that the best performance is achieved when we have used 16 processes. Indeed, given that, there is no shared memory between the nodes, launching a huge number of processes can degrade the search performance because of the large number of lists of results returned by each process that are merged by the master process in a sequential way. Therefore, the number of processes must be chosen so that the search process would be distributed enough without bringing too much burden to the master process. Figure \ref{diag1} shows that this solution gives a better performance than the multi-core based solution by reaching an acceleration around 4.5x.

\begin{figure}
\centering
\includegraphics[width=9cm,keepaspectratio]{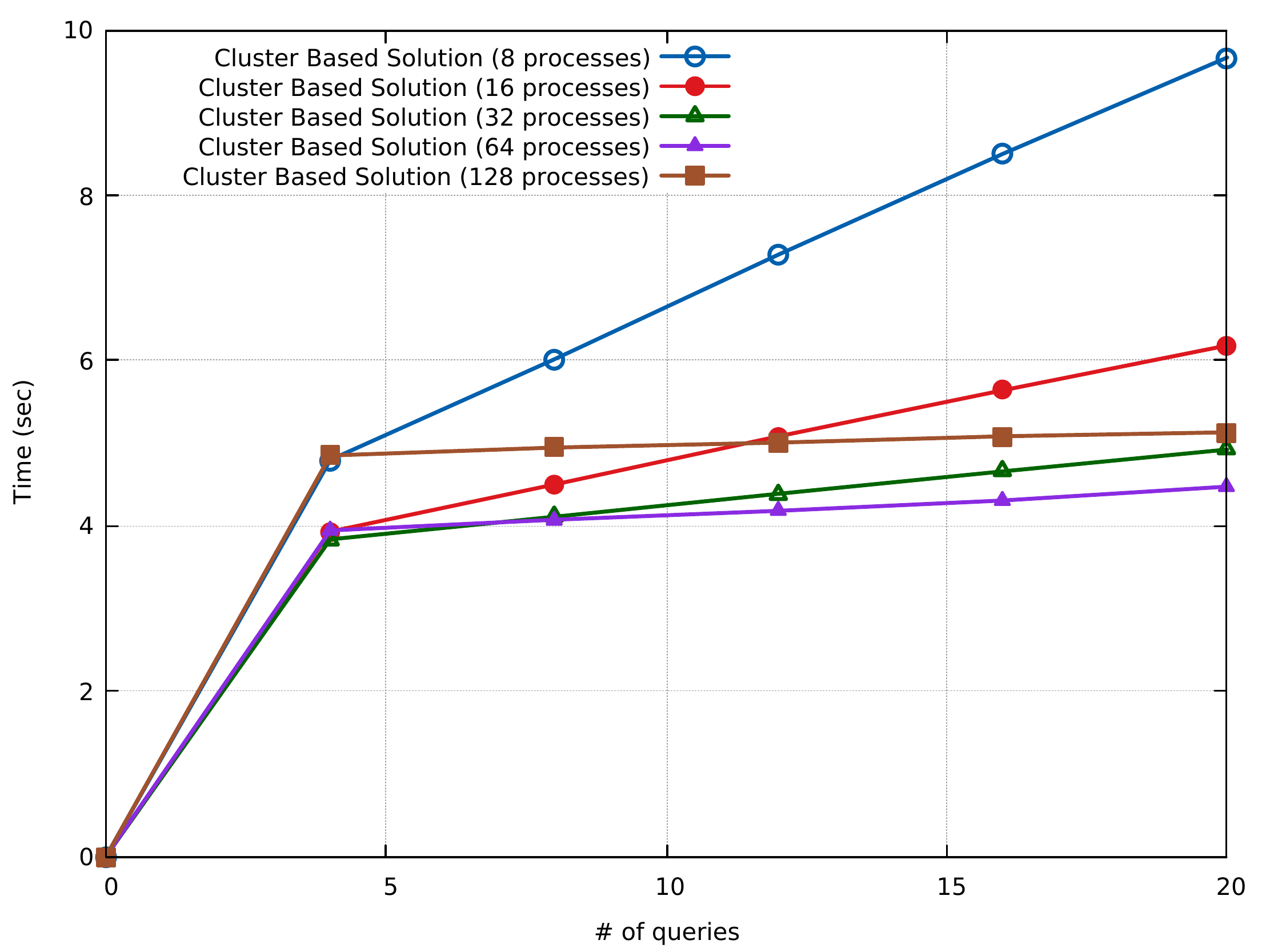}
\caption[Cluster based solution, experiment 2]{Different number of processes used in cluster based solution (experiment 2)}
\label{cluster2}
\end{figure}

Likewise, we also tested several numbers of processes in the second experiment. Figure \ref{cluster2} shows that the best performance was achieved when we used 64 processes. In addition, Figure \ref{diag2} shows that there is no a significant increase in the computing time, even if the number of queries increases. Therefore, this solution allows to simultaneously process a huge number of queries without any degradation in the performance. Thus, cluster based solution gives the best performance among all proposed solutions by achieving an acceleration around 46x. Moreover, the acceleration can be enhanced, if there are more documents or queries to be processed.

\subsection{Synthesis}
The aim of this experimental study is to show the interest of the different techniques that we have proposed. On one hand, the first experiment shows the benefit of distributing the documents between multiple threads during the search process. This experiment has validated the different solutions that we have proposed by achieving accelerations around a factor of 4, 4.5, 5.5 and 4.5 for multi-core based solution, GPU based solution, the hybrid solution, and cluster based solution, respectively. On the other hand, the second experiment has proved the advantage of processing multiple queries simultaneously. This new method has improved the accelerations from 4.5x to 8.5x for GPU based solution, from 5.5x to 12x for the hybrid solution, and from 4.5x to 46x for cluster based solution.

Moreover, if we compare the four solutions that we have proposed, we notice that multi-core based solution is the cheapest, but it allows to achieve an acceleration around 4x. This solution is useful for an individual who outsources a small data collection, whereas, GPU based solution requires a graphic processor which is not really expensive, in order to achieve an acceleration around 8.5x. Likewise, the hybrid solution exploits both CPU and GPU to achieve an acceleration around 12x, so the hybrid solution is definitely better than GPU based solution. Moreover, the hybrid solution is sufficient when the collection of documents is not too big and the number of users is small. Thus, this solution is adequate for a small company. However, the cluster solution is the most expensive, but it achieves very good accelerations and can treat a large number of queries simultaneously over a large collection of documents without any degradation in the search performance. Therefore, it is adequate for a big company.

\section{Summary}

In this chapter, we dealt with the performance issue faced in the search over encrypted data. Indeed, for security and feasibility purposes, most approaches that have been proposed in the literature opt for the choice of the vector model during the indexing process. However, this representation causes the degradation of the search performance since the query vector must be compared with each document vector. To overcome this problem, we have proposed four techniques that use some high performance computing architectures in order to accelerate any approach based on the vector model. We have conducted an experimental study which confirmed that the distribution of documents between different processes as well as the processing of several queries simultaneously allow to drastically improve the search performance.

\chapter{Searchable Encryption with User Access Rights Management}

\label{Chapter6}

\section{Motivation}

Most {\color{black}of the} prior approaches \cite{yu:toward, cao2014privacy, li:efficient, wang2010secure, xia2016secure} that have been proposed in the literature are based on the vector model. Unfortunately, approaches based on the vector model have two major disadvantages. 1) The search process becomes inefficient when the data collection grows considerably. Indeed, in this case, both the number of vectors and their sizes increase, leading to a very big index table. During the search process, the server calculates the scalar product between each document vector and the query vector. For that, it has to load and browse through the entire index which makes the search inefficient when the index table is very large. 2) The update of the index is not supported when new terms appear in the data collection which makes this kind of {\color{black}approach} ineffective in practice.

Although the inverted index speeds up the search process through a direct access to documents containing the query terms, it remains largely unexplored in the literature because of its vulnerability. Indeed, an inverted index is constructed such that each term appearing in the collection points to the set of documents that contain it. Unfortunately, this structure offers an adversary the relationship between a set of terms and a given document. Consequently, the keyword privacy constraint\footnote{The scheme must be able to prevent an attacker from making a link between a set of terms and a document.} is not respected which may cause sensitive information leakage.

 The first aim of this work is to propose a new construction of a secure inverted index by exploiting and proposing several techniques such as homomorphic encryption, the dummy documents technique, the compressed table technique, and the double score formula. The second aim of this work is to take into consideration the users' access rights to the data. For this purpose, it is necessary to apply an access control policy that enables the data owner to grant or revoke access for any data to any user.

In the rest of the chapter, we first present the problem formulation. After that, we explain our proposed solution, and we finish by a detailed experimental study.

\section{Problem Formulation}

In this section, we present the challenges that we have to overcome.

\subsection{Index Structure}

When using the vector model, each document (resp. query) is represented by a vector of size\footnote{{\color{black}$n$ is the number of terms in the data collection.}} $n$. The $j^{th}$ field of the vector contains the weight of the term $j$ of the dictionary\footnote{A dictionary contains all terms of the data collection.} in the document (resp. query). This weight is equal to zero {\color{black}in the case where the term} does not belong to the document (resp.  query). All vectors are encrypted using an appropriate encryption method. Unlike the inverted index, the vector representation prevents the server from making any link between a set of terms and a document (resp. query). Therefore, the keyword privacy constraint is respected. In addition to the security aspect, this model is quite simple to implement. indeed, during the search process, the cloud server calculates the scalar product between each document vector and the query vector, sorts the scores, and returns the top-$k$ documents to the user. However, the drawback is that any approach based on the vector model is a time consuming process, given that the server must load and browse through the entire index {\color{black}during the search process}. The other disadvantage of using the vector model is its inability to manage the data update, since each vector of the index has a size equal to the number of terms in the data collection and the same goes for the trapdoors. Thus, when new terms appear in the data collection, the document vectors as well as the trapdoors must be recalculated and re-encrypted. Consequently, the search process will be inefficient and impractical.

Some approaches \cite{xu:two, sun:secure, wang2015inverted} have tried to exploit an inverted index to speed up the search process. The drawback of using an inverted index in a searchable encryption approach is the difficulty {\color{black}of respecting} some security constraints such as the keyword privacy. Indeed, unlike the vector model, the  inverted index structure does not hide the relationship between terms and documents even after encryption. This lack of security is explained by the fact that each entry of the inverted index corresponds to a term in the data collection, and each entry leads to a set of documents that contain this term. Thus, it is easy for an adversary to link between a set of terms and a document. This drawback explains the little use of the inverted index in the searchable encryption area despite that it is the most appropriate way to perform an effective search. Thus, the aim of this work is to overcome the issues encountered when using an inverted index, with particular focus on the security aspect. For this purpose, we exploit a homomorphic encryption cryptosystem \cite{brakerski2012leveled} to encrypt the scores of the index. Homomorphic encryption enables {\color{black}performing} calculations using encrypted data without any need of decryption in order to obtain an encrypted result. This property is very useful in the searchable encryption area since it protects the scores and the obtained results. However, even if homomorphic encryption protects the scores of the index, it is still not sufficient to ensure the index security because of its inability to hide the relationship between terms and documents, which makes it easy for an adversary to deduce whether a term belongs to a given document. Therefore, in order to enhance the index security, we took inspiration from the work of "Cao et al." \cite{cao2014privacy} to propose a technique that allows to respect the keyword privacy constraint {\color{black}by preventing adversaries from deducing whether a term belongs to a given document}. We call our technique the dummy documents technique. This technique consists in adding randomly selected {\color{black}document IDs} to each entry of the index {\color{black}and} assigning each of them (the {\color{black}document IDs}) a score zero encrypted using homomorphic encryption. The number of dummy documents depends on the system security level. However, the techniques {\color{black}that are} used in our approach, namely, homomorphic encryption and the dummy documents technique present some drawbacks that have to be overcome. 

On one hand, homomorphic encryption generates very large ciphertexts that are 3 orders of magnitude larger than the plaintext size. Consequently, the encrypted index will be very large and heavy to manage which may cause the system slow-down. To overcome this problem we propose a technique that we call the compressed table of encrypted scores. This technique consists in limiting the number of scores within a given interval, for instance, integers from zero to one hundred. After that, each score is encrypted a number of times using homomorphic encryption and each ciphertext is represented by an ID. These IDs are used within the inverted index instead of using ciphertexts. In addition, a table containing all ciphertexts with their corresponding IDs is outsourced with the index in order to enable the cloud server to do the necessary calculations during the search process. This technique considerably reduces the number of ciphertexts, which enables compressing the encrypted index. However, in practice, it is important to pay attention to the score distribution. For example, given that some dummy documents are added to the index, score zero will be the most frequently used among all the scores present in the index. Thus, if an adversary notices that some ciphertexts are more repeated than others, it can easily conclude that they represent the encryption of zero. To correct this security breach, we propose a formula that allows to assign each score an appropriate number of ciphertexts so that all ciphertexts are used equitably. 

On the other hand, the use of homomorphic encryption cryptosystem involves that the similarity scores obtained during the search process are encrypted which disables the server's ability to sort the results. Consequently, the server returns to the user the documents that contain at least one query term instead of returning a list of top-$k$ documents. In addition, the use of dummy documents causes some false positives. Since the similarity scores are encrypted, the irrelevant documents cannot be filtered by the server. Thus, to allow the server to return top-$k$ documents and to {\color{black}discard} a great part of irrelevant documents, we use the double score weighting (DSW) formula (see Subsection \ref{sec_DSW}) which requires the use of an ontology such as Wikipedia that maps between terms and concepts. Notice that the entries of the inverted index in our proposal correspond to concepts instead of terms. DSW formula consists in assigning two scores to each document, a primary score and a secondary score. The primary score represents the number of terms in the document that are associated with the concept corresponding to the entry, whereas, the secondary score is calculated by applying the TFIDF formula. Both primary and secondary scores are encrypted using homomorphic encryption. When building the trapdoor, the user's query will be represented by a set of concepts. During the search process, the server chooses the documents appearing the most frequently in the selected entries of the index (i.e. the documents {\color{black}that contain} the greatest number of concepts in common with the query). Finally, after the decryption of scores in the user side, the documents will be sorted based on their primary scores, then on their secondary scores in the case of equality. Therefore, DSW formula {\color{black}filters} a great part of irrelevant documents even if their similarity scores are encrypted. Still, some irrelevant documents may be returned to the user. Fortunately, it is easy to filter them given that their similarity scores are equal to zero after decryption.

The purpose of this subsection is to present our motivations of proposing and using some techniques such as the inverted index, homomorphic encryption, the dummy documents technique, the compressed table of encrypted scores technique, and the double score weighting formula. In Section \ref{sec_SIIS}, we explain in more detail our proposed approach and the techniques that we have presented in this subsection.

\subsection{User Access Rights}

Cloud computing is a technology that allows a data owner to store his data in a remote server and share the outsourced data with other users. However, it happens that the data owner does not want to grant the same access rights to all users. In this case, it is necessary to apply an access control policy. For that, CP-ABE method was proposed \cite{bethencourt2007ciphertext}.

However, even if the CP-ABE method allows to apply an access policy on encrypted data while protecting the data privacy, another problem arises when a user performs a search on the data collection. The issue is that the cloud server {\color{black}cannot recognize} which documents the user has the right to access, since the access structure is encrypted and stored in the ciphertext and the attributes are included in the user's private key. Therefore, during the search process, the cloud server is forced to take into consideration the entire data collection rather than the part of data which is accessible by the user. This has the drawback of returning a large number of documents that the user cannot decrypt. "Bouabana-Tebibel and Kaci" \cite{bouabana2015parallel} tried to solve this problem by sending the encrypted private key with the trapdoor to the cloud so that the server can identify the documents that are accessible by the user and hence, be able to perform the search on this dataset. Unfortunately, this method is not secure given that it is risky to send the private key to the cloud server which is considered as semi-trusted\footnote{Honest but curious.}. To solve this problem, we propose a method that exploits an additional secure index. The role of this additional index is to reduce the search space to keep only the documents which are accessible by the user, in addition to some dummy documents\footnote{The user does not have the right to access these documents.} {\color{black}that enhance the security}. A great part of the dummy documents will be {\color{black}discarded} when applying the double score formula. The entries of this second index correspond to the user IDs and each entry leads to the documents which are accessible by the user in addition to some dummy documents. Each document has a random access score between 1 and 100, whereas, each dummy document has an access score equal to zero. The scores of both indexes are encrypted using homomorphic encryption. These access scores will be used on the user side after decryption to {\color{black}filter} the dummy documents that have an access score equal to zero.

\section{Proposed Scheme}\label{sec_SIIS}

As explained above, one of the aims of this work is to propose a secure inverted index dedicated to the search over encrypted cloud data. The use of an inverted index is {\color{black}an appropriate} way to perform a search because it enables a direct access to the relevant documents. However, it is essential to secure this index before using it in order to protect the data privacy and prevent any sensitive information leakage. In this section, we give a detailed explanation of the proposed approach. For this purpose, we start by presenting an overview of our approach called "secure inverted index based searchable encryption" (SIIS). Then, we explain step by step the construction of the proposed secure inverted index. After that, we present a new method that enables to manage the users' access rights to the data by exploiting a second secure inverted index. Finally, we present other details of the proposed scheme.

\subsection{Overview of the Proposed Approach (SIIS)}
\begin{figure}
\centering
\includegraphics[width=11cm,keepaspectratio]{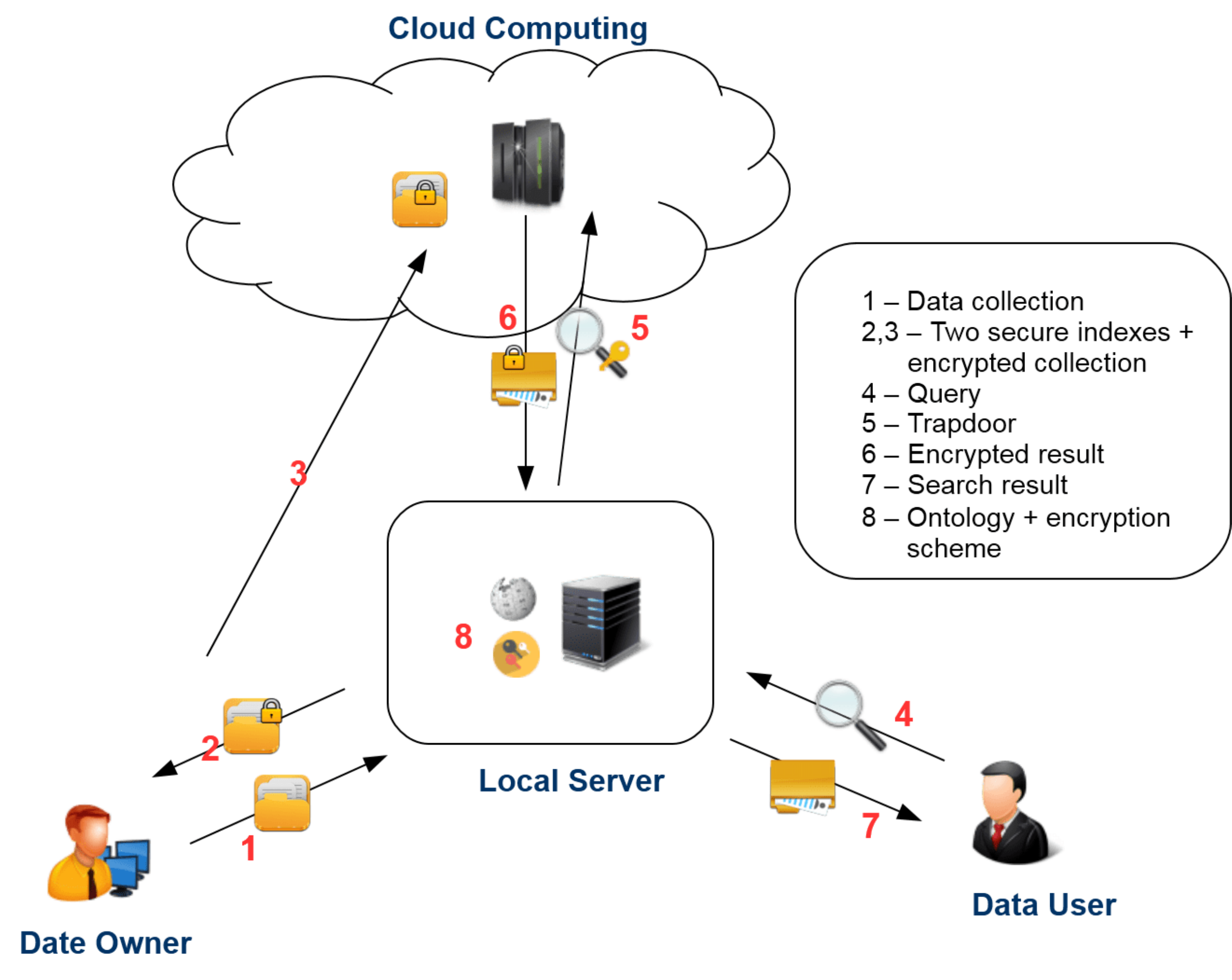}
\decoRule
\caption{General architecture of SIIS approach}
\label{fig_archi}
\end{figure}

The aim of our work is to propose a semantic search approach on encrypted data that should be practical and secure. For this purpose, we have exploited the inverted index which is rarely used in the literature because of its vulnerability. In order to enhance its security, we have opted for the use of few techniques such as homomorphic encryption and the dummy documents technique. Indeed, in our proposal, before outsourcing documents, the data owner builds a concept based inverted index using the Wikipedia ontology. Then, some random documents are added to each entry of the index and a weight equal to zero is associated to each of these documents. After that, a second inverted index is constructed in order to manage the users' access rights to the data collection. The entries of this second index correspond to the user IDs and each entry leads to a set of documents that the user has the right to access, in addition to some dummy documents that enhance the index security. Both indexes are then encrypted using homomorphic encryption, whereas, the data collection is encrypted using the CP-ABE method which guarantees the access policy applied by the data owner. Then, the encrypted data collection as well as the two secure indexes are sent to the cloud server. Finally, the data owner constructs a set of public keys whose role is to build trapdoors, and private keys for decrypting documents. These keys will be shared with authorized users. To perform a search, a user starts by formulating a query, builds a trapdoor using a public key, and sends the trapdoor to the cloud server. Upon receiving the trapdoor, the cloud utilizes the first index to retrieve the relevant documents and simultaneously, it exploits the second index to get the list of documents that the user has the right to access. After that, it keeps the documents belonging to both lists before applying the double score formula to obtain the top-$k$ documents that will be returned to the user. Finally, the returned results will be decrypted, filtered, and sorted at the user side via a trusted authority (Figure \ref{fig_archi}).

\subsection{Construction of a Secure Inverted Index}

In this subsection we present the four techniques necessary for the construction of a secure inverted index $I_1$ from the original unencrypted index $I$.

\subsubsection{Homomorphic Encryption to Encrypt the Index}

The leveled homomorphic encryption \cite{brakerski2012leveled} (see Subsection \ref{sec_leveled}) is an encryption method with the particularity of enabling the calculations on ciphertexts without any need of decryption in order to obtain an encrypted result. This feature is very useful in our case since it allows the data owner to encrypt the scores of the inverted index while enabling the cloud server to exploit these scores during the search process. However, the drawback of this method is that it produces very large ciphertexts that exceed thousands of times the size of the corresponding plaintext, resulting in a heavy index which can cause performance issues during the search and
the update processes. To solve this problem, we propose the compressed table of encrypted scores technique.

\subsubsection{Compressed Table of Encrypted Scores} \label{compressed_sec}

\begin{figure*}
\centering
\includegraphics[width = 17cm, height = 20cm]{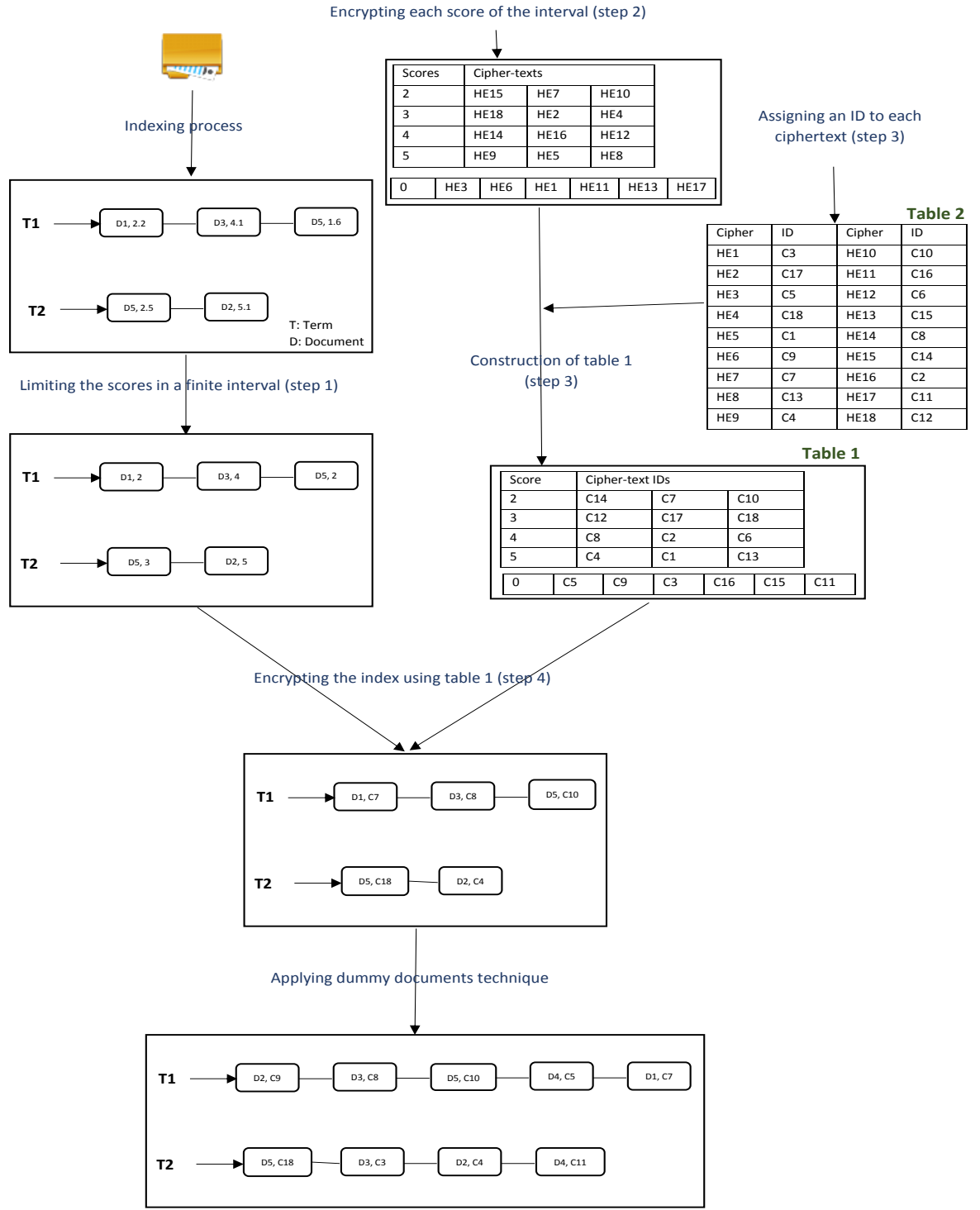}
\caption[The compressed table of encrypted scores]{The compressed table of encrypted scores and the dummy documents techniques.}
\label{Compressed}
\end{figure*}

The aim of this technique is to reduce the inverted index size in order to make it more practical and easily exploitable. This technique is summarized in the following steps (Figure \ref{Compressed}).

\begin{enumerate}

\item First, before the encryption step and during the indexing process, the scores of documents in the inverted index are limited in a finite interval $Inv$. For instance, $Inv$ corresponds to positive integers from zero to one hundred ($Inv = [0, 100]$).
\item Then, for each score of the interval $Inv$, some corresponding ciphertexts are produced using the leveled homomorphic encryption method, for example, producing of one thousand ciphertexts for the score zero, one hundred ciphertexts for the score one, one hundred ciphertexts for the score two, and so on. However, for security reasons, a large number of dummy documents having a weight equal to zero are injected into the inverted index. The number of ciphertexts corresponding to zero must be chosen appropriately in order to be hidden (the ciphertexts corresponding to zero) among the other ciphertexts. For this purpose, we have proposed a formula that enables to efficiently find the appropriate number of ciphertexts corresponding to zero (Formula \ref{occ}).

\begin{equation}\label{occ}
NC_0 = \frac{NC \times NS}{2} \times  \frac{K}{10}
\end{equation}

Where $NC_0$ is the number of ciphertexts corresponding to zero, $NC$ (chosen by the data owner) is the number of ciphertexts corresponding to any other score, $NS$ is the number of scores belonging to the interval $Inv$ {\color{black}($NS$ is equal to 101 if $Inv = [0, 100]$)}, $K$ {\color{black}(also chosen by the data owner)} is a security parameter belonging to the interval $[1, 10]$ and such that $K = 10$ corresponds to the highest security level.

When the security parameter is equal to 10, then, the number of ciphertexts corresponding to zero is equal to half of the number of ciphertexts corresponding to the remaining scores. This formula is based on Formula \ref{DDN} in which the number of dummy documents when $K$ is  equal to 10 is roughly equal to half of documents in the index. The Formula \ref{DDN} is presented in the next subsection.

\item After that, a unique identifier is assigned to each ciphertext produced in the previous step. Then, a table $T_1$ containing all scores of the interval $Inv$ is constructed, where each score is associated with its corresponding ciphertexts represented by their identifiers. This table will be used by the data owner when encrypting the inverted index.
\item During the encryption of the inverted index, each score\footnote{Two scores (a primary and a secondary score) are assigned to each document of the index $I_1$ as explained in Subsection \ref{sec_DSW}.} $S_i$ will be replaced by the identifier of {\color{black}a} corresponding ciphertext $C_j$ by using the table $T_1$. The ciphertext $C_j$ is randomly chosen among the set of ciphertexts corresponding to the score $S_i$.
\item Finally, a second table $T_2$ containing all ciphertexts produced with their associated identifiers is constructed. The table $T_2$ will be used by the cloud server during the search process, whereas, the table $T_1$ is encrypted (using AES) and stored in the data owner side.
\end{enumerate}

\subsubsection{Dummy Documents Technique} \label{DDT_section}

The hardest hurdle of using an inverted index in a search approach on encrypted cloud data is the security aspect. To overcome this problem, we have tried to exploit homomorphic encryption. However, even by encrypting the index scores, it is still very easy for an attacker to deduce the relationship between a set of terms and a document. {\color{black}In fact}, the inverted index structure is made so that each entry of the index (concept) leads to the set of documents containing that concept. Therefore, the keyword privacy constraint is not respected. In spite of the use of the concept identifiers (resp. document identifiers) in the inverted index rather than the concepts themselves (resp. document titles), the inverted index remains vulnerable if no further improvements are made. To solve this problem, we took inspiration from a technique proposed by "Cao et al." \cite{cao2014privacy} which consists in adding dummy elements to the vectors. Similarly, our technique consists in adding for each entry of the index some documents not containing the concept corresponding to that entry. Then, a score zero is assigned to each dummy document. These scores are encrypted using the table $T_1$ (last step of Figure \ref{Compressed}). The number of dummy documents added to an entry $E_i$ is calculated by the following formula.
\begin{equation}\label{DDN}
DDN_i = \frac{DN_i \times Random(1, 10 \times K)}{100}
\end{equation}
Where $DDN_i$ is the number of dummy documents to be added, $DN_i$ is the number of documents in the entry $E_i$, $K$ is a security parameter belonging to the interval $[1, 10]$ and such that $K = 10$ corresponds to the highest security level. Finally, $Random (X, Y)$ is a function that returns a random value belonging to the interval $[X, Y]$.

\subsubsection{Double Score Formula}\label{double2}

Even if the dummy documents technique enhances the index security by respecting the keyword privacy constraint, it has the drawback of causing many false positives in the search results. Indeed, during the search process, the dummy documents belonging to the selected entries of the index cannot be {\color{black}filtered} by the cloud server and will be returned to the user. Furthermore, the similarity scores between documents and queries are encrypted due to the use of homomorphic encryption. Consequently, the cloud server cannot sort the selected documents and hence, it is forced to return every document belonging to the selected entries rather than returning top-$k$ documents. To overcome these two problems, we took inspiration from the DSW formula that we have proposed in Chapter \ref{Chapter4} (see Subsection \ref{sec_DSW}). The search process when applying the double score formula is accomplished as follows.

\begin{enumerate}
\item Upon receiving the encrypted query $Q$, the cloud server selects the entries of the index corresponding to the concepts of the query $Q$.
\item After that, the cloud server classifies the selected documents into categories. A document $D$ belongs to a category {\color{black}$G_i$} if it appears in $i$ different selected entries of the index.
\item Then, it selects $k$ documents belonging to the highest categories\footnote{A category {\color{black}$G_{i+1}$} is higher than a category {\color{black}$G_i$}.}, calculates the similarity scores by applying Formula \ref{sim}, and returns the identifiers of the selected documents as well as their encrypted scores to the user.
\item Once the user receives the search result, he decrypts the similarity scores. Then, he {\color{black}discards} any dummy document that {\color{black}has} a score equal to zero. Finally, he sorts the remaining documents based on their primary scores and then on their secondary scores in the case of equality.
\end{enumerate}

The experimental study that we have conducted shows that it is unlikely that the same dummy document appears in several selected entries, for this reason that the double score formula is able to ignore a great number of dummy documents.

\begin{equation}\label{sim}
\begin{aligned}
D_i(x_i, y_i) = \underset{C_j \in E_j \wedge C_j \in Q}\sum{(CP_j \times DP_{ij}, CS_j \times DS_{ij})} 
\end{aligned}
\end{equation}

Where $D_i$ is the document number $i$, $C_j$ is the concept number $j$, $E_j$ is the $j^{th}$ entry of the index, $Q$ is the encrypted query, $CP_j$ (resp. $CS_j$) is the primary (resp. secondary) score of the concept $C_j$ in the query $Q$, $DP_{ij}$ (resp. $DS_{ij}$)  is the primary (resp. secondary) score of the concept $C_j$ in the document $D_i$. Finally, $x_i$ (resp. $y_i$) is the primary (resp. secondary) similarity score between the document $D_i$ and the query $Q$. Notice that the scores $DP_{ij}$ and $DS_{ij}$ are obtained from the index $I_1$, and the scores $CP_j$ And $CS_j$ are obtained from the trapdoor $Q$.

\subsection{User Access Rights}

The second goal of our approach is to apply an access control to the data collection. {\color{black}In fact}, the data owner must be able to grant the access for a given document to a set of users and to revoke the access to others. However, this feature should not cause any lack of security and must be done in complete safety. For this purpose, we have exploited two techniques. The first one is CP-ABE \cite{bethencourt2007ciphertext} which is used to encrypt the data collection, whereas, the second technique consists in using a second secure inverted index $I_2$ to filter the documents that the user does not have the right to access in order to perform the search only on the accessible documents.

\subsubsection{CP-ABE to Encrypt the Data}
We decided to use CP-ABE method to allow the data owner to grant or revoke the access to a given document for any user. Furthermore, this encryption method is an additional security layer that protects the data collection in case a document is returned to an unauthorized user. Indeed, the data collection is stored at the cloud side and hence, during the search process, the cloud is supposed to search only on the dataset that the user is allowed to access. Unfortunately, there are cases where the server returns documents that the user does not have the right to access. Therefore, the CP-ABE method guarantees the data protection and hence, the documents can be decrypted only by authorized users.

\subsubsection{Use of a Second Secure Inverted Index}\label{index2}

The second technique consists in using a second inverted index $I_2$ whose entries correspond to the identifiers of users that point to the set of documents accessible by the corresponding user. The remainder of the index will be constructed as follows.

\begin{enumerate}
\item First, a random access score\footnote{The scores of the index $I_2$ are called access scores.} between $1$ and $100$ is assigned to each document of the index $I_2$.
\item Then, for each entry of the index $I_2$, some dummy documents are added. Each dummy document has an access score equal to zero. The number of dummy documents is calculated by applying Formula \ref{DDN}.
\item Finally, the table $T_1$ is used to encrypt each access score by randomly choosing a ciphertext of the corresponding score. Notice that it is the identifier that will be used rather than the ciphertext itself as explained in Subsection \ref{compressed_sec}.
\end{enumerate}

During the search process, both indexes $I_1$ and $I_2$ are used simultaneously. The first index returns relevant documents with some false positives for security purpose. The second index returns the documents that the user has the right to access with some false positives. Then, the server gets the intersection of the two results in order to obtain a list of relevant documents accessible by the user with some false positives. After that, the server applies the double score formula as explained in Subsection \ref{double2} in order to keep the top-$k$ documents and ignore a large part of the false positives. Upon receiving the list of results, the data user decrypts the scores. Then, any dummy document that {\color{black}has} a similarity score\footnote{The similarity scores are obtained from the index $I_1$.} equal to zero will be {\color{black}discarded}. Similarly, any document having an access score equal to zero will be {\color{black}discarded} since it is a document that the user cannot decrypt. The remaining documents will be sorted according to their similarity scores.  Notice that our experimental study shows that there are very few dummy documents which are returned by the server.

\subsection{Access Pattern Hiding}\label{sad}

Most approaches that have been proposed in the literature do not hide from the server the returned results and the selected documents. This negligence allows the server to deduce some valuable information during its statistical analyses by linking the trapdoor or the selected entries of the index to the set of documents satisfying the user's need which is inconsistent with the access pattern constraint (see Subsection \ref{sec_threat}). Unfortunately, this lack of security may cause sensitive information leakage as demonstrated by "Islam et al." \cite{islam2012access}.

Therefore, to prevent the server from inferring any information about the inverted index, we propose a technique able to obtain documents securely by preventing any attacker from knowing neither about the search result nor about the chosen documents. This technique {\color{black}that} we call the "access pattern hiding" (APH) is an improvement of an existing technique called "blind storage" \cite{naveed2014dynamic} which consists in splitting each document into several blocks in order to prevent the server from knowing the number of documents or distinguishing between them. {\color{black}The proposed technique} is a combination of the following four solutions.

\subsubsection{Separating Technique}

This technique aims to break any link between the index and the data collection in the cloud side. For this purpose, each document ID in the index should be different from that identifying the same document in the data collection (Figure \ref{separating}). Consequently, the correspondence between the index and the data collection should be done in the user side. The separating technique is performed as follows:

\begin{figure}
\centering
\includegraphics[width=10cm,keepaspectratio]{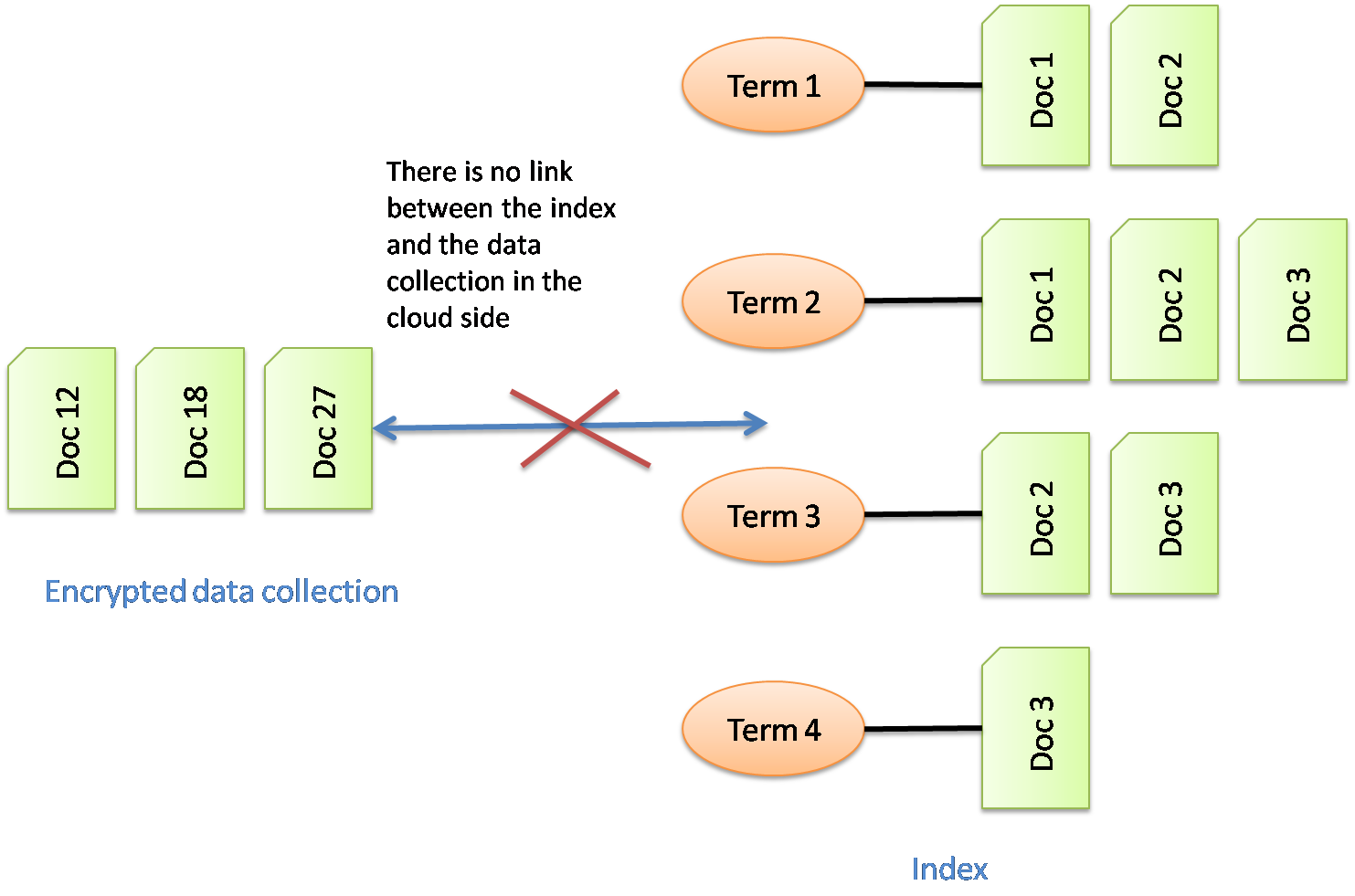}
\caption[The separating technique:]{The separating technique: there is no link between the index and the data collection in the cloud side.}
\label{separating}
\end{figure}

\begin{enumerate}
    \item Before the indexing process, two identifiers called "document ID" and "data ID" are affected to each document of the data collection. The data IDs are used to represent the encrypted documents of the data collection, whereas, the document IDs are used to represent the documents in the index. The relationship between a data ID and a document ID can be revealed using a table stored in the user side (Table \ref{correspondance}).
    
\begin{table}
\centering
\setlength\extrarowheight{2pt}
\begin{tabular}{c|c} 
\hline
   \rowcolor{Gray}
   \textbf{Data ID} & \textbf{Document ID} \\ \hline
   $Doc_{12}$ & $D_3$ \\
   \rowcolor{Gray}
   $Doc_{18}$ & $D_1$ \\
   $Doc_{27}$ & $D_2$ \\ \hline
 \end{tabular}
 \caption[The correspondence table]{The correspondence table linking between data IDs and documents IDs (example of Figure \ref{separating}).}
  \label{correspondance}  
\end{table}

    \item When receiving a search result, the "correspondence table" is exploited in the user side to link between the document IDs that have been returned by the cloud and the data IDs. The latter are used to request the necessary documents stored in the cloud server.
\end{enumerate}

The separating technique is able to prevent the server from making links between the data collection and the index as long as no search has been done. However, after few searches, the server can easily deduce the relationship between some document IDs and data IDs. Let us take the example of Figure \ref{separating} and suppose that after a first search, the server returns a list of results $R_1 = \{d_1, d_3\}$ and the user requests the document $Doc_{18}$. Then, after a second search, the server returns a list $R_2 = \{d_1, d_2\}$ and the user requests again the document $Doc_{18}$. Therefore, the server can easily deduce that the data ID $Doc_{18}$ in the collection corresponds to the document ID $D_1$ in the index. To overcome this vulnerability, we propose a second technique called the "splitting technique".

\subsubsection{Splitting Technique}

\begin{figure}
\centering
\includegraphics[width=10cm,keepaspectratio]{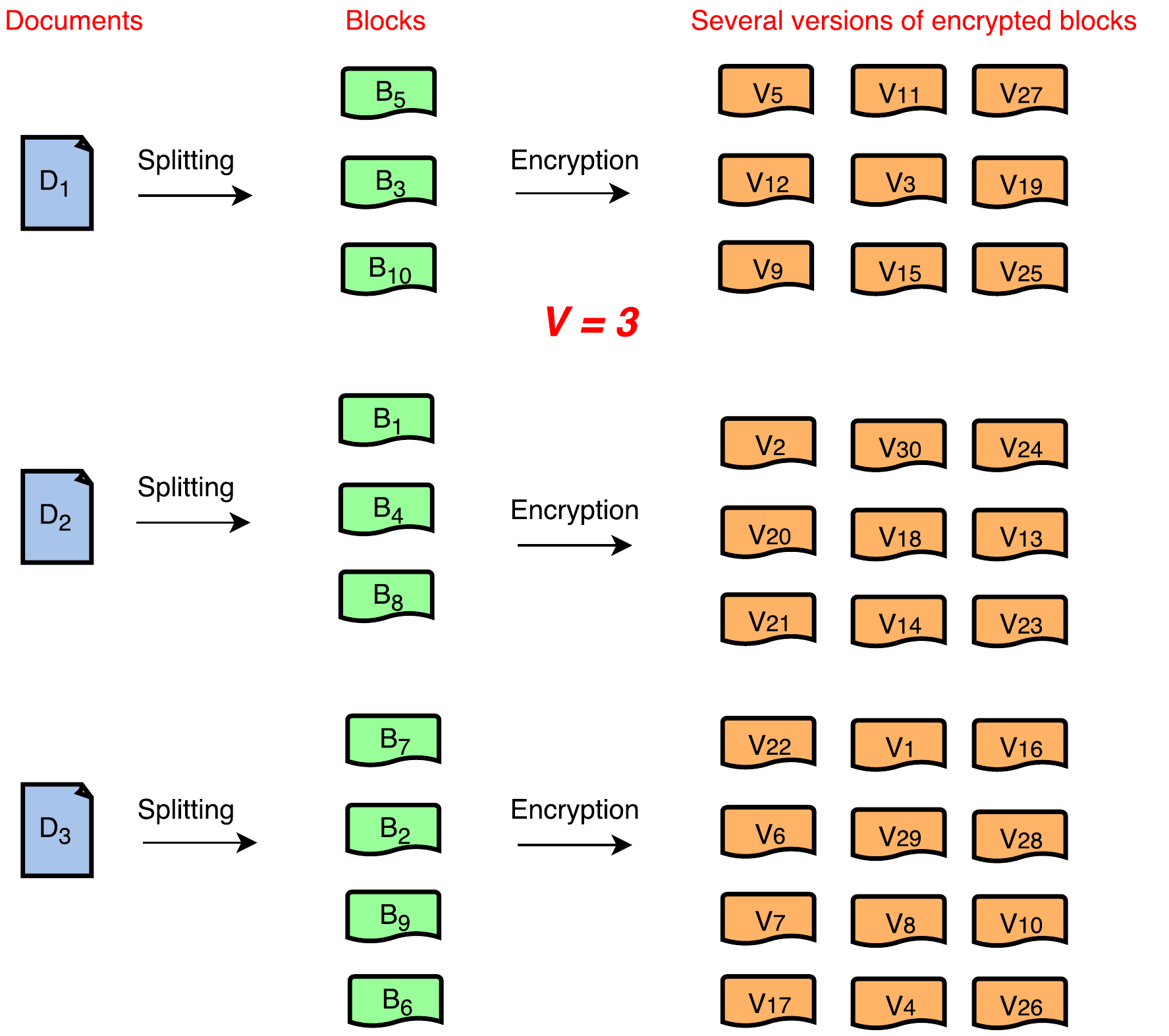}
\caption[The splitting technique]{Splitting of three documents into several blocks encrypted in different versions.}
\label{splitting}
\end{figure}

This technique consists in splitting each document into several blocks to avoid the server from distinguishing between the documents of the data collection. Then, in order to make the server task even more complicated, we propose to generate during the encryption process several versions of each block (Figure \ref{splitting}). To request a document, only one version of each block is randomly chosen in the user side. Consequently, different blocks are requested by the data users for different accesses to the same document. The number of versions should be chosen by the data owner before the encryption process. The splitting technique is done in the data owner side as follows.

\begin{table}
\centering
\setlength\extrarowheight{2pt}
\begin{tabular}{c|c|c}
\hline
    \rowcolor{Gray}
    \textbf{Document ID} & \textbf{Block ID} & \textbf{Rank} \\ \hline
    \multirow{3}{*}{$D_1$} & $B_5$ & 1 \\
    & $B_3$ & 2 \\
    & $B_{10}$ & 3 \\ \hline
    \multirow{3}{*}{$D_2$} & $B_1$ & 1 \\
    & $B_4$ & 2 \\
    & $B_8$ & 3 \\ \hline
    \multirow{4}{*}{$D_3$} & $B_7$ & 1 \\
    & $B_2$ & 2 \\
    & $B_9$ & 3 \\
    & $B_6$ & 4 \\ \hline
 \end{tabular}
 \caption[The table of blocks]{The table of blocks (TB) which is generated from the example of Figure (\ref{splitting}).}
  \label{tab_blocks}  
\end{table}

\begin{table*}
\centering
\setlength\extrarowheight{2pt}
\begin{tabular}{c|c||c|c}
\hline
    \rowcolor{Gray}
    \textbf{Block ID} & \textbf{Version ID} & \textbf{Block ID} & \textbf{Version ID} \\ \hline
    \multirow{3}{*}{$B_5$} & $V_5$ & \multirow{3}{*}{$B_3$} & $V_{12}$ \\
    & $V_{11}$ & & $V_3$ \\
    & $V_{27}$ & & $V_{19}$ \\ \hline
      \multirow{3}{*}{$B_{10}$} & $V_9$ & \multirow{3}{*}{$B_1$} & $V_2$ \\
    & $V_{15}$ & & $V_{30}$ \\ 
    & $V_{25}$ & & $V_{24}$ \\ \hline
     \multirow{3}{*}{$B_4$} & $V_{20}$ & \multirow{3}{*}{$B_8$} & $V_{21}$ \\
    & $V_{18}$ & & $V_{14}$ \\
    & $V_{13}$ & & $V_{23}$ \\ \hline
     \multirow{3}{*}{$B_7$} & $V_{22}$ & \multirow{3}{*}{$B_2$} & $V_{6}$ \\
    & $V_{1}$ & & $V_{29}$  \\
    & $V_{16}$ & & $V_{28}$  \\ \hline
      \multirow{3}{*}{$B_9$} & $V_{7}$ & \multirow{3}{*}{$B_6$} & $V_{17}$ \\
    & $V_{8}$ & & $V_{4}$  \\
    & $V_{10}$ & & $V_{26}$  \\ \hline
 \end{tabular}
 \caption[The table of versions]{The table of versions (TV) which is generated from the example of Figure (\ref{splitting}).}
 \label{tab_versions}  
\end{table*}

\begin{enumerate}
    \item The data owner starts by choosing the block size $S$ (for example, $S = 4$kb for each block), and a parameter $V$ (for example, $V = 10$) that corresponds to the number of versions of each block. 
    \item Then, each document of the data collection is split into several blocks of size $S$. A rank $r_i$ is affected to each block $b_i$ to indicate its position when merging back the blocks.
    \item After that, each block $b_i$ is encrypted several times\footnote{The initialization vector (IV) should be different at each new encryption to obtain different ciphertexts.} to obtain different versions $\{v_{i1}, v_{i2}, v_{i3}, ...\}$. The number of versions is equal to the parameter $V$.
    \item Then, the data owner generates two tables. The first one that we call the table of blocks (TB) (Table \ref{tab_blocks}) associates each document to its different blocks with their associated ranks, whereas, the second one that we call the table of versions (TV) (Table \ref{tab_versions}) associates each block to its different encrypted versions. Note that the different versions of the same block have the same rank. Both tables are encrypted and stored in the user side.
    \item To request a document, different set of blocks are selected for each new access to the document which prevents the server from knowing either the same document has been accessed.
\end{enumerate}

This technique is able to prevent the server from making links between the data collection and the index as long as the number of accesses to a given document $d$ does not exceed a security threshold $\theta_1(d)$. The latter varies from one document to another and is calculated by the following formula:

\begin{equation}\label{threshold}
   \theta_1(d) = V^{\beta(d)}
\end{equation}

Where $d$ is a document, $\theta_1(d)$ is the security threshold of the document $d$, $\beta(d)$ is the number of unencrypted blocks constituting the document $d$ and $V$ is a parameter indicating the number of versions of each block.

Let us take the example of Figure \ref{splitting} and suppose that after a first search, the server returns a list of results $R_1 = \{d_1,d_3\}$ and the user requests the blocks $\{v_5, v_3, v_9\}$ (document $d_1$). Then, after a second search, the server returns a list $R_2 = \{d_1, d_2\}$ and the user requests the blocks $\{v_{11}, v_{19}, v_{25}\}$ (document $d_1$) and so on until the $27^{th}$ search\footnote{Given that the document is split into 3 blocks and V is equal to 3, therefore, $\theta_1(d_1) = 3^3 = 27$.} (we suppose that after every search, the user requests the document $d_1$). During the $28^{th}$ search, let us suppose that the server returns the list $R_{28} = \{d_1, d_2\}$ and the user requests the blocks $\{v_5, v_3, v_9\}$. In this case, the server can easily conclude that the blocks $\{v_5, v_3, v_9\}$ correspond to the document $d_1$.

Therefore, the document identity becomes vulnerable after being accessed a number of times higher than the security threshold. To overcome this problem, we propose an additional solution that we call the \emph{scrambling} technique.

\subsubsection{Scrambling Technique}\label{scr}

\begin{figure}
\centering
\includegraphics[width=7cm,keepaspectratio]{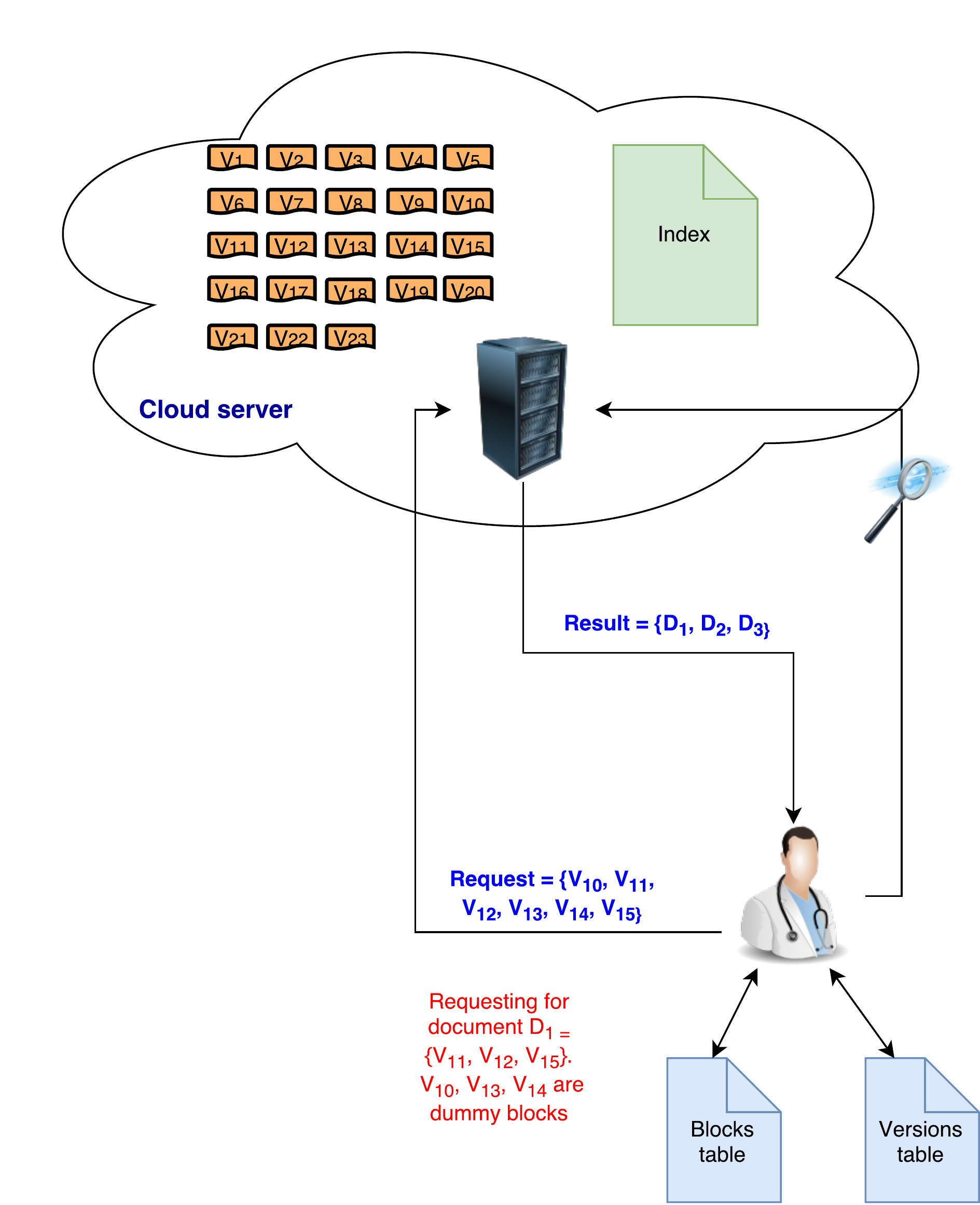}
\caption[The scrambling technique]{The scrambling technique: dummy blocks are added before requesting the server.}
\label{scrambling}
\end{figure}

This technique consists in requesting the server other blocks that we call the "dummy blocks" in addition to those constituting the wanted document (Figure \ref{scrambling}). The aim of these blocks is to add noise and confuse the issue in order to prevent the server from making a link between the true blocks and the corresponding document. The "scrambling technique" is performed in the user side as follows.

\begin{enumerate}
    \item A parameter $\lambda$ is randomly chosen from an interval $]x, y]$ in order to determine the number of dummy blocks $N_{db}(d)$ which is calculated by the following formula:
    
       \begin{equation}\label{dummy_blocks}
           N_{db}(d) = \lambda \times \beta(d)
       \end{equation}
       
       Where $\beta(d)$ is the number of blocks composing the wanted document $d$, the interval $]x, y]$ is set by the data owner.
       
    \item To obtain a document $d$, the authorized user has to request the encrypted blocks constituting this document. Furthermore, additional dummy blocks are also requested in order to add noise and prevent the server from knowing whether a requested block is a part of those composing the wanted document. The dummy blocks are randomly chosen from all blocks of the data collection and their number is calculated by applying Formula (\ref{dummy_blocks}).
    \item Finally, after receiving the requested blocks, the dummy ones are dropped, whereas, the true blocks are decrypted and merged in order to recover the wanted document that has been sent to the user.
\end{enumerate}

Let us take again the example of Figure \ref{splitting} and suppose that after a first search, the server returns a list of results $R_1 = \{d_1,d_3\}$ and the user requests the blocks $\{v_5, v_3, v_9, v_2, v_{30}, v_{19}\}$ (in this example, the first three blocks are true and the rest are dummy). Then, after a second search, the server returns a list $R_2 = \{d_1, d_2\}$ and the user requests the blocks $\{v_{11}, v_{19}, v_{25}, v_{15}, v_{26}, v_{27}\}$ and so on until the $27^{th}$ search (we suppose that the user always requests the document $d_1$). During the $28^{th}$ search, let us suppose that the server returns the list $R_{28} = \{d_1, d_2\}$ and the user requests the blocks $\{v_5, v_3, v_9, v_{21}, v_4, v_{29}\}$. In this case, the server can conclude that the blocks $E = \{v_5, v_3, v_9\}$ correspond \emph{probably} to the document $d_1$ with a probability $P_k$, where $P_k$ is the probability that at most $k$ blocks can constitute a document (in this example $k$ is equal to 3 blocks). After a while, if the server does not find any set of blocks that appears at least twice and includes $E$, it can conclude that the blocks $\{v_5, v_3, v_9\}$ correspond to the document $d_1$. Therefore, the identity of documents is protected as long as the threshold $\theta_2(d)$ is not exceeded. 

\begin{equation}
    \theta_2(d) = A \times \theta(d)
\end{equation}

Where $A \in \mathbb{N}_{>1}$ and it increases when $\beta(d)$ or/and $N_{db}(d)$ increase.

This example shows that the scrambling technique is able to prevent the server from linking a set of blocks $E$ to a document $d$ even if the threshold $\theta_1(d)$ is exceeded. However, even though this technique complicates the server task to disclose the access pattern, it cannot permanently hide the identity of the documents frequently accessed. To achieve this, we propose an additional solution that we call the "grouping technique". 

\subsubsection{Grouping Technique}
The dummy blocks are randomly selected from the whole blocks constituting the data collection, which explains the fact that there is almost no repeat when requesting them. Contrary to the true blocks that are repeated after the threshold $\theta_1(d)$ is exceeded. Therefore, after requesting a document several times\footnote{A number of time much more than the threshold $\theta_1(d)$.}, the server would be able to distinguish the true blocks from the dummy ones (see Subsection \ref{scr}). To overcome this security breach, our idea is to divide the blocks constituting the data collection into several groups. When requesting the blocks of a document, the dummy ones are selected from the same groups of the true blocks. This technique allows to repeat a certain number of dummy blocks which avoids the attacker from distinguishing between the true blocks and the dummy ones. The grouping technique is performed as follows:

\begin{enumerate}
    \item The blocks constituting the data collection are divided into several groups. Two blocks of the same document should not appear in the same group. The blocks should have the same size $Sz$, where:
    \begin{equation}\label{eq:sz}
        Sz > y + 1
    \end{equation}
    With $y$ is the parameter of the interval $]x, y]$ that have been used in the scrambling technique (see Subsection \ref{scr}).
    \item During the scrambling technique, after the number of dummy blocks $N_{db}(d)$ is calculated, the dummy blocks are selected from the groups to which the true blocks belong. For example, if the user wants to request three true blocks belonging to the groups $A$, $B$, and $C$, respectively. If the number of dummy blocks calculated by Formula \ref{dummy_blocks} is equal to six. Therefore, two dummy blocks should be randomly selected from the group $A$, two dummy blocks should be selected from the group $B$, and two dummy blocks should be selected from the group $C$.
\end{enumerate}

To show the benefits of the grouping technique, we take the example of Figure \ref{splitting}. First, we suppose that the grouping is made as shown in Table \ref{group}. Now, Let us suppose that after a first search, the server returns a list of results $R_1 = \{d_1,d_3\}$ and the user requests the blocks $\{v_5, v_3, v_9, v_{21}, v_{28}, v_{32}\}$ (in this example, the first three blocks are true and the rest are dummy blocks belonging to the same groups as the first three blocks). Then, after a second search, the server returns a list $R_2 = \{d_1, d_2\}$ and the user requests the blocks $\{v_{11}, v_{19}, v_{25}, v_{4}, v_{33}, v_{29}\}$ and so on until the $27^{th}$ search (the user always requests the document $d_1$). During the $28^{th}$ search, let us suppose that the server returns the list $R_{28} = \{d_1, d_2\}$ and the user requests the blocks $\{v_5, v_3, v_9, v_{21}, v_{28}, v_{13}\}$. Given that some dummy blocks are repeated with the true blocks, the server task becomes more difficult to disclose the document identity. Therefore, the identities of documents are protected as long as the threshold $\theta_3(d)$ is not exceeded.

\begin{table*}
\centering
\setlength\extrarowheight{2pt}
\begin{tabular}{c|c||c|c}
\hline
    \rowcolor{Gray}
   \textbf{Group ID} & \textbf{Blocks} & {Group ID} & \textbf{Blocks} \\ \hline
   $G_1$ & $V_5$, $V_{21}$, $V_7$ & $G_2$ & $V_{23}$, $V_{10}$, $V_{35}$ \\
   \rowcolor{Gray}
   $G_3$ & $V_{11}$, $V_{20}$, $V_4$ & $G_4$ & $V_{24}$, $V_6$, $V_{26}$ \\
   $G_5$ & $V_{16}$, $V_{12}$, $V_{36}$ & $G_6$ & $V_{19}$, $V_{1}$, $V_{33}$ \\
   \rowcolor{Gray}
   $G_7$ & $V_{15}$, $V_{2}$, $V_{17}$ & $G_8$ & $V_{27}$, $V_{8}$, $V_{34}$ \\
   $G_9$ & $V_{18}$, $V_{22}$, $V_{31}$ & $G_{10}$ & $V_{25}$, $V_{30}$, $V_{29}$ \\
   \rowcolor{Gray}
   $G_{11}$ & $V_{3}$, $V_{14}$, $V_{28}$ & $G_{12}$ & $V_{9}$, $V_{13}$, $V_{32}$ \\ \hline
 \end{tabular}
 \caption[The grouping table]{The grouping table (example of Figure \ref{splitting}).}
  \label{group}  
\end{table*}

\begin{equation}
    \theta_3(d) = B \times \theta_2(d)
\end{equation}

Where $B \in \mathbb{Q}_{>1}$ and it increases when $\beta(d)$, $N_{db}(d)$ and/or $Sz$ increase. 

\subsubsection{Aggregation}

The APH technique is the aggregation of the four proposed methods previously introduced, namely, the separating technique, the splitting technique, the scrambling technique, and the grouping technique. The APH technique allows to keep the SE scheme respecting the access pattern constraint when requesting the needed documents after a search. In the following we explain how to exploit the APH technique.

\begin{enumerate}
    \item Before the encryption of the data collection, the data owner applies the splitting technique. For that, each document is split into several blocs of size $S$. A rank $r_i$ is attributed to each block $b_i$ in order to keep the ability to reconstruct the document later. Finally, the data owner generates the table of blocks ($TB$) to link between a document and its different blocks with their associated ranks.
    
    \item After that, each block is encrypted into several versions. The number of versions is equal to the parameter $V$ chosen by the data owner. The latter constructs the table of versions ($TV$) in order to link between a block and its different encrypted versions. Finally, the table of blocks and the table of versions are encrypted and stored in the user side (internal server), whereas, the encrypted blocks are outsourced to the cloud server. 
    
    \item Then, the data owner applies the grouping technique. For that, he chooses the interval $]x, y]$ and uses the parameter $y$ in order to calculate the size of the groups ($Sz$) by applying Formula \ref{eq:sz}. After that, he divides the blocks into several groups of size $Sz$. The groups should have the same size and the blocks of the same document should appear in different groups.
    
    \item During the indexing process, the data owner applies the separating technique to obtain an index with document identifiers different from those of the data collection. For that, the document IDs are used in the index, whereas, the version IDs are used in the data collection. The correspondence between the index and the data collection should be made in the user side using both $TB$ and $TV$. Finally, the index should be encrypted and outsourced to the cloud server.
    
    \item During the search, the data user formulates a query and encrypts it to construct a trapdoor that is sent to the cloud server. Upon receiving the trapdoor, the cloud server launches the search and returns a list of top-$k$ document IDs to the user. The latter chooses the document IDs that he needs from the returned results and uses the table of blocks to obtain the block IDs corresponding to the document IDs. After that, the table of versions is also exploited to randomly obtain an encrypted version ID corresponding to each block ID. Then, the scrambling technique is applied to add some dummy block IDs. The dummy blocks should be selected from the same groups as the true blocks. Finally, the dummy and the true blocks are requested from the server.
    
    \item After receiving the requested blocks from the server, the dummy ones are dropped, whereas, the true blocks are exploited to reconstruct the needed documents. Note that the ranks of blocks stored in the table of blocks are primordial in the reconstruction process. Finally, the documents are decrypted and exploited by the user.
\end{enumerate}

\subsection{Functions of the SIIS Approach}

The proposed scheme is composed of seven functions (\textit{KeyGen}, \textit{TabGen}, \textit{BuidOnto}, \textit{BuildIndexes}, \textit{Trapdoor}, \textit{Search}, and \textit{Sort}) and two main phases (\textit{Initialization} phase and \textit{Retrieval} phase). We start by presenting the seven functions of the proposed scheme.

\begin{itemize}
\item \textit{KeyGen.} It generates a private key $SK_{he}$ and a public key $PK_{he}$
by exploiting the leveled fully homomorphic encryption method \cite{brakerski2012leveled}. The private key $SK_{he}$ will be used by the authorized users to decrypt the search results (the similarity and the access scores), whereas, the public key $PK_{he}$ will be used by the data owner to construct the two tables $T_1$ and $T_2$ and by the users to generate encrypted queries (trapdoors). Furthermore, a set of private keys ${SK}_{cp} = \{SK^1_{cp}, SK^2_{cp}, SK^3_{cp}, ..., SK^n_{cp}\}$ and a public key $PK_{cp}$ are generated by exploiting the CP-ABE methode \cite{bethencourt2007ciphertext}. The public key $PK_{cp}$ will be used by the data owner to encrypt the data collection, whereas, each private key $SK^i_{cp}$ will be used to decrypt a part of the data collection that the user $U_i$ has the right to access.
\item \textit{TabGen ($Inv$, $PK_{he}$).} {\color{black}Steps 2, 3 and 5 of the compressed table technique are exploited in this function} to construct both tables $T_1$ and $T_2$. The table $T_1$ contains the set of scores belonging to the interval $Inv$ used by the SIIS approach, where each score $S_i$  is related to a set of ciphertexts. These ciphertexts are obtained by encrypting the score $S_i$ several times using the public key $PK_{he}$. {\color{black}The initialization vector (\textit{IV}) should be different at each new encryption to obtain different ciphertexts}. This set of ciphertexts is associated with the score $S_i$ through the table $T_1$. Each ciphertext in the table $T_1$ is represented by an ID. Table $T_2$ is used to associate ciphertexts with their identifiers (see Figure \ref{Compressed}). {\color{black} Table $T_1$ will be used by the BuildIndexes function to obtain an encrypted and compressed indexes, whereas, table $T_2$ will be exploited by the \textit{Search} function when performing a search.}
\item \textit{BuildOnto.} The proposed approach exploits an ontology built from Wikipedia as follows. First, we suppose that every Wikipedia page corresponds to a concept. After that, the English pages of Wikipedia are indexed such that each page is represented by a vector of weighted terms by applying the TFIDF formula. Finally, in order to accelerate the mapping between terms and concepts, an inverted index $I_{wiki}$ is created where each term is represented by a set of weighted concepts (see Subsection \ref{sec_wiki}).
\item \textit{BuildIndexes ($F$, $T_1$).} It constructs both indexes $I_1$ and $I_2$ from the data collection $F$. The first index $I_1$ is used by the cloud server during the search process to retrieve the most relevant documents. The entries of the index $I_1$ correspond to the concepts of the Wikipedia ontology. Each entry $E_i$ leads to a set of documents containing the concept $C_i$ in addition to some dummy documents that are used to enhance the index security. The second index $I_2$ is used by the cloud server to reduce the search space to the set of documents accessible by the user. The index $I_2$ is constructed in the same way with the index $I_1$ except that the entries of the index $I_2$ correspond to the user IDs and each entry $E^\prime_i$ leads to the documents accessible by the user $U_i$ in addition to some dummy documents. The scores of the index $I_1$ called similarity scores are calculated by applying the DSW Formula (see Subsection \ref{sec_DSW}), whereas, the scores of the index $I_2$ called access scores are calculated as explained in Subsection \ref{index2}. Both similarity scores and access scores are encrypted using the {\color{black}table $T_1$ (see step 4 of Subsection \ref{compressed_sec})}. Notice that it is the identifiers of ciphertexts that are used in both indexes instead of the ciphertexts themselves.
\item \textit{Trapdoor ($W$, $PK_{he}$).} When a user formulates a query, the Wikipedia ontology is used to map between the query terms $W$ and a set of concepts $C$. Then, each concept of the query is weighted by applying the DSW formula. After that, a subset of concepts $C^\prime$ is randomly chosen from the set $C$ to represent the query\footnote{In our experimental study we have chosen 10 concepts among the best 15 concepts representing the query.}. This method of choosing the concepts is used to ensure the query unlinkability constraint. Finally, each concept of the query is represented by an ID and the scores of each concept are encrypted using the leveled homomorphic encryption.
\item \textit{Search ($Q$, $I_1$, $I_2$, $k$, $U_i$).} Upon receiving the encrypted query $Q$, the cloud server selects the entries of the index $I_1$ that correspond to the concepts of the query. At the same time, it selects the entry of the index $I_2$ that corresponds to the user $U_i$ who launched the search. Then, the cloud server selects the documents belonging to both the selected entries of the index $I_1$ and the selected entry of the index $I_2$. After that, it uses the double score formula to keep the top-$k$ documents (see Subsection \ref{double2}). Finally, each selected document is assigned three scores calculated by Formula \ref{three} which are then returned to the user.

\begin{equation}\label{three}
D_i(x_i, y_i, z_i) = \left(\underset{C_j \in E_j \wedge C_j \in Q}\sum{(CP_j \times DP_{ij}, CS_j \times DS_{ij})}, DA_{il}\right)
\end{equation}
Where $D_i$ is the document number $i$, $C_j$ is the concept number $j$, $E_j$ is the $j^{th}$ entry of the index $I_1$, $Q$ is the encrypted query, $CP_j$ (resp. $CS_j$) is the primary (resp. secondary) score of the concept $C_j$ in the query $Q$ and $DP_{ij}$ (resp. $DS_{ij}$)  is the primary (resp. secondary) score of the concept $C_j$ in the document $D_i$. Finally, $x_i$ (resp. $y_i$) is the primary (resp. secondary) similarity score between the document $D_i$ and the query $Q$, and $z_i$ ($DA_{il}$) is the access score of the user $U_l$ to the document $D_i$. Notice that the score $DA_{il}$ is obtained from the index $I_2$.

\item \textit{Sort ($R$, $SK_{he}$).} Upon receiving the search result $R$, the user decrypts the scores using the private key $Sk_{he}$. Then, any dummy document having a similarity score equal to zero is ignored. Similarly, any inaccessible document having an access score equal to zero is ignored. Finally, The remaining documents are sorted based on their primary scores and their secondary scores in the case of equality.
\end{itemize}

The search process consists in two main steps.
\begin{itemize}
\item \textit{Initialization phase}. In this phase, the data owner prepares the search environment as follows.
\begin{enumerate}
\item First, the \textit{KeyGen} function is called to generate the different keys that will be shared with authorized users using a secure communication protocol.
\item Then, the \textit{BuildOnto} function is called to build an ontology from Wikipedia. This ontology will be stored in an internal server and will be accessible by authorized users only.
\item After that, the \textit{TabGen} function is called to generate both tables $T_1$ and $T_2$ that will be used during the construction of the index and the search process, respectively.
\item Finally, from a data collection $F$, the data owner builds the secure indexes $I_1$ and $I_2$ by calling the \textit{BuildIndexes} function. The secure indexes as well as the encrypted data collection\footnote{The data collection is encrypted using the CP-ABE method} are then sent to the cloud server.
\end{enumerate}
\item \textit{Retrieval phase.} In this phase, a search process is performed as follows.
\begin{enumerate}
\item First, an authorized user builds an encrypted query by calling the \textit{Trapdoor} function.
\item Upon receiving the encrypted query, the server launches the \textit{Search} function and returns to the user the document IDs with their corresponding encrypted scores.
\item Finally, in the user side, the returned documents are filtered and sorted using the \textit{Sort} function.
\end{enumerate}
\end{itemize}

\section{Analyses and Comparison}

\subsection{Security Analyses}

In order to analyse the security aspect of the SIIS scheme, we will verify whether the security constraints presented in Subsection \ref{sec_constraintes} are respected. In addition, we will analyse the security of the encrypted inverted index and the users' access rights.

\begin{itemize}
\item \textit{Protected content.} This constraint consists in encrypting any data passing through the cloud server, namely, the indexes, the data collections, and the queries. Firstly, scores in both $I_1$ and $I_2$ indexes used in our approach are encrypted using homomorphic encryption, whereas, concepts and documents are represented by identifiers. Secondly, a trapdoor is represented by a set of concepts with their associated weights. Each concept is represented by an identifier and each weight is encrypted using homomorphic encryption. Finally, the data collection is encrypted using the CP-ABE method which allows to protect the content of documents and enables to apply an access control policy. Conclusively, the protected content constraint is respected by the proposed approach since the indexes, the data collections and the queries are encrypted. 

\item \textit{Keyword privacy.} This constraint consists in preventing the server from making any link between terms and documents. For this purpose, two properties must be hidden, namely, the term distribution and the inter-distribution. On one hand, the inter-distribution consists in the distribution of scores of terms in a given document. This property is hidden by homomorphic encryption which allows to encrypt the scores in the inverted index. On the other hand,  the term distribution consists in the frequencies of a given term in each document of the collection. This property is hidden through the dummy documents technique that prevents the server from knowing whether a term belongs to the documents to which the corresponding entry leads. We conclude that the keyword privacy constraint is respected in our approach.

\item \textit{Trapdoor unlinkability.} This constraint consists in preventing the server from making links between different trapdoors. For this purpose, a subset of concepts {\color{black}($x$ concepts)} is randomly chosen from the set of concepts {\color{black}($y$ concepts)} representing the query, for example, {\color{black}choosing} 10 concepts among the {\color{black}100} concepts representing the query. {\color{black}Note that} each concept {\color{black}is} represented by an ID and {\color{black}its weight is} encrypted using homomorphic encryption. {\color{black}This construction allows to get $C_y^x$ different trapdoors for the same query  ($C_y^x = \frac{y!}{(y-x)! \times x!}$). In the above example where $x=10$ and $y=100$, $1.7 \times 10^{13}$ different trapdoors could be generated for the same query. Therefore, the SIIS approach provides a non-deterministic encryption scheme which allows to respect the trapdoor unlinkability constraint.}
 
\item \textit{Access pattern}. {\color{black}This constraint consists in hiding from the server the result returned to the user. When performing a search, a set of dummy document identifiers is always returned with the right result. These false positives allow to hide the correct result from the server. Furthermore, the APH technique (see Subsection \ref{sad}) allows the user to access the documents that he needs without disclosing their identities which prevents the server from recognizing the false positives in the search result. Therefore, the access pattern constraint is respected in the SIIS approach.}

\item \textit{Index security.} The index security is ensured by the use of some techniques that enable to wrap the index in four security layers. The first technique consists in using identifiers to represent concepts and documents. These identifiers are used when creating the index instead of using the entitles of concepts and documents. This technique is very simple and {\color{black}straightforward} but it enables to implement a first security layer. The second technique consists in exploiting homomorphic encryption to encrypt the scores of documents. {\color{black}It enables the association degrees between concepts and documents to be hidden}. Then, the third security layer is ensured by the dummy documents technique which consists in adding a noise to each entry of the index in order to prevent any attacker from linking between a concept and a document. Finally, the last security consists in the APH technique which allows to break any link between the index and the data collection. This feature is very important, given that it allows to respect one of the most complicated constraints, namely, the access pattern.  

\item \textit{Access rights privacy.} The users' access rights are stored in a second inverted index. Therefore, their privacy is based on the safety of the secure inverted index that we have proposed which is based on four security layers. 
In addition, we have preferred to expand the interval of the access scores from the set\footnote{The score 1 corresponds to the access right and 0 is assigned to the dummy documents.} \{0,1\} to the set $Inv$\footnote{Any score except 0 corresponds to the access right.} containing all exploited scores. This trick is introduced in order to protect the table of encrypted scores by preventing any attacker from linking between the set \{0,1\} and the set of the corresponding encrypted scores. 

\end{itemize}

\subsection{Performance Analysis}

We performed our experiments on the "Yahoo! Answers"\footnote{\url{https://answers.yahoo.com/}, Feb.2019.} collection which contains \numprint{142627} queries and \numprint{962,232} documents. We carried out several experiments in order to measure the performance of the proposed approach. The experiments were performed on a computer equipped with an Intel Xeon 2.13 GHz processor. The prototype of the proposed approach which is available on-line\footnote{\url{https://www.dropbox.com/sh/ju2wzqa56x5c8pm/AADfFZE3POdxxr7YsHjULcx0a?dl=0}, Feb.2019.} was developed using the C programming language. In addition, we have used HElib library\footnote{\url{https://github.com/shaih/HElib}, Feb.2019.} in order to exploit the various functions of the leveled homomorphic encryption approach.
{\color{black}By default, the following parameters are used: top-$k$ = top-100; $Inv = [0, 100]$; $NC$ = \numprint{10000}, where $NC$ is the number of different ciphertexts generated for each score (see Subsection \ref{compressed_sec}) and $K = 10$ is the highest security level.}

\subsubsection{Size of the Encrypted Index}

\begin{figure}
\centering
\includegraphics[width=9cm,keepaspectratio]{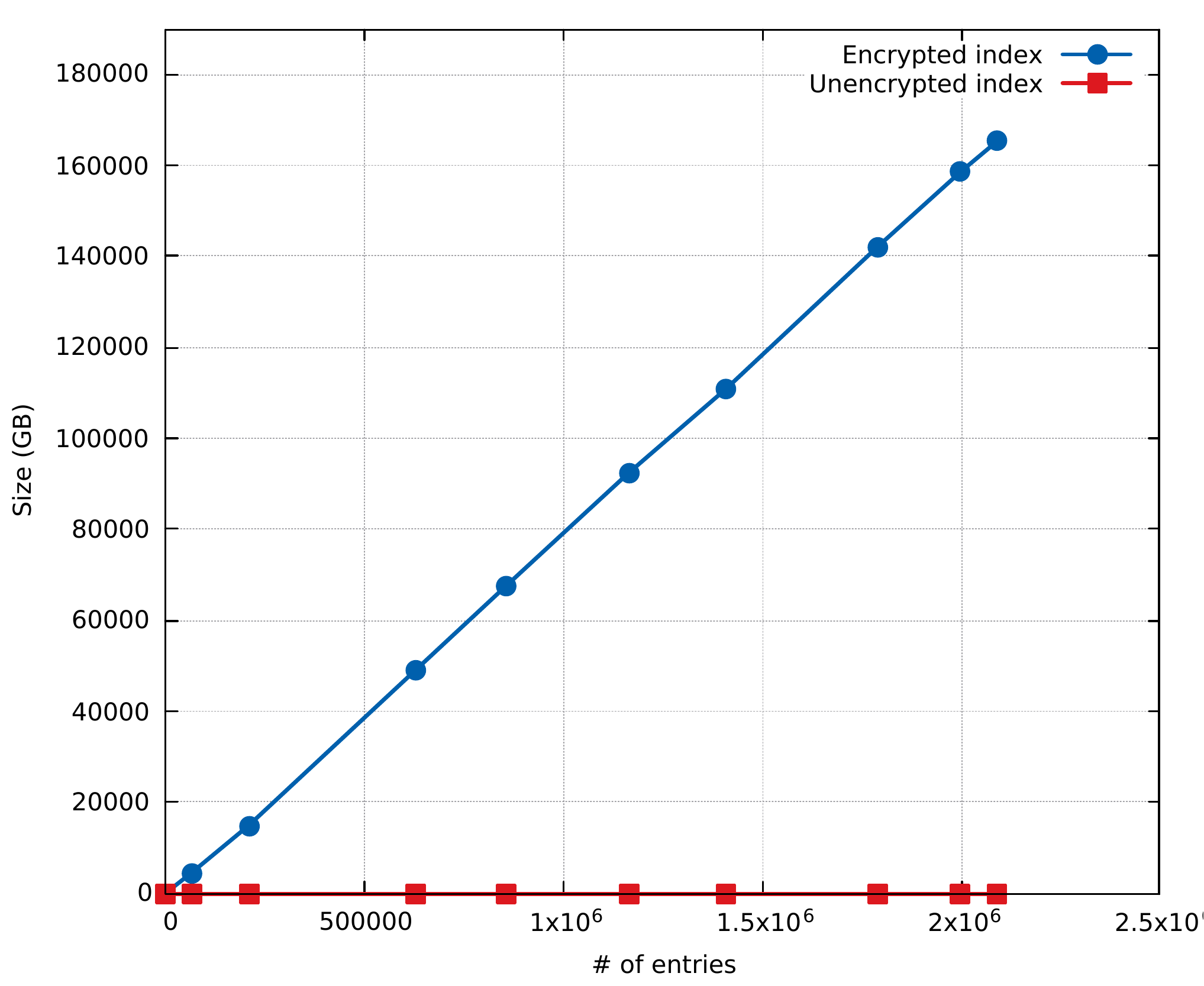}
\caption[The index size when using homomorphic encryption]{The size of the generated index using homomorphic encryption.}\label{HE}
\end{figure}

\begin{figure}
\centering
\includegraphics[width=9cm,keepaspectratio]{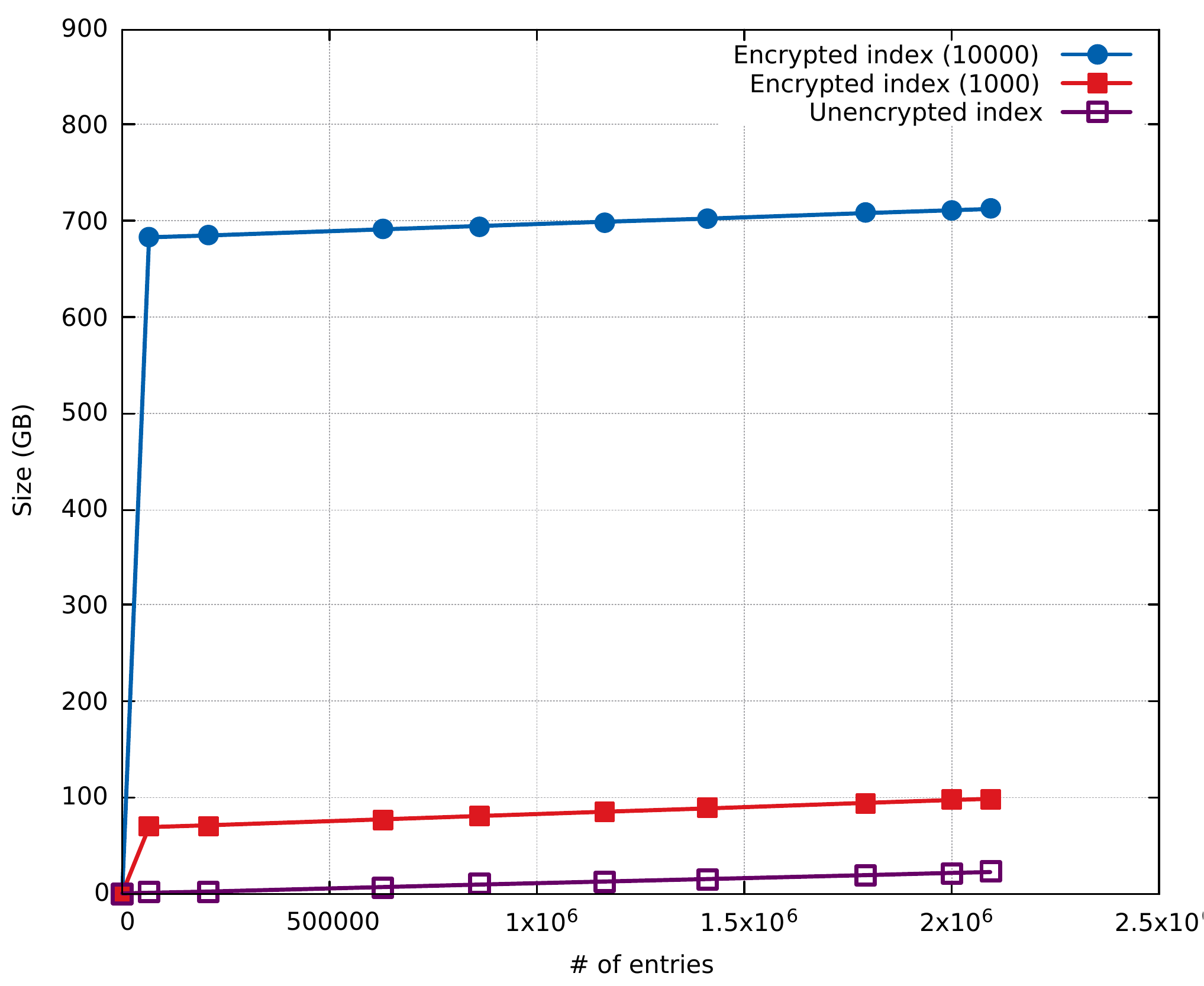}
\caption[The index size when using the compressed table technique]{The size of the generated index using the compressed table of encrypted scores technique ({\color{black}NC =} \numprint{1000} and {\color{black}NC =} \numprint{10000} correspond to the number of ciphertexts generated for each score).}\label{IHE}
\end{figure}

We have performed several experiments in order to validate the effectiveness of the proposed approach (SIIS scheme) in compressing the encrypted index. For this purpose, we started by measuring the size of the secure index generated in our approach without using the compressed table of encrypted scores technique (Figure \ref{HE}). After that, we have carried out the same experiment using the compressed table technique (Figure \ref{IHE}).

\vspace{3mm}

\vspace{3mm}

Figure \ref{HE} shows that the size of the index increases considerably when creating the inverted index using homomorphic encryption. Indeed, the size increased from \numprint{22,2} GB before encryption to 161,45 TB after encryption which corresponds to a \numprint{7447}-fold increase.

To overcome this problem, we have proposed the compressed table of encrypted scores technique. Figure \ref{IHE} shows that this technique allows to considerably reduce the size of the index. {\color{black}In fact}, the size decreases from 161,45 TB to 712,54 GB after using the compressed table technique (for each score, \numprint{10000} corresponding ciphertexts are produced (see step 2 of Subsection \ref{compressed_sec} for more details)). In addition, the size of the index does not depend on the number of documents to be indexed, but depends on the security level, which allows the data owner to choose the appropriate balance between the security and the performance.

\begin{figure}
\centering
\includegraphics[width=9cm,keepaspectratio]{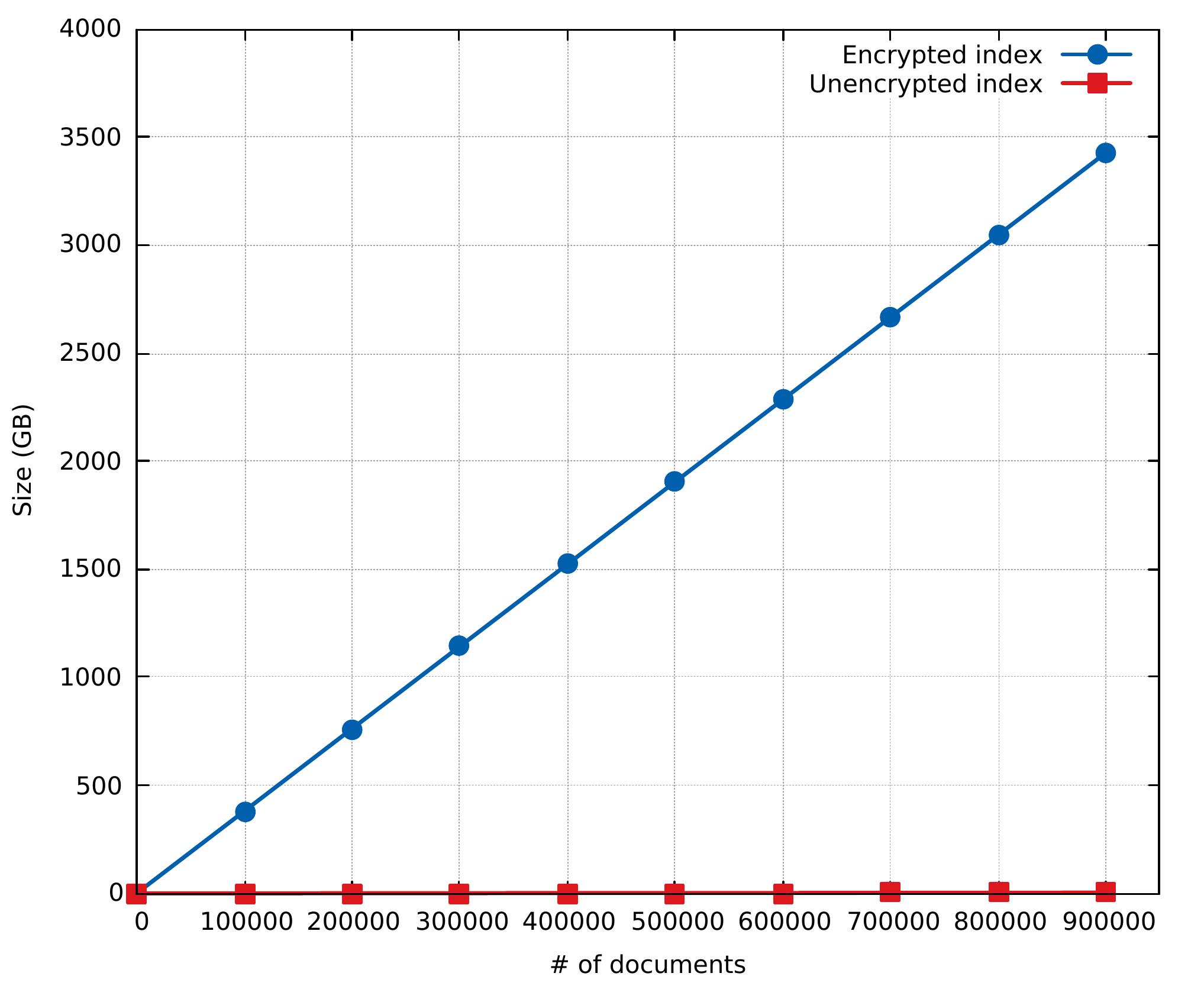}
\caption[The index size in the MRSE appraoch]{The size of the generated index in the MRSE approach.}\label{size_mrse}
\end{figure}

In order to compare the size of the encrypted index generated in the proposed approach with an index generated in a method based on the vector model, we have performed the same experiment on the MRSE approach \cite{cao2014privacy} which is based on the vector model (Figure \ref{size_mrse}).

\vspace{3mm}

Figure \ref{size_mrse} shows that the encrypted index of the MRSE approach (encrypted by the S$k$NN algorithm) increases linearly with the number of documents in the data collection. Indeed, the size of the index increases from 3,5 GB before encryption to \numprint{3427,74} GB after encryption which corresponds to a 979-fold increase.

In comparison with our proposed approach, the index increases from 22,2 GB before encryption to 712,54 GB after encryption which corresponds to a 32-fold increase. Moreover, in our approach, the size of the encrypted index is almost constant even if the number of documents increases.

\subsubsection{Relevance of Results}

\begin{figure}
\centering
\includegraphics[width=9cm,keepaspectratio]{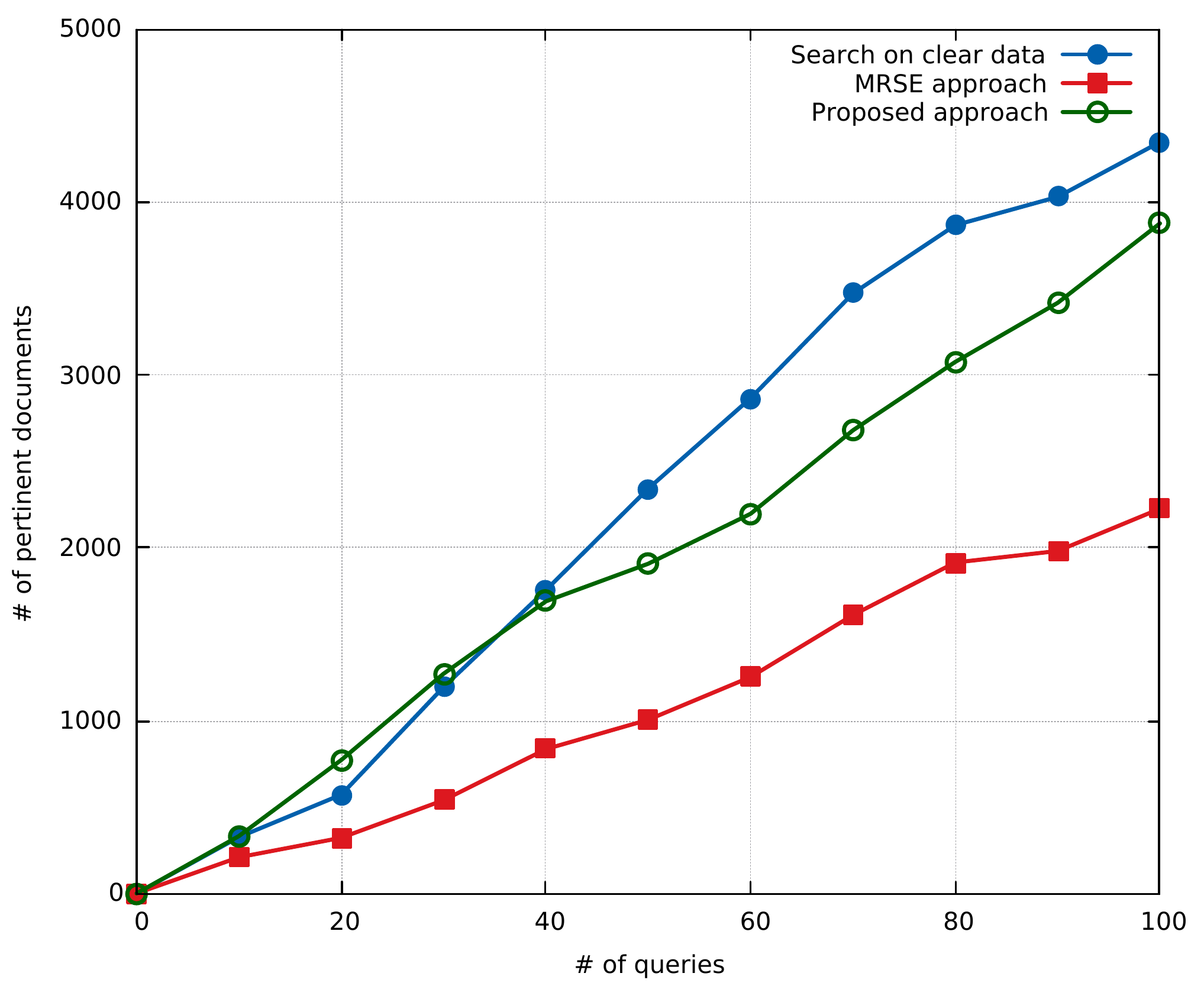}
\caption[The relevance of results in three different approaches]{Relevance of the returned results in three different approaches.}\label{pertinence}
\end{figure}

We have compared the proposed SIIS approach with two other approaches based on the relevance of the returned results. The first one is a traditional approach which enables performing a semantic search on an unencrypted index. The second approach is called MRSE and is based on the vector model. The MRSE approach allows to perform a syntactic search over an encrypted data collection (see Figure \ref{pertinence}).

\vspace{3mm}

Unsurprisingly, the semantic search on clear data gives the best results by reaching an accuracy of 43,46\%. The relevance of results returned by our approach is not far from those returned in the semantic search on clear data. Indeed, the SIIS approach that performs a semantic search through a secure inverted index is able to reach an accuracy of 38,76\%. Finally, the relevance of results returned by the MRSE method which performs a syntactic search over encrypted index based on the vector model does not exceed 23\%. This experiment clearly demonstrates that the proposed approach enables returning a very good quality of results which are not far in terms of accuracy from the results returned in a semantic search on clear data. 

\begin{figure}
\centering
\includegraphics[width=9cm,keepaspectratio]{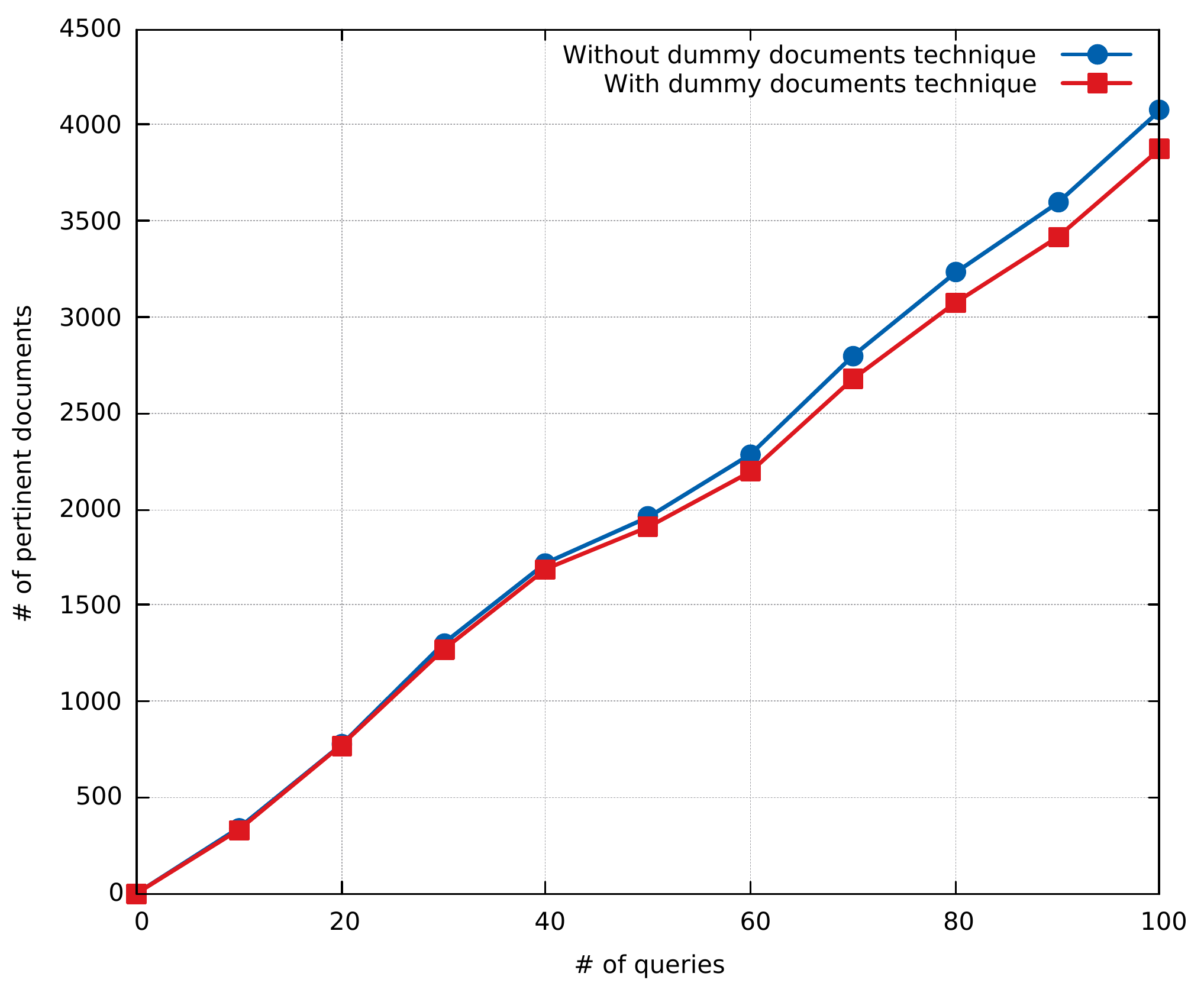}
\caption[The dummy documents technique]{Quality of the returned results in the proposed approach with and without applying the dummy documents technique.}\label{dummy}
\end{figure}

\begin{figure}
\centering
\includegraphics[width=9cm,keepaspectratio]{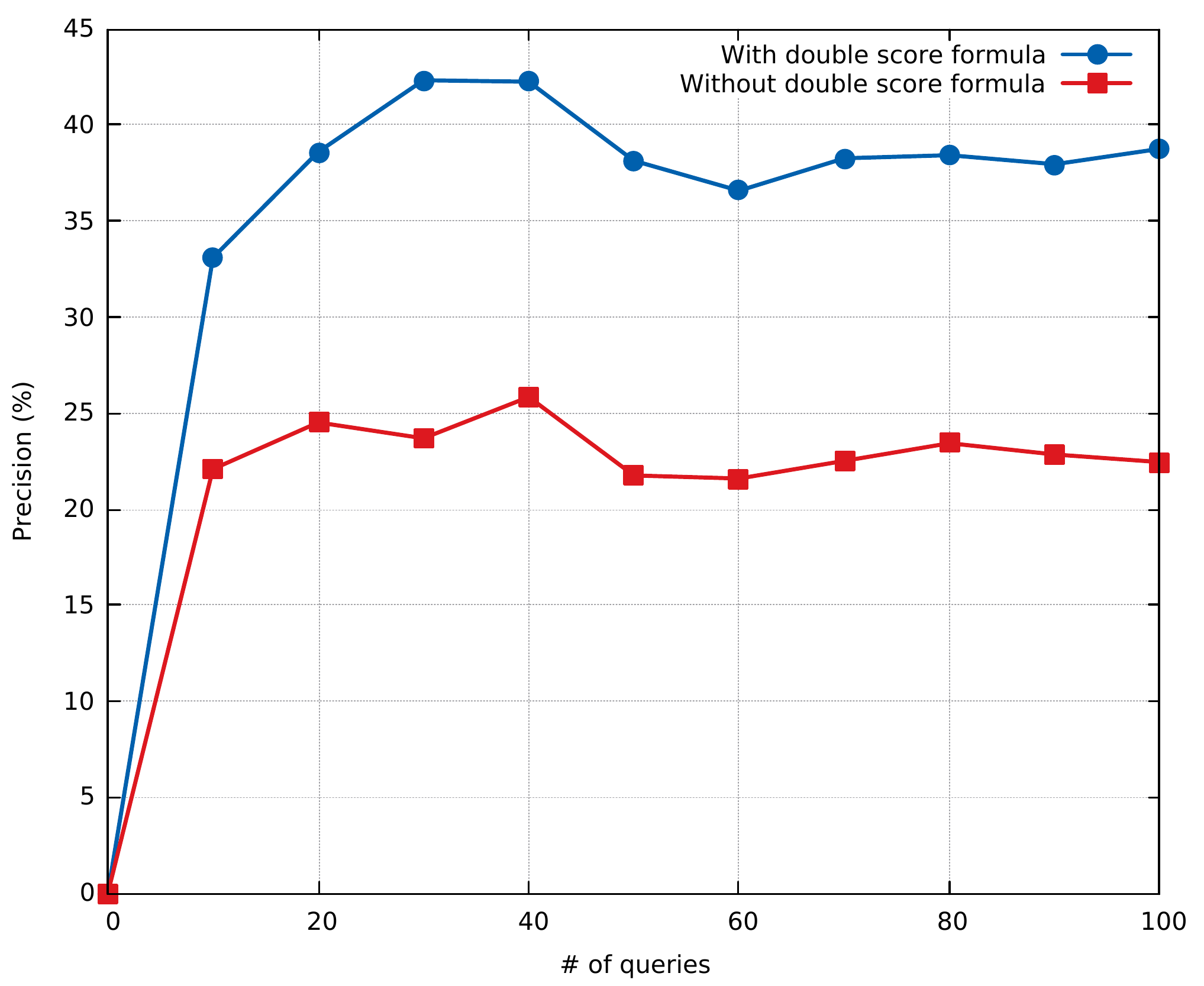}
\caption[The double score formula]{Quality of the returned results by the proposed approach with and without applying the double score formula.}\label{score}
\end{figure}

Then, we performed a second experiment in order to test the effect of the dummy documents technique on the quality of the returned results. For that, we tested the SIIS approach with the minimal security level ($K$ = 0) so that no dummy document is injected in the index. After that, we tested it with the maximal security level ($K$ = 10). Figure \ref{dummy} shows that the number of relevant documents decreases only shortly after the injection of dummy documents. In fact, the accuracy decreases from 40,76\% when $K$ is equal to 0, to 38,76\% when $K$ is equal to 10. The reason why the precision only decreases very shortly is due to the use of the double score formula which enables to {\color{black}discard} a very large number of irrelevant documents including the dummy documents (see Figure \ref{score}).

Finally, to measure the effectiveness of the double score formula, we carried out a third experiment where we tested two versions of our proposed approach. In the first version, we have tested the SIIS approach without using the double score formula, whereas, in the second version, the formula was exploited during the search process. Figure \ref{score} shows that the double score formula brings an improvement in the quality of the returned results by increasing the accuracy from 22,45\% to 38,76\%. This improvement is justified by the ability of the double score formula to {\color{black}filter} a great part of irrelevant documents.

\vspace{3mm}

\vspace{3mm}

\subsubsection{Search Time}

Concerning the search time, we tried to compare our approach with the MRSE approach which is based on the vector model. We tested our approach on the "Yahoo! Answers" collection which contains \numprint{962323} documents. However, we tested the MRSE approach only on \numprint{5000} documents due to insufficient memory space (we could not load more than \numprint{5000} document vectors of size equal to \numprint{409334}).

\begin{figure}
\centering
\includegraphics[width=9cm,keepaspectratio]{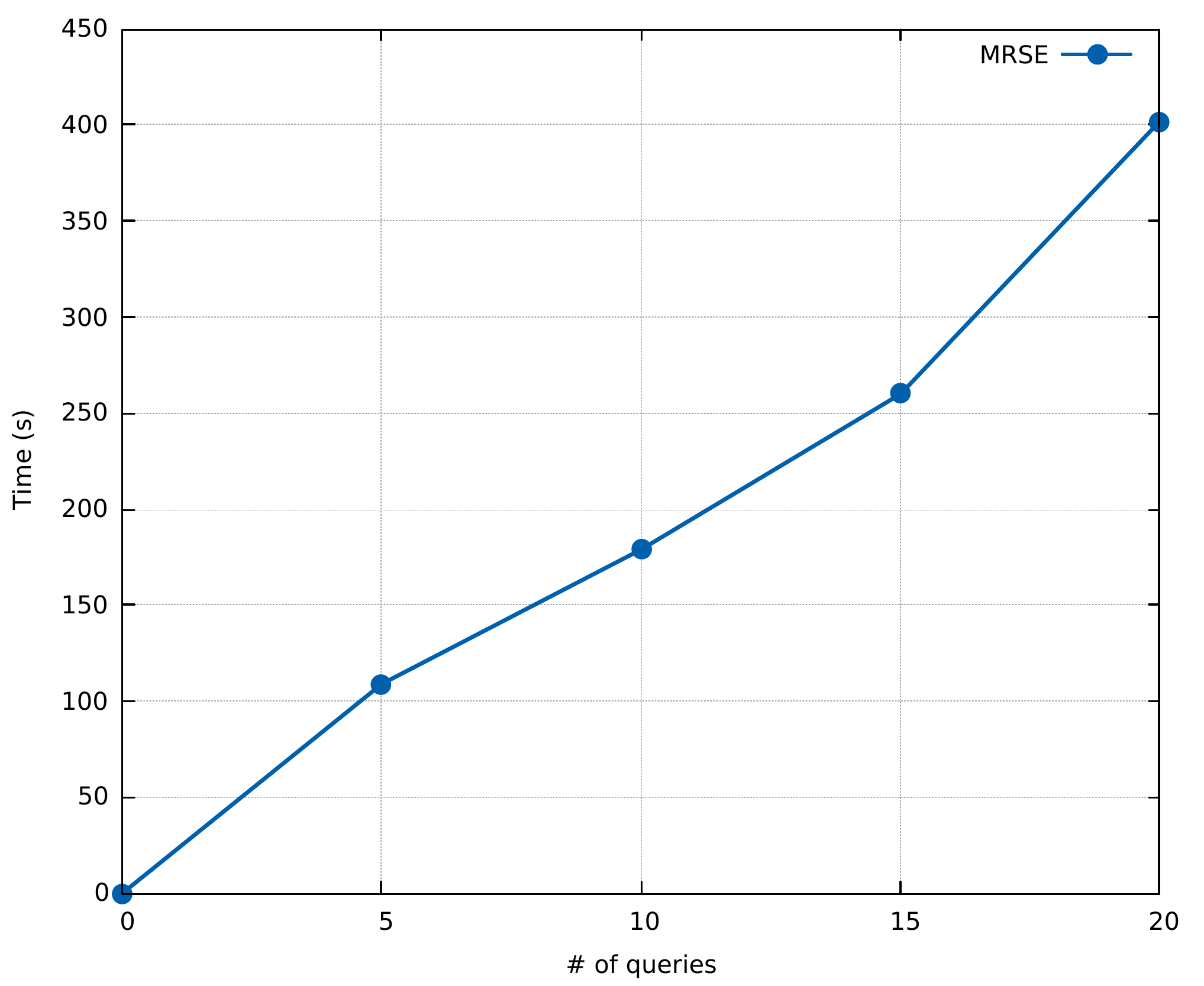}
\caption[The search time in the MRSE approach]{Time needed to perform 20 consecutive searches on a collection of \numprint{5000} documents using the MRSE method.}\label{mrse_search}
\end{figure}

{\color{black}On} one hand, Figure \ref{mrse_search} shows that in the MRSE approach, 402 seconds are necessary to perform 20 sequential searches on a collection of \numprint{5000} documents. {\color{black}Note that} the calculation of the similarity scores takes 84\% of the search time while sorting the documents takes the remaining time, therefore, the time required to perform 20 consecutive searches using the MRSE approach on a collection of \numprint{962323} documents is theoretically equal to 21,5 hours (\numprint{77371}s).

\vspace{3mm}

\begin{figure}
\centering
\includegraphics[width=9cm,keepaspectratio]{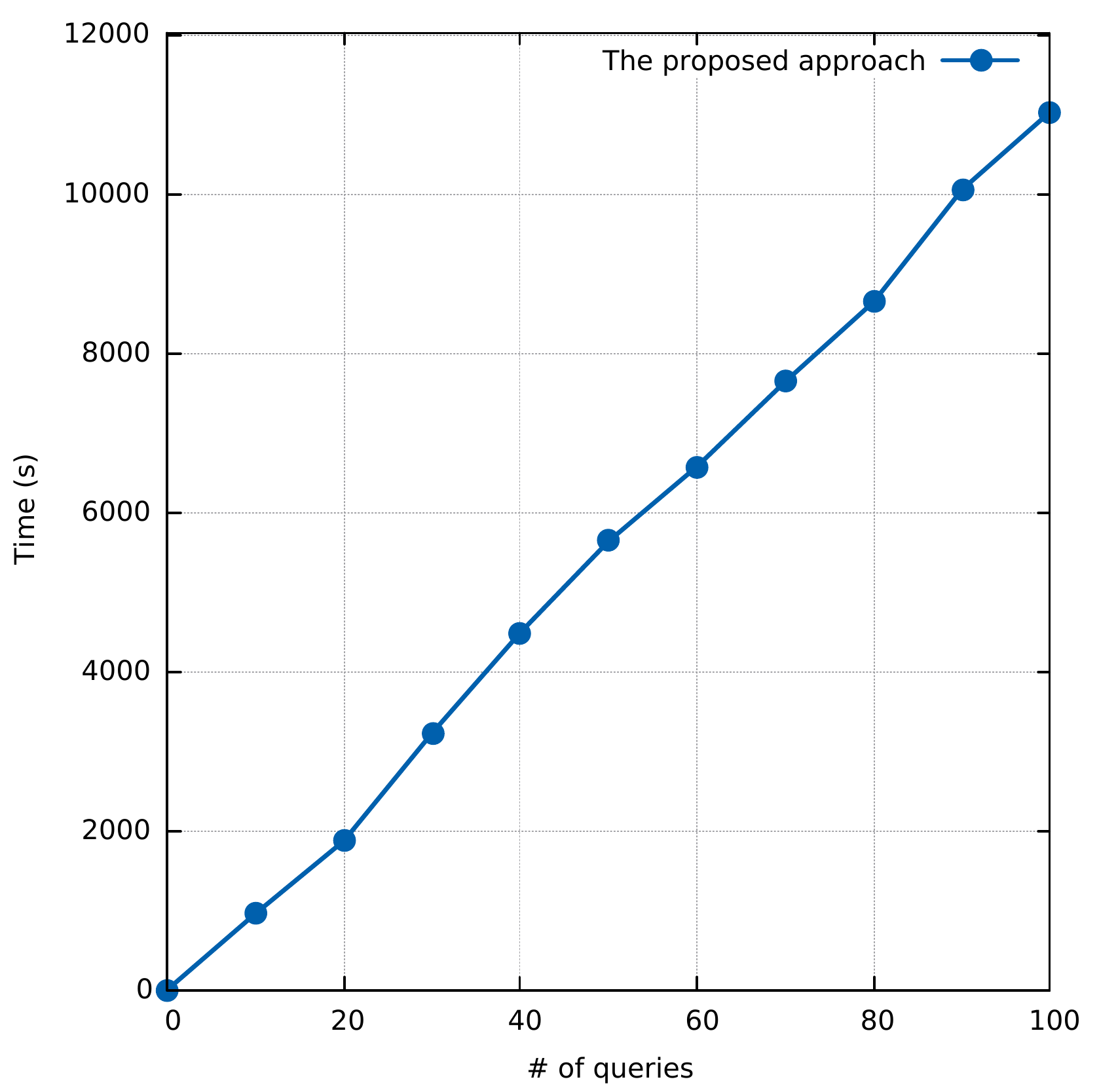}
\caption[The search time in the proposed approach]{Time needed to perform searches on a collection of 962,323 documents using our proposed approach.}\label{our_search}
\end{figure}

{\color{black}On} the other hand, Figure \ref{our_search} shows that the time required to perform 20 consecutive searches on a collection of 962,323 documents using our proposed approach is equal to 31 minutes (\numprint{1873}s) which makes the SIIS approach 41 times faster than the MRSE approach. Notice that the experiments were performed in a sequential way without any parallelism.

\vspace{3mm}

\begin{figure}
\centering
\includegraphics[width=9cm,keepaspectratio]{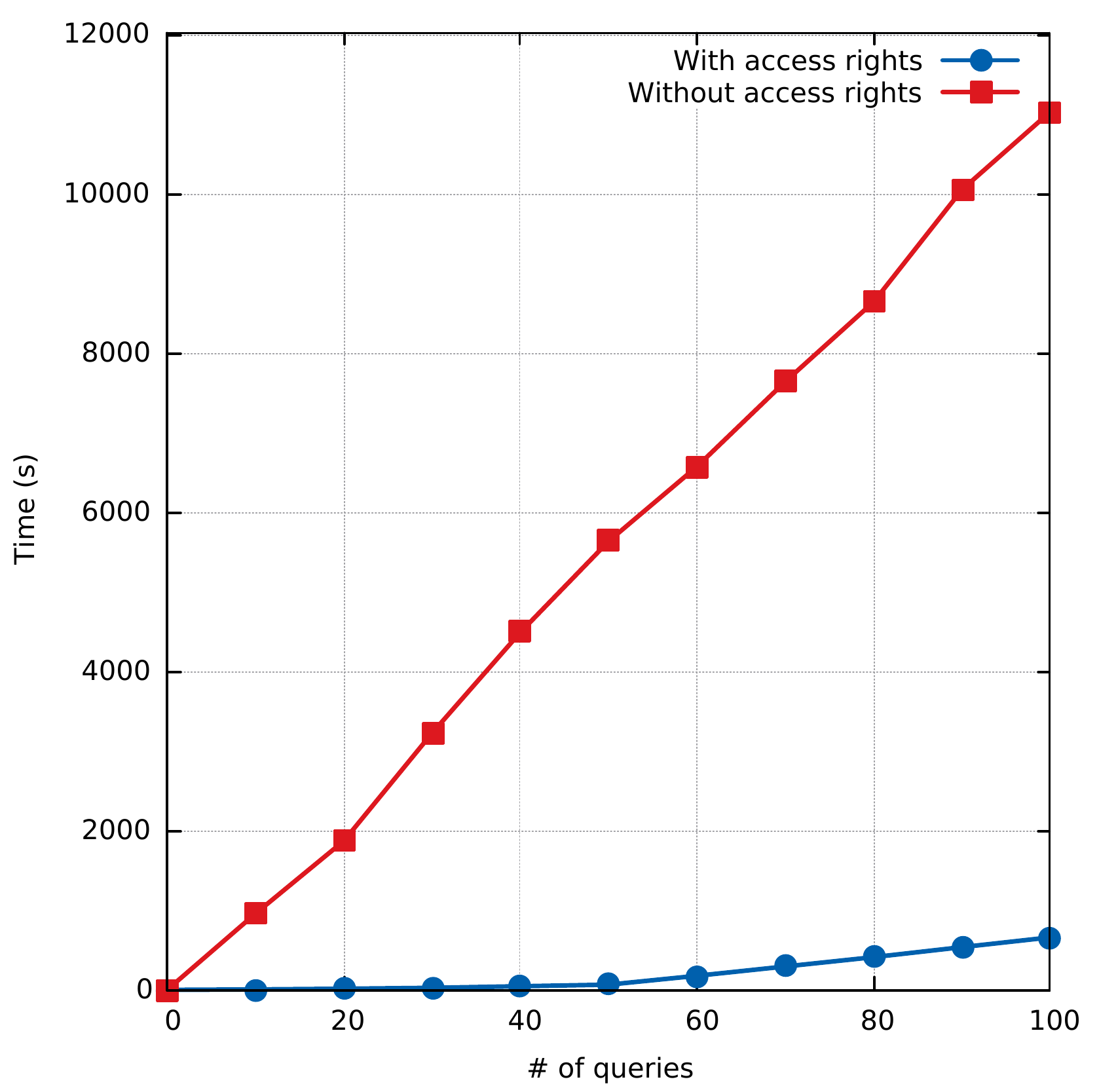}
\caption[The search time when considering the user access rights]{Time needed to perform searches with and without considering the user access rights.}\label{access}
\end{figure}

Finally, we tested the search time when exploiting a second index that is used to take into consideration the users' access rights. This experiment was performed on a user {\color{black}who has} the right to access 16 documents. In addition, 8 dummy documents were added to these documents after encrypting the index. Unlike Bouabana's approach \cite{bouabana2015parallel} where the server verifies the documents one by one to retrieve those that the user can access, in the SIIS approach, a second index is exploited which enables a direct access to these documents. Figure \ref{access} shows that the use of a second index brings an improvement in the search performance since the search space is reduced by keeping only the documents that the user has the right to access. The time saved during the search depends on the number of documents accessible by the user. In this experiment, we have compared two versions of our approach. In the first one, the users' access rights are not taken into account, whereas, in the second version, a second index is used to take into consideration the users' access rights. Figure \ref{access} shows that the search process is 16 times faster when considering the access rights.

\section{Summary}

The inverted index is barely used in the literature for security reasons and most approaches prefer to exploit the vector model. However, the later has a lot of drawbacks such as the slowdown of the search process, difficulty in updating data and generating very large indexes. In order to improve the search performance and make the search process more natural, we have proposed a method that enables constructing a secure inverted index using several techniques such as homomorphic encryption, the dummy documents technique, the compressed table of encrypted scores, and the double score formula. The proposed secure inverted index is the basis of the SIIS approach. After that, we have exploited a second secure inverted index in order to efficiently and securely manage the users' access rights to the data. The experimental study that we have performed demonstrates the benefit of the proposed approach in terms of performance in comparison with the approaches based on the vector model.

\def\baselinestretch{1}
\chapter{Conclusions}
\label{chap:conclusions}

\section{Summary}

In this thesis, we dealt with the problem of searchable encryption over encrypted cloud data. The challenge in this research area is how to retrieve the best results (accuracy and recall), within an acceptable time frame, while respecting the security constraints, and without causing any sensitive information leakage. Our first contribution consists in improving the quality of the search result in terms of recall and precision by adapting a semantic search to an encrypted environment. To make it possible, we have proposed a new weighting formula called the double score weighting (DSW) formula. This proposal allows to considerably improve the search accuracy and recall. In the second contribution, we improved the scheme performance by reducing the search time. In fact, we proposed 4 techniques that allow to perform the search process in a parallel way, and to deal with multiple queries simultaneously. Different HPC architectures are exploited by these techniques, such as the GPU, and the computer cluster. Applying these techniques enable to accelerate the SE scheme around a factor of forty.

Finally, our third contribution is a compromise between the quality of the results, the search performance, and the security of the scheme. First, the scheme performance in terms of search time is guaranteed through the use of the inverted index. Indeed, this index structure allows a direct access to the relevant documents without browsing through the entire index as is done in the vector space model which is widely used in the searchable encryption area (for security reasons). Second, the quality of the results is ensured by the use of the semantic search proposed in the first contribution. In fact, each entry of the inverted index corresponds to a concept\footnote{The concepts of Wikipedia ontology are exploited by the index instead of using the set of terms that belong to the data collection.} of Wikipedia ontology, and each concept leads to a set of documents that contain it. Finally, two techniques were exploited to secure the inverted index, namely, homomorphic encryption, and the dummy documents technique. Furthermore, we have proposed two additional techniques. The first one is used to compress the encrypted index, whereas, the second technique is necessary to filter the false positives. In addition, we have taken advantage of the secure inverted index, by exploiting a second index that allows to manage the users' access rights to the outsourced data.

\section{Perspectives}

As a perspective, we intend to take into consideration the data updates. In fact, when some documents are updated, inserted, or removed from the data collection, the encrypted index should be updated as well. However, the index update may cause sensitive information leakage. To avoid this problem, two new constraints have been proposed in the literature. The first constraint called "forward privacy" guarantees that old queries cannot be performed on inserted documents, whereas, the second one is called "backward privacy" and it makes sure that new queries cannot be applied on deleted data (see Section \ref{sec_constraintes}). Therefore, our upcoming goal is to adapt the SIIS approach (See Chapter \ref{Chapter6}) so that, backward privacy and forward privacy constraints will be respected when the secure inverted index is updated.

Our second perspective consists in proposing a personalized searchable encryption scheme. Indeed, our aim is to adapt the search result to the user's preferences, such as the languages, the document formats, and the area of interest. For this purpose, we think about implicitly creating encrypted profiles for each user by the cloud server from the search history. During the search process, the user profile is exploited in addition to the trapdoor in order to return the most appropriate search result. These profiles should be securely updated after each search operation.

\clearpage

\begin{center}
    {\LARGE \textbf{List of Publications}}
\end{center}

\begin{enumerate}
    \item Boucenna, F., Nouali, O., Kechid, S., \& Kechadi, M. T. (2019). Secure Inverted Index Based Search over Encrypted Cloud Data with User Access Rights Management. Journal of Computer Science and Technology, 34(1), 133-154.
    \item Boucenna, F., Nouali, O., Dabah, A., \& Kechid, S. (2017, February). Accelerated search over encrypted cloud data. In 2017 IEEE International Conference on Big Data and Smart Computing (BigComp) (pp. 170-177). IEEE.
    \item Boucenna, F., Nouali, O., \& Kechid, S. (2016, April). Concept-based Semantic Search over Encrypted Cloud Data. In WEBIST (2) (pp. 235-242).
\end{enumerate}

\bibliographystyle{ieeetr}
\renewcommand{\bibname}{References}           
\bibliography{References/references}          

\begin{thebibliography}{100}

\bibitem{boucenna2016concept}
F.~Boucenna, O.~Nouali, and S.~Kechid, ``Concept-based semantic search over
  encrypted cloud data.,'' in {\em WEBIST (2)}, pp.~235--242, 2016.

\bibitem{boucenna2017accelerated}
F.~Boucenna, O.~Nouali, A.~Dabah, and S.~Kechid, ``Accelerated search over
  encrypted cloud data,'' in {\em 2017 IEEE International Conference on Big
  Data and Smart Computing (BigComp)}, pp.~170--177, IEEE, 2017.

\bibitem{boucenna2019secure}
F.~Boucenna, O.~Nouali, S.~Kechid, and M.~T. Kechadi, ``Secure inverted index
  based search over encrypted cloud data with user access rights management,''
  {\em Journal of Computer Science and Technology}, vol.~34, no.~1,
  pp.~133--154, 2019.

\bibitem{wang:privacy}
B.~Wang, S.~Yu, W.~Lou, and Y.~T. Hou, ``Privacy-preserving multi-keyword fuzzy
  search over encrypted data in the cloud,'' in {\em INFOCOM, 2014 Proceedings
  IEEE}, pp.~2112--2120, IEEE, 2014.

\bibitem{cao2014privacy}
N.~Cao, C.~Wang, M.~Li, K.~Ren, and W.~Lou, ``Privacy-preserving multi-keyword
  ranked search over encrypted cloud data,'' {\em IEEE Transactions on parallel
  and distributed systems}, vol.~25, no.~1, pp.~222--233, 2014.

\bibitem{li:efficient}
K.~Li, W.~Zhang, K.~Tian, R.~Liu, and N.~Yu, ``An efficient multi-keyword
  ranked retrieval scheme with johnson-lindenstrauss transform over encrypted
  cloud data,'' in {\em 2013 International Conference on Cloud Computing and
  Big Data}, pp.~320--327, IEEE, 2013.

\bibitem{xia2016secure}
Z.~Xia, X.~Wang, X.~Sun, and Q.~Wang, ``A secure and dynamic multi-keyword
  ranked search scheme over encrypted cloud data.,'' {\em IEEE Trans. Parallel
  Distrib. Syst.}, vol.~27, no.~2, pp.~340--352, 2016.

\bibitem{song:practical}
D.~X. Song, D.~Wagner, and A.~Perrig, ``Practical techniques for searches on
  encrypted data,'' in {\em Security and Privacy, 2000. S\&P 2000. Proceedings.
  2000 IEEE Symposium on}, pp.~44--55, IEEE, 2000.

\bibitem{curtmola:searchable}
R.~Curtmola, J.~Garay, S.~Kamara, and R.~Ostrovsky, ``Searchable symmetric
  encryption: improved definitions and efficient constructions,'' in {\em
  Proceedings of the 13th ACM conference on Computer and communications
  security}, pp.~79--88, ACM, 2006.

\bibitem{kamara2013parallel}
S.~Kamara and C.~Papamanthou, ``Parallel and dynamic searchable symmetric
  encryption,'' in {\em International Conference on Financial Cryptography and
  Data Security}, pp.~258--274, Springer, 2013.

\bibitem{sharma2013survey}
M.~Sharma and R.~Patel, ``A survey on information retrieval models, techniques
  and applications,'' {\em International Journal of Emerging Technology and
  Advanced Engineering}, vol.~3, no.~11, pp.~542--545, 2013.

\bibitem{baeza2011modern}
R.~Baeza-Yates, B.~d. A.~N. Ribeiro, {\em et~al.}, {\em Modern information
  retrieval}.
\newblock New York: ACM Press; Harlow, England: Addison-Wesley,, 2011.

\bibitem{singhal2001modern}
A.~Singhal {\em et~al.}, ``Modern information retrieval: A brief overview,''
  {\em IEEE Data Eng. Bull.}, vol.~24, no.~4, pp.~35--43, 2001.

\bibitem{mogotsi2010christopher}
I.~Mogotsi, ``Christopher d. manning, prabhakar raghavan, and hinrich
  sch{\"u}tze: Introduction to information retrieval,'' 2010.

\bibitem{mcnamee2004character}
P.~Mcnamee and J.~Mayfield, ``Character n-gram tokenization for european
  language text retrieval,'' {\em Information retrieval}, vol.~7, no.~1-2,
  pp.~73--97, 2004.

\bibitem{wilbur1992automatic}
W.~J. Wilbur and K.~Sirotkin, ``The automatic identification of stop words,''
  {\em Journal of information science}, vol.~18, no.~1, pp.~45--55, 1992.

\bibitem{korenius2004stemming}
T.~Korenius, J.~Laurikkala, K.~J{\"a}rvelin, and M.~Juhola, ``Stemming and
  lemmatization in the clustering of finnish text documents,'' in {\em
  Proceedings of the thirteenth ACM international conference on Information and
  knowledge management}, pp.~625--633, ACM, 2004.

\bibitem{aizawa2003information}
A.~Aizawa, ``An information-theoretic perspective of tf--idf measures,'' {\em
  Information Processing \& Management}, vol.~39, no.~1, pp.~45--65, 2003.

\bibitem{croft2010search}
W.~B. Croft, D.~Metzler, and T.~Strohman, {\em Search engines: Information
  retrieval in practice}, vol.~283.
\newblock Addison-Wesley Reading, 2010.

\bibitem{alshari2015semantic}
E.~M. Alshari, ``Semantic arabic information retrieval framework,'' {\em arXiv
  preprint arXiv:1512.03165}, 2015.

\bibitem{nomula2010system}
J.~R. Nomula and C.~Stelzmuller, ``System and method for enhanced text
  matching,'' 2010.
\newblock US Patent 7,783,660.

\bibitem{ismalon2014techniques}
E.~Ismalon, ``Techniques for ranking search results,'' 2014.
\newblock US Patent 8,903,810.

\bibitem{salton1975vector}
G.~Salton, A.~Wong, and C.-S. Yang, ``A vector space model for automatic
  indexing,'' {\em Communications of the ACM}, vol.~18, no.~11, pp.~613--620,
  1975.

\bibitem{choi2010survey}
S.-S. Choi, S.-H. Cha, and C.~C. Tappert, ``A survey of binary similarity and
  distance measures,'' {\em Journal of Systemics, Cybernetics and Informatics},
  vol.~8, no.~1, pp.~43--48, 2010.

\bibitem{ponte1998language}
J.~M. Ponte and W.~B. Croft, ``A language modeling approach to information
  retrieval,'' in {\em Proceedings of the 21st annual international ACM SIGIR
  conference on Research and development in information retrieval},
  pp.~275--281, ACM, 1998.

\bibitem{wiki:xxx}
W.~Commons, ``File:precisionrecall.svg --- wikimedia commons{,} the free media
  repository,'' 2016.
\newblock [Online; accessed 19-December-2018].

\bibitem{muller2004textpresso}
H.-M. M{\"u}ller, E.~E. Kenny, and P.~W. Sternberg, ``Textpresso: an
  ontology-based information retrieval and extraction system for biological
  literature,'' {\em PLoS biology}, vol.~2, no.~11, p.~e309, 2004.

\bibitem{castells2007adaptation}
P.~Castells, M.~Fern{\'a}ndez~S{\'a}nchez, and D.~J. Vallet~Weadon, ``An
  adaptation of the vector-space model for ontology based information
  retrieval,'' {\em IEEE transactions on knowledge and data engineering}, 2007.

\bibitem{fernandez2011semantically}
M.~Fern{\'a}ndez, I.~Cantador, V.~L{\'o}pez, D.~Vallet, P.~Castells, and
  E.~Motta, ``Semantically enhanced information retrieval: An ontology-based
  approach,'' {\em Web semantics: Science, services and agents on the world
  wide web}, vol.~9, no.~4, pp.~434--452, 2011.

\bibitem{ogden1923meaning}
C.~K. Ogden, I.~A. Richards, B.~Malinowski, and F.~G. Crookshank, {\em The
  meaning of meaning}.
\newblock Kegan Paul London, 1923.

\bibitem{gabrilovich2006overcoming}
E.~Gabrilovich and S.~Markovitch, ``Overcoming the brittleness bottleneck using
  wikipedia: Enhancing text categorization with encyclopedic knowledge,'' in
  {\em AAAI}, vol.~6, pp.~1301--1306, 2006.

\bibitem{miller1998wordnet}
G.~Miller, {\em WordNet: An electronic lexical database}.
\newblock MIT press, 1998.

\bibitem{finkelstein2002placing}
L.~Finkelstein, E.~Gabrilovich, Y.~Matias, E.~Rivlin, Z.~Solan, G.~Wolfman, and
  E.~Ruppin, ``Placing search in context: The concept revisited,'' {\em ACM
  Transactions on information systems}, vol.~20, no.~1, pp.~116--131, 2002.

\bibitem{liu2004effective}
S.~Liu, F.~Liu, C.~Yu, and W.~Meng, ``An effective approach to document
  retrieval via utilizing wordnet and recognizing phrases,'' in {\em
  Proceedings of the 27th annual international ACM SIGIR conference on Research
  and development in information retrieval}, pp.~266--272, ACM, 2004.

\bibitem{kruse2005clever}
P.~M. Kruse, A.~Naujoks, D.~R{\"o}sner, and M.~Kunze, ``Clever search: A
  wordnet based wrapper for internet search engines,'' {\em arXiv preprint
  cs/0501086}, 2005.

\bibitem{varelas2005semantic}
G.~Varelas, E.~Voutsakis, P.~Raftopoulou, E.~G. Petrakis, and E.~E. Milios,
  ``Semantic similarity methods in wordnet and their application to information
  retrieval on the web,'' in {\em Proceedings of the 7th annual ACM
  international workshop on Web information and data management}, pp.~10--16,
  ACM, 2005.

\bibitem{mell2011nist}
P.~Mell, T.~Grance, {\em et~al.}, ``The nist definition of cloud computing,''
  2011.

\bibitem{wiki:cloud}
W.~Commons, ``File:cloud computing types.svg --- wikimedia commons{,} the free
  media repository,'' 2019.
\newblock [Online; accessed 15-May-2019].

\bibitem{zissis2012addressing}
D.~Zissis and D.~Lekkas, ``Addressing cloud computing security issues,'' {\em
  Future Generation computer systems}, vol.~28, no.~3, pp.~583--592, 2012.

\bibitem{xu:two}
J.~Xu, W.~Zhang, C.~Yang, J.~Xu, and N.~Yu, ``Two-step-ranking secure
  multi-keyword search over encrypted cloud data,'' in {\em Cloud and Service
  Computing (CSC), 2012 International Conference on}, pp.~124--130, IEEE, 2012.

\bibitem{yu:toward}
J.~Yu, P.~Lu, Y.~Zhu, G.~Xue, and M.~Li, ``Toward secure multikeyword top-k
  retrieval over encrypted cloud data,'' {\em Dependable and Secure Computing,
  IEEE Transactions on}, vol.~10, no.~4, pp.~239--250, 2013.

\bibitem{xia2018towards}
Z.~Xia, Y.~Zhu, X.~Sun, Z.~Qin, and K.~Ren, ``Towards privacy-preserving
  content-based image retrieval in cloud computing,'' {\em IEEE Transactions on
  Cloud Computing}, no.~1, pp.~276--286, 2018.

\bibitem{gentry2009fully}
C.~Gentry and D.~Boneh, {\em A fully homomorphic encryption scheme}, vol.~20.
\newblock Stanford University Stanford, 2009.

\bibitem{wong2009secure}
W.~K. Wong, D.~W.-l. Cheung, B.~Kao, and N.~Mamoulis, ``Secure knn computation
  on encrypted databases,'' in {\em Proceedings of the 2009 ACM SIGMOD
  International Conference on Management of data}, pp.~139--152, ACM, 2009.

\bibitem{islam2012access}
M.~S. Islam, M.~Kuzu, and M.~Kantarcioglu, ``Access pattern disclosure on
  searchable encryption: Ramification, attack and mitigation.,'' in {\em Ndss},
  vol.~20, p.~12, 2012.

\bibitem{liu2014search}
C.~Liu, L.~Zhu, M.~Wang, and Y.-a. Tan, ``Search pattern leakage in searchable
  encryption: Attacks and new construction,'' {\em Information Sciences},
  vol.~265, pp.~176--188, 2014.

\bibitem{stefanov2014practical}
E.~Stefanov, C.~Papamanthou, and E.~Shi, ``Practical dynamic searchable
  encryption with small leakage.,'' in {\em NDSS}, vol.~71, pp.~72--75, 2014.

\bibitem{bosch2015survey}
C.~B{\"o}sch, P.~Hartel, W.~Jonker, and A.~Peter, ``A survey of provably secure
  searchable encryption,'' {\em ACM Computing Surveys (CSUR)}, vol.~47, no.~2,
  p.~18, 2015.

\bibitem{goh2003secure}
E.-J. Goh {\em et~al.}, ``Secure indexes.,'' {\em IACR Cryptology ePrint
  Archive}, vol.~2003, p.~216, 2003.

\bibitem{chang2005privacy}
Y.-C. Chang and M.~Mitzenmacher, ``Privacy preserving keyword searches on
  remote encrypted data,'' in {\em International Conference on Applied
  Cryptography and Network Security}, pp.~442--455, Springer, 2005.

\bibitem{curtmola2011searchable}
R.~Curtmola, J.~Garay, S.~Kamara, and R.~Ostrovsky, ``Searchable symmetric
  encryption: improved definitions and efficient constructions,'' {\em Journal
  of Computer Security}, vol.~19, no.~5, pp.~895--934, 2011.

\bibitem{boneh2004public}
D.~Boneh, G.~Di~Crescenzo, R.~Ostrovsky, and G.~Persiano, ``Public key
  encryption with keyword search,'' in {\em International conference on the
  theory and applications of cryptographic techniques}, pp.~506--522, Springer,
  2004.

\bibitem{raykova2009secure}
M.~Raykova, B.~Vo, S.~M. Bellovin, and T.~Malkin, ``Secure anonymous database
  search,'' in {\em Proceedings of the 2009 ACM workshop on Cloud computing
  security}, pp.~115--126, ACM, 2009.

\bibitem{bloom1970space}
B.~H. Bloom, ``Space/time trade-offs in hash coding with allowable errors,''
  {\em Communications of the ACM}, vol.~13, no.~7, pp.~422--426, 1970.

\bibitem{bellare2007deterministic}
M.~Bellare, A.~Boldyreva, and A.~O’Neill, ``Deterministic and efficiently
  searchable encryption,'' in {\em Annual International Cryptology Conference},
  pp.~535--552, Springer, 2007.

\bibitem{wang2010secure}
C.~Wang, N.~Cao, J.~Li, K.~Ren, and W.~Lou, ``Secure ranked keyword search over
  encrypted cloud data,'' in {\em Distributed Computing Systems (ICDCS), 2010
  IEEE 30th International Conference on}, pp.~253--262, IEEE, 2010.

\bibitem{boldyreva2009order}
A.~Boldyreva, N.~Chenette, Y.~Lee, and A.~O’neill, ``Order-preserving
  symmetric encryption,'' in {\em Annual International Conference on the Theory
  and Applications of Cryptographic Techniques}, pp.~224--241, Springer, 2009.

\bibitem{van2010fully}
M.~Van~Dijk, C.~Gentry, S.~Halevi, and V.~Vaikuntanathan, ``Fully homomorphic
  encryption over the integers,'' in {\em Annual International Conference on
  the Theory and Applications of Cryptographic Techniques}, pp.~24--43,
  Springer, 2010.

\bibitem{elmehdwi2014secure}
Y.~Elmehdwi, B.~K. Samanthula, and W.~Jiang, ``Secure k-nearest neighbor query
  over encrypted data in outsourced environments,'' in {\em Data Engineering
  (ICDE), 2014 IEEE 30th International Conference on}, pp.~664--675, IEEE,
  2014.

\bibitem{paillier1999public}
P.~Paillier, ``Public-key cryptosystems based on composite degree residuosity
  classes,'' in {\em International Conference on the Theory and Applications of
  Cryptographic Techniques}, pp.~223--238, Springer, 1999.

\bibitem{wang2015inverted}
B.~Wang, W.~Song, W.~Lou, and Y.~T. Hou, ``Inverted index based multi-keyword
  public-key searchable encryption with strong privacy guarantee,'' in {\em
  Computer Communications (INFOCOM), 2015 IEEE Conference on}, pp.~2092--2100,
  IEEE, 2015.

\bibitem{li2010fuzzy}
J.~Li, Q.~Wang, C.~Wang, N.~Cao, K.~Ren, and W.~Lou, ``Fuzzy keyword search
  over encrypted data in cloud computing,'' in {\em Infocom, 2010 proceedings
  ieee}, pp.~1--5, IEEE, 2010.

\bibitem{levenshtein1966binary}
V.~I. Levenshtein, ``Binary codes capable of correcting deletions, insertions,
  and reversals,'' in {\em Soviet physics doklady}, vol.~10, pp.~707--710,
  1966.

\bibitem{sun:secure}
X.~Sun, Y.~Zhu, Z.~Xia, J.~Wang, and L.~Chen, ``Secure keyword-based ranked
  semantic search over encrypted cloud data,'' {\em Proceedings of the Advanced
  Science and Technology Letters (MulGraB 2013)}, vol.~31, pp.~271--283, 2013.

\bibitem{fu2014semantic}
Z.~Fu, J.~Shu, X.~Sun, and D.~Zhang, ``Semantic keyword search based on trie
  over encrypted cloud data,'' in {\em Proceedings of the 2nd international
  workshop on security in cloud computing}, pp.~59--62, ACM, 2014.

\bibitem{porter1980algorithm}
M.~F. Porter, ``An algorithm for suffix stripping,'' {\em Program}, vol.~14,
  no.~3, pp.~130--137, 1980.

\bibitem{fu2016enabling}
Z.~Fu, F.~Huang, X.~Sun, A.~Vasilakos, and C.-N. Yang, ``Enabling semantic
  search based on conceptual graphs over encrypted outsourced data,'' {\em IEEE
  Transactions on Services Computing}, 2016.

\bibitem{fu2016towards}
Z.~Fu, X.~Sun, S.~Ji, and G.~Xie, ``Towards efficient content-aware search over
  encrypted outsourced data in cloud,'' in {\em Computer communications, IEEE
  INFOCOM 2016-the 35th annual IEEE international conference on}, pp.~1--9,
  IEEE, 2016.

\bibitem{kamara2012dynamic}
S.~Kamara, C.~Papamanthou, and T.~Roeder, ``Dynamic searchable symmetric
  encryption,'' in {\em Proceedings of the 2012 ACM conference on Computer and
  communications security}, pp.~965--976, ACM, 2012.

\bibitem{hahn2014searchable}
F.~Hahn and F.~Kerschbaum, ``Searchable encryption with secure and efficient
  updates,'' in {\em Proceedings of the 2014 ACM SIGSAC Conference on Computer
  and Communications Security}, pp.~310--320, ACM, 2014.

\bibitem{bao2008private}
F.~Bao, R.~H. Deng, X.~Ding, and Y.~Yang, ``Private query on encrypted data in
  multi-user settings,'' in {\em International Conference on Information
  Security Practice and Experience}, pp.~71--85, Springer, 2008.

\bibitem{rhee2010trapdoor}
H.~S. Rhee, J.~H. Park, W.~Susilo, and D.~H. Lee, ``Trapdoor security in a
  searchable public-key encryption scheme with a designated tester,'' {\em
  Journal of Systems and Software}, vol.~83, no.~5, pp.~763--771, 2010.

\bibitem{baek2008public}
J.~Baek, R.~Safavi-Naini, and W.~Susilo, ``Public key encryption with keyword
  search revisited,'' in {\em International conference on Computational Science
  and Its Applications}, pp.~1249--1259, Springer, 2008.

\bibitem{byun2006off}
J.~W. Byun, H.~S. Rhee, H.-A. Park, and D.~H. Lee, ``Off-line keyword guessing
  attacks on recent keyword search schemes over encrypted data,'' in {\em
  Workshop on Secure Data Management}, pp.~75--83, Springer, 2006.

\bibitem{li2011authorized}
M.~Li, S.~Yu, N.~Cao, and W.~Lou, ``Authorized private keyword search over
  encrypted data in cloud computing,'' in {\em Distributed Computing Systems
  (ICDCS), 2011 31st International Conference on}, pp.~383--392, IEEE, 2011.

\bibitem{okamoto2009hierarchical}
T.~Okamoto and K.~Takashima, ``Hierarchical predicate encryption for
  inner-products,'' in {\em International Conference on the Theory and
  Application of Cryptology and Information Security}, pp.~214--231, Springer,
  2009.

\bibitem{zhao2011multi}
F.~Zhao, T.~Nishide, and K.~Sakurai, ``Multi-user keyword search scheme for
  secure data sharing with fine-grained access control,'' in {\em International
  Conference on Information Security and Cryptology}, pp.~406--418, Springer,
  2011.

\bibitem{bethencourt2007ciphertext}
J.~Bethencourt, A.~Sahai, and B.~Waters, ``Ciphertext-policy attribute-based
  encryption,'' in {\em 2007 IEEE symposium on security and privacy (SP'07)},
  pp.~321--334, IEEE, 2007.

\bibitem{bouabana2015parallel}
T.~Bouabana-Tebibel and A.~Kaci, ``Parallel search over encrypted data under
  attribute based encryption on the cloud computing,'' {\em Computers \&
  Security}, vol.~54, pp.~77--91, 2015.

\bibitem{goyal2006attribute}
V.~Goyal, O.~Pandey, A.~Sahai, and B.~Waters, ``Attribute-based encryption for
  fine-grained access control of encrypted data,'' in {\em Proceedings of the
  13th ACM conference on Computer and communications security}, pp.~89--98,
  Acm, 2006.

\bibitem{yuan2015seisa}
J.~Yuan, S.~Yu, and L.~Guo, ``Seisa: Secure and efficient encrypted image
  search with access control,'' in {\em Computer Communications (INFOCOM), 2015
  IEEE Conference on}, pp.~2083--2091, IEEE, 2015.

\bibitem{deng2016multi}
Z.~Deng, K.~Li, K.~Li, and J.~Zhou, ``A multi-user searchable encryption scheme
  with keyword authorization in a cloud storage,'' {\em Future Generation
  Computer Systems}, 2016.

\bibitem{johnson1984extensions}
W.~B. Johnson and J.~Lindenstrauss, ``Extensions of lipschitz mappings into a
  hilbert space,'' {\em Contemporary mathematics}, vol.~26, no.~189-206, p.~1,
  1984.

\bibitem{fu2015achieving}
Z.~Fu, X.~Sun, Q.~Liu, L.~Zhou, and J.~Shu, ``Achieving efficient cloud search
  services: multi-keyword ranked search over encrypted cloud data supporting
  parallel computing,'' {\em IEICE Transactions on Communications}, vol.~98,
  no.~1, pp.~190--200, 2015.

\bibitem{xia2017epcbir}
Z.~Xia, N.~N. Xiong, A.~V. Vasilakos, and X.~Sun, ``Epcbir: An efficient and
  privacy-preserving content-based image retrieval scheme in cloud computing,''
  {\em Information Sciences}, vol.~387, pp.~195--204, 2017.

\bibitem{van2010computationally}
P.~Van~Liesdonk, S.~Sedghi, J.~Doumen, P.~Hartel, and W.~Jonker,
  ``Computationally efficient searchable symmetric encryption,'' in {\em
  Workshop on Secure Data Management}, pp.~87--100, Springer, 2010.

\bibitem{bosch2012selective}
C.~B{\"o}sch, Q.~Tang, P.~Hartel, and W.~Jonker, ``Selective document retrieval
  from encrypted database,'' in {\em International Conference on Information
  Security}, pp.~224--241, Springer, 2012.

\bibitem{brakerski2011efficient}
Z.~Brakerski and V.~Vaikuntanathan, ``Efficient fully homomorphic encryption
  from (standard) lwe,'' in {\em Proceedings of the 2011 IEEE 52nd Annual
  Symposium on Foundations of Computer Science}, pp.~97--106, IEEE Computer
  Society, 2011.

\bibitem{cash2013highly}
D.~Cash, S.~Jarecki, C.~Jutla, H.~Krawczyk, M.-C. Ro{\c{s}}u, and M.~Steiner,
  ``Highly-scalable searchable symmetric encryption with support for boolean
  queries,'' in {\em Advances in cryptology--CRYPTO 2013}, pp.~353--373,
  Springer, 2013.

\bibitem{sahai2005fuzzy}
A.~Sahai and B.~Waters, ``Fuzzy identity-based encryption,'' in {\em Annual
  International Conference on the Theory and Applications of Cryptographic
  Techniques}, pp.~457--473, Springer, 2005.

\bibitem{emura2009ciphertext}
K.~Emura, A.~Miyaji, A.~Nomura, K.~Omote, and M.~Soshi, ``A ciphertext-policy
  attribute-based encryption scheme with constant ciphertext length,'' in {\em
  International Conference on Information Security Practice and Experience},
  pp.~13--23, Springer, 2009.

\bibitem{ibraimi2009efficient}
L.~Ibraimi, Q.~Tang, P.~Hartel, and W.~Jonker, ``Efficient and provable secure
  ciphertext-policy attribute-based encryption schemes,'' in {\em International
  Conference on Information Security Practice and Experience}, pp.~1--12,
  Springer, 2009.

\bibitem{smart2010fully}
N.~P. Smart and F.~Vercauteren, ``Fully homomorphic encryption with relatively
  small key and ciphertext sizes,'' in {\em International Workshop on Public
  Key Cryptography}, pp.~420--443, Springer, 2010.

\bibitem{brakerski2012leveled}
Z.~Brakerski, C.~Gentry, and V.~Vaikuntanathan, ``(leveled) fully homomorphic
  encryption without bootstrapping,'' in {\em Proceedings of the 3rd
  Innovations in Theoretical Computer Science Conference}, pp.~309--325, ACM,
  2012.

\bibitem{okamoto1998new}
T.~Okamoto and S.~Uchiyama, ``A new public-key cryptosystem as secure as
  factoring,'' in {\em International conference on the theory and applications
  of cryptographic techniques}, pp.~308--318, Springer, 1998.

\bibitem{gentry2011implementing}
C.~Gentry and S.~Halevi, ``Implementing gentry’s fully-homomorphic encryption
  scheme,'' in {\em Annual international conference on the theory and
  applications of cryptographic techniques}, pp.~129--148, Springer, 2011.

\bibitem{naehrig2011can}
M.~Naehrig, K.~Lauter, and V.~Vaikuntanathan, ``Can homomorphic encryption be
  practical?,'' in {\em Proceedings of the 3rd ACM workshop on Cloud computing
  security workshop}, pp.~113--124, ACM, 2011.

\bibitem{brakerski2011fully}
Z.~Brakerski and V.~Vaikuntanathan, ``Fully homomorphic encryption from
  ring-lwe and security for key dependent messages,'' in {\em Annual cryptology
  conference}, pp.~505--524, Springer, 2011.

\bibitem{brakerski2012fully}
Z.~Brakerski, ``Fully homomorphic encryption without modulus switching from
  classical gapsvp,'' in {\em Advances in cryptology--crypto 2012},
  pp.~868--886, Springer, 2012.

\bibitem{gentry2013homomorphic}
C.~Gentry, A.~Sahai, and B.~Waters, ``Homomorphic encryption from learning with
  errors: Conceptually-simpler, asymptotically-faster, attribute-based,'' in
  {\em Advances in Cryptology--CRYPTO 2013}, pp.~75--92, Springer, 2013.

\bibitem{howgrave2001approximate}
N.~Howgrave-Graham, ``Approximate integer common divisors,'' in {\em
  Cryptography and Lattices}, pp.~51--66, Springer, 2001.

\bibitem{lyubashevsky2010ideal}
V.~Lyubashevsky, C.~Peikert, and O.~Regev, ``On ideal lattices and learning
  with errors over rings,'' in {\em Annual International Conference on the
  Theory and Applications of Cryptographic Techniques}, pp.~1--23, Springer,
  2010.

\bibitem{yang2015attribute}
Y.~Yang, ``Attribute-based data retrieval with semantic keyword search for
  e-health cloud,'' {\em Journal of Cloud Computing}, vol.~4, no.~1, p.~10,
  2015.

\bibitem{egozi:concept}
O.~Egozi, S.~Markovitch, and E.~Gabrilovich, ``Concept-based information
  retrieval using explicit semantic analysis,'' {\em ACM Transactions on
  Information Systems (TOIS)}, vol.~29, no.~2, p.~8, 2011.

\bibitem{cuda:programming}
NVIDIA, ``Cuda toolkit documentation - nvidia.''
  \url{http://docs.nvidia.com/cuda/cuda-c-programming-guide/}.
\newblock Accessed: 2018-10-30.

\bibitem{yang2014secure}
Y.~Yang, H.~Li, W.~Liu, H.~Yao, and M.~Wen, ``Secure dynamic searchable
  symmetric encryption with constant document update cost,'' in {\em Global
  Communications Conference (GLOBECOM), 2014 IEEE}, pp.~775--780, IEEE, 2014.

\bibitem{wang2018searchable}
Q.~Wang, M.~He, M.~Du, S.~S. Chow, R.~W. Lai, and Q.~Zou, ``Searchable
  encryption over feature-rich data,'' {\em IEEE Transactions on Dependable and
  Secure Computing}, vol.~15, no.~3, pp.~496--510, 2018.

\bibitem{li2018personalized}
H.~Li, D.~Liu, Y.~Dai, T.~H. Luan, and S.~Yu, ``Personalized search over
  encrypted data with efficient and secure updates in mobile clouds,'' {\em
  IEEE Transactions on Emerging Topics in Computing}, vol.~6, no.~1,
  pp.~97--109, 2018.

\bibitem{naveed2014dynamic}
M.~Naveed, M.~Prabhakaran, and C.~A. Gunter, ``Dynamic searchable encryption
  via blind storage,'' in {\em 2014 IEEE Symposium on Security and Privacy},
  pp.~639--654, IEEE, 2014.

\bibitem{regev2009lattices}
O.~Regev, ``On lattices, learning with errors, random linear codes, and
  cryptography,'' {\em Journal of the ACM (JACM)}, vol.~56, no.~6, p.~34, 2009.

\end{thebibliography}
\addcontentsline{toc}{chapter}{References}    
\nocite{*}
\end{document}